\documentclass[twocolumn,showpacs,amsmath,amssymb]{revtex4}

\usepackage{graphicx}
\usepackage{dcolumn}
\usepackage{bm}
\newcommand{\etal}{\emph{et al.}}
\newcommand{\be}{\begin{equation}}
\newcommand{\ee}{\end{equation}}
\newcommand{\bfig}{\begin{figure}}
\newcommand{\efig}{\end{figure}}
\newcommand{\incl}{\includegraphics}

\begin{document}


\title{The Nernst effect in high-$T_c$ superconductors}

\author{Yayu Wang$^*$, Lu Li and N. P. Ong}

\affiliation{Department of Physics, Princeton University, Princeton, New Jersey 08544.}

\date{\today}


\begin{abstract}
The observation of a large Nernst signal $e_N$ in an extended region above 
the critical temperature $T_c$ in hole-doped cuprates provides evidence that 
vortex excitations survive above $T_c$.  The results support the scenario that 
superfluidity vanishes because long-range phase coherence is destroyed by
thermally-created vortices (in zero field), and that the pair condensate extends 
high into the pseudogap state in the underdoped (UD) regime.  We present a series 
of measurements to high fields $H$ which provide strong evidence for this 
phase-disordering scenario.  Measurements of $e_N$ in fields $H$ up to
45 T reveal that the vortex Nernst signal has a characteristic ``tilted-hill" profile, 
which is qualitatively distinct from that of quasi-particles.  The hill profile, 
which is observed above and below $T_c$, underscores the continuity between the 
vortex-liquid state below $T_c$ and the Nernst region above $T_c$.  The upper 
critical field (depairing field) $H_{c2}$ determined by the hill profile (in slightly 
UD to overdoped samples) displays an anomalously weak $T$ dependence,
which is consistent with the phase-disordering 
scenario.  We contrast the Nernst results and $H_{c2}$ behavior in hole-doped and 
electron-doped cuprates.  Contour plots of $e_N(T,H)$ in the $T$-$H$ plane clearly 
bring out the continuous extension of the low-$T$ vortex liquid state into the the 
high-$T$ Nernst region in hole-doped cuprates (but not in the electron-doped cuprate).  
The existence of an enhanced diamagnetic magnetization $M$ that survives to 
intense $H$ above $T_c$ is obtained from torque magnetometry.  The observed $M$ 
scales accurately like $e_N$ above $T_c$, confirming that the large Nernst signal is 
associated with local diamagnetic supercurrents that persist above $T_c$.  
We emphasize implications of the new features in the phase diagram 
implied by the high-field results, and discuss several theories.
\end{abstract}


\maketitle


\section{Introduction}\label{intro}

In the quest to understand high-$T_c$ superconductivity in the cuprates, two 
related important issues are the nature of the pseudogap state, which appears 
at the temperature $T^*$~\cite{Timusk,PALee,Kivelson}, 
and the nature of the superconducting transition at the critical temperature $T_c$.   Does 
the transition follow the familiar ``gap-closing" BCS (Bardeen-Cooper-Schrieffer) scenario,
or the phase-disordering scenario in which thermally generated vortices destroy long-range 
phase coherence~\cite{Emery}?  The former case would imply that the pseudogap 
state is inherently antagonistic to $d$-wave superconductivity and competes with it.  
In the latter case, by contrast, the pair condensate, bereft of phase rigidity, 
extends high above $T_c$ into the pseudogap state.  The two states are 
closely related, differing in a subtle way that is fundamental to the pairing mechanism.

The phase-disordering scenario, which lately has gained increased 
theoretical interest~\cite{Emery,PALee,Kivelson,Baskaran,Dorsey,Larkin,Franz},
is a three-dimensional (3D)
version of the well-known two-dimensional (2D) Kosterlitz-Thouless (KT) 
transition~\cite{KT,Villain,Beasley,Doniach,Halperin}.  There are many investigations 
of KT physics in 2D cuprates realized in ultra-thin films or superlattices~\cite{KTcuprate}.  
A notable result is the detection of kinetic inductance above $T_c$ 
at THz frequencies~\cite{Corson}.  

In bulk cuprates, the Uemura plot~\cite{Uemura} provided early evidence that 
$T_c$ scales with the superfluid density inferred from muon spin relaxation ($\mu$SR), 
consistent with the phase disordering scenario.
Direct evidence for this scenario has been obtained from Nernst experiments 
on single crystals~\cite{Xu}. 
When a flow of vortices is induced in a superconductor, an electric field appears transverse to the 
flow direction because of the Josephson effect~\cite{Kim}.  The Nernst experiment, which
exploits an unusual symmetry of the vortex-current response, is capable of detecting 
vorticity with high sensitivity.  A large Nernst signal $e_N$ extending from 
below $T_c$ to a broad interval above has been detected in many hole-doped cuprates.  
The results have been interpreted as evidence for vortices existing above $T_c$ and --
by direct implication -- the phase-disordering 
scenario~\cite{Xu,WangPRB,WangPRL,WangSci,OngRio,OngAnn}. 
See also Refs. \cite{Capan,Wen}.  

In defining an extended region above the ``$T_c$ dome'' in which vorticity exists
(which we call the ``Nernst region"), the Nernst results are increasingly 
influencing the ongoing pseudogap 
debate~\cite{PALee,Kivelson,Lee1,Tesanovic,Weng,Honerkamp,Zhang,Sachdev,Balents,Anderson05}.
Nonetheless, acceptance of a vortex origin for $e_N$ above $T_c$ is by no 
means unanimous; several models interpreting the Nernst results strictly in terms of 
quasiparticles have appeared~\cite{Kontani,Hu2,Dora,Levin,Alexandrov04}.
The difficulties may arise because the Nernst experiment is a relatively unfamiliar probe 
of superconductivity, with a checkered theoretical history~\cite{Caroli,Maki68,Hu1,Iddo}.  
Moreover, the notion that vortex excitations exist high above $T_c$ in \emph{bulk} samples goes
against deeply entrenched ideas of the superconducting state derived from 
BCS superconductors.  In this paper, we lay out in some detail the 
reasoning and evidence that have guided our thinking, with focus on 
recent measurements in intense fields.

The organization of the paper is as follows.  Section \ref{nernst} explains our notation and 
concepts relevant to the vortex-Nernst effect.  
Section \ref{vortex} sketches the phase-disordering scenario and the role of singular phase fluctuations.
In Secs. \ref{optimal} and \ref{under}, we give an introductory overview of Nernst results 
on optimally-doped and underdoped cuprates, respectively.
Section \ref{profile} describes the characteristic hill profile of the vortex signal and the 
continuity of the vortex liquid phase across $T_c$, while Sec. \ref{hc2}
discusses the anomalous $T$ dependence of the depairing field $H_{c2}$.  The phase diagram is 
discussed in Sec. \ref{phase}.  Recent corroboration of this interpretation from magnetization is
summarized in Sec. \ref{magnetization}.  Nernst results in the electron-doped cuprate are 
described in Sec. \ref{ncco}.  Theoretical issues are surveyed in Sec. \ref{discussion}.
Finally, in Sec. \ref{summary} we summarize the results and conclusions.

The standard acronyms are used to identify the cuprates: 
LSCO for $\rm La_{2-x}Sr_xCuO_4$, 
Bi 2201 for $\rm Bi_2Sr_{2-y}La_yCuO_6$, 
Bi 2212 for $\rm Bi_2Sr_2CaCu_2O_{8+\delta}$, 
Bi 2223 for $\rm Bi_2Sr_2Ca_2Cu_3O_{10+\delta}$, 
YBCO for $\rm YBa_2Cu_3O_y$, and 
NCCO for $\rm Nd_{2-x}Ce_xCuO_4$.
For brevity, qp, UD, OP and OV stand for quasi-particle, underdoped, optimally-doped 
and overdoped, respectively.  Where necessary, we distinguish vortex and qp terms 
by the superscripts $``{s}"$ and $``{n}"$, respectively.


\section{The vortex-Nernst experiment}\label{nernst}
The Nernst effect in a solid is the detection of an electric field ${\bf E}$ (along $\pm{\bf \hat{y}}$, say) 
when a temperature gradient $-\nabla T||{\bf \hat{x}}$ is applied in the presence of a 
magnetic field $\bf H ||\hat{z}$ ($\bf E$ is antisymmetric in $\bf H$).  
The Nernst \emph{signal}, defined as $E$ per unit gradient, viz. $e_N(H,T) = E/|\nabla T|$, 
is generally much larger in ferromagnets and superconductors than in non-magnetic 
normal metals.  Where $e_N$ is linear in $H$ (conventional metals), it is customary 
to define the Nernst coefficient $\nu = e_N/B$ with $B = \mu_0H$.  Our focus here, 
however, is on the vortex-Nernst effect in type II superconductors, where $e_N$ 
is intrinsically strongly nonlinear in $H$.  Instead of $\nu$, it is more useful for 
our purpose to discuss the Nernst signal $e_N(T,H)$.

We remark that, to produce in a ferromagnetic conductor a signal $e_N$ of the 
magnitude reported here ($\sim 0.1-1\;\mu$V/K), one would need a 
magnetization $M$ of $10^4$ to $10^5$ A/m~\cite{WLee}.  
This is very far from the case in cuprates, where $|M|$ = 5-50 A/m in the Nernst region.  
Hereafter, we focus on the vortex mechanism.

\bfig
\incl[width=5cm]{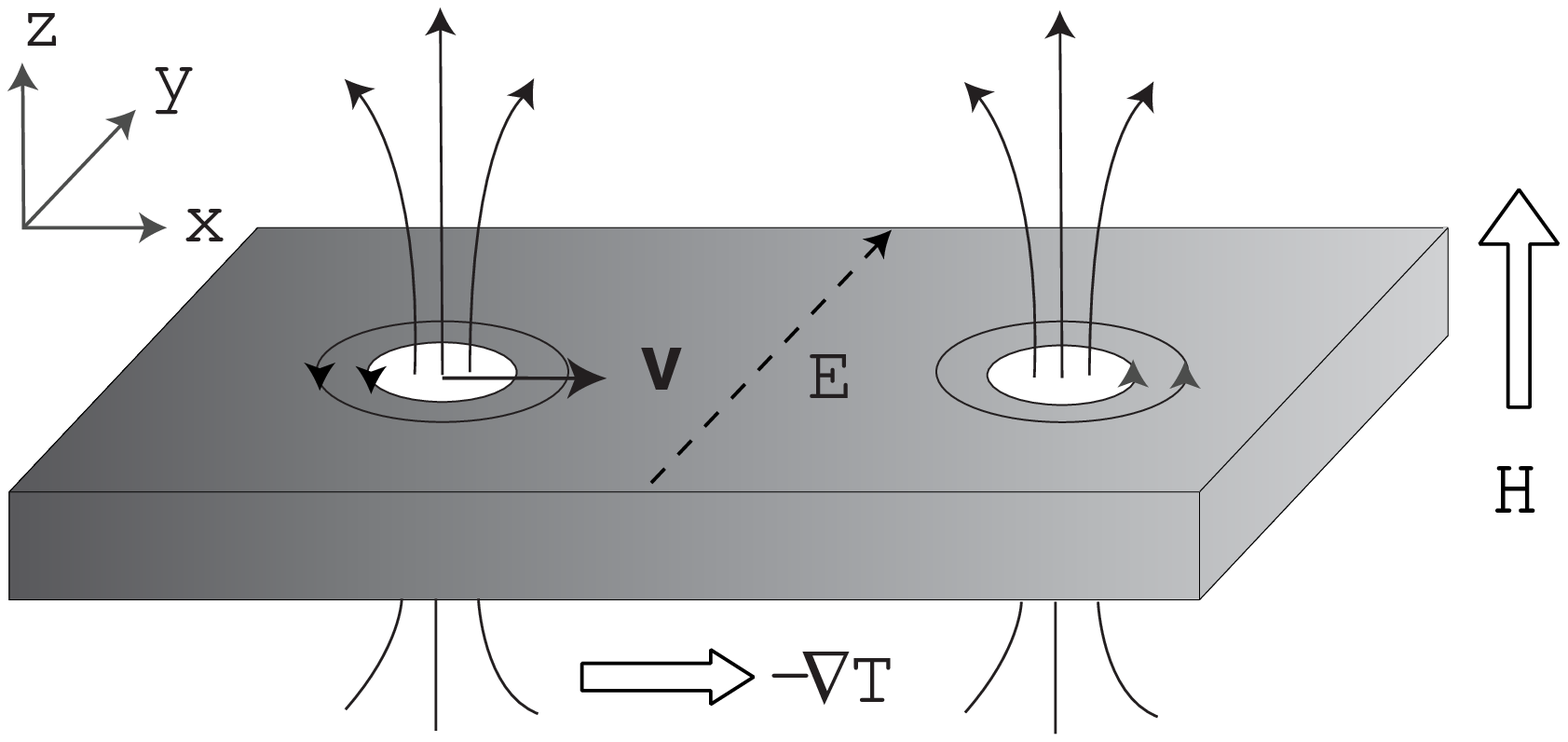}
\caption {\label{nernstexpt} The vortex-Nernst effect in a type II superconductor. Concentric
circles represent vortices.}
\efig

In the vortex-liquid state, a gradient $-\nabla T$ drives the vortices to the cooler 
end of the sample because a normal vortex core has a finite amount of entropy relative 
to the zero-entropy condensate (Fig. \ref{nernstexpt})~\cite{Kim,Serin,Huebener}. 
Because of the 2$\pi$ phase singularity at each vortex core, vortex motion induces 
phase slippage~\cite{Anderson66}.  
By the Josephson equation $2eV_J = \hbar\dot{\theta}$, the time 
derivative of the phase $\dot{\theta}$ produces 
an electrochemical potential difference $V_J$.  We have $\dot{\theta} = 2\pi \dot{N}_v$, where 
$\dot{N}_v$ is the number of vortices crossing a line $||{\bf \hat{y}}$ per second.  
The Josephson voltage $V_J$ may be expressed as a transverse electric 
field $\bf E = B\times v$ which is detected as the Nernst signal.  

The peculiar symmetry here, in which a driving force along $\bf\hat{x}$ produces -- 
as the leading response -- a conjugate current along $\bf\hat{y}$ that is 
antisymmetric in $\bf H$, is specific to vortex currents.  
The Nernst effect is particularly suited to its observation.
The sign of the Nernst signal is not intrinsically related to the 
sign of a charge (unlike the Hall effect).  Fortunately, the Josephson 
equation, which dictates that ${\bf E\,||\, H}\times(-\nabla T)$, provides a sign 
convention for the Nernst experiment~\cite{WangPRB}.  
We regard the Nernst signal as positive if it is consistent with vortex flow.

Generally, because $e_N$ is difficult to calculate from a microscopic model, a 
phenomenological description is often used~\cite{Kim,Serin,Huebener}.  The force 
exerted by the gradient on the vortex (per unit length) is 
${\bf f} = s_{\phi}(-\nabla T)$ where $s_{\phi}$ is called the ``transport entropy'' (per length).
Balancing this against the frictional force in steady state, we
have $\eta{\bf v} = s_{\phi}(-\nabla T)$, where the damping viscosity $\eta$ may be
inferred from the flux-flow resistivity $\rho = B\phi_0/\eta$ with $\phi_0 = h/2e$ 
the superconducting flux quantum.  The Nernst signal is then
\be
e_N =  \frac{Bs_{\phi}}{\eta} = \frac{\rho s_{\phi}}{\phi_0}.
\label{sf}
\ee
We may extract $s_{\phi}$ by measuring $e_N$ and $\rho$, but
now all the difficulties attendant to $e_N$ reside in $s_{\phi}$.

[In the vortex solid state (when $H$ is below the melting field $H_m$), the force due to the gradient
$\bf f$ is too feeble to cause vortex motion, and $e_N$ is rigorously zero.  In low-$T_c$ type
II superconductors, it is more practical to employ the Ettingshausen effect~\cite{Kim}, 
which is related to the Nernst effect by reciprocity.  In the Ettingshausen experiment, 
a current density $\bf J$ is applied $||\,\bf\hat{x}$ 
with $\bf H||\hat{z}$.  Vortex motion transverse to $\bf J$ produces a heat current which leads
to a gradient $\nabla T\,||\bf\hat{y}$ detected as the Ettingshausen signal.  
The Ettingshausen coefficient is ${\cal Q}_E = |\nabla T|/JH$.  The advantage of the 
Ettingshausen experiment is that a large $\bf J$ may be used to depin the vortex lattice 
below $H_m$~\cite{Vidal,Serin}.]

\bfig
\incl[width=4cm]{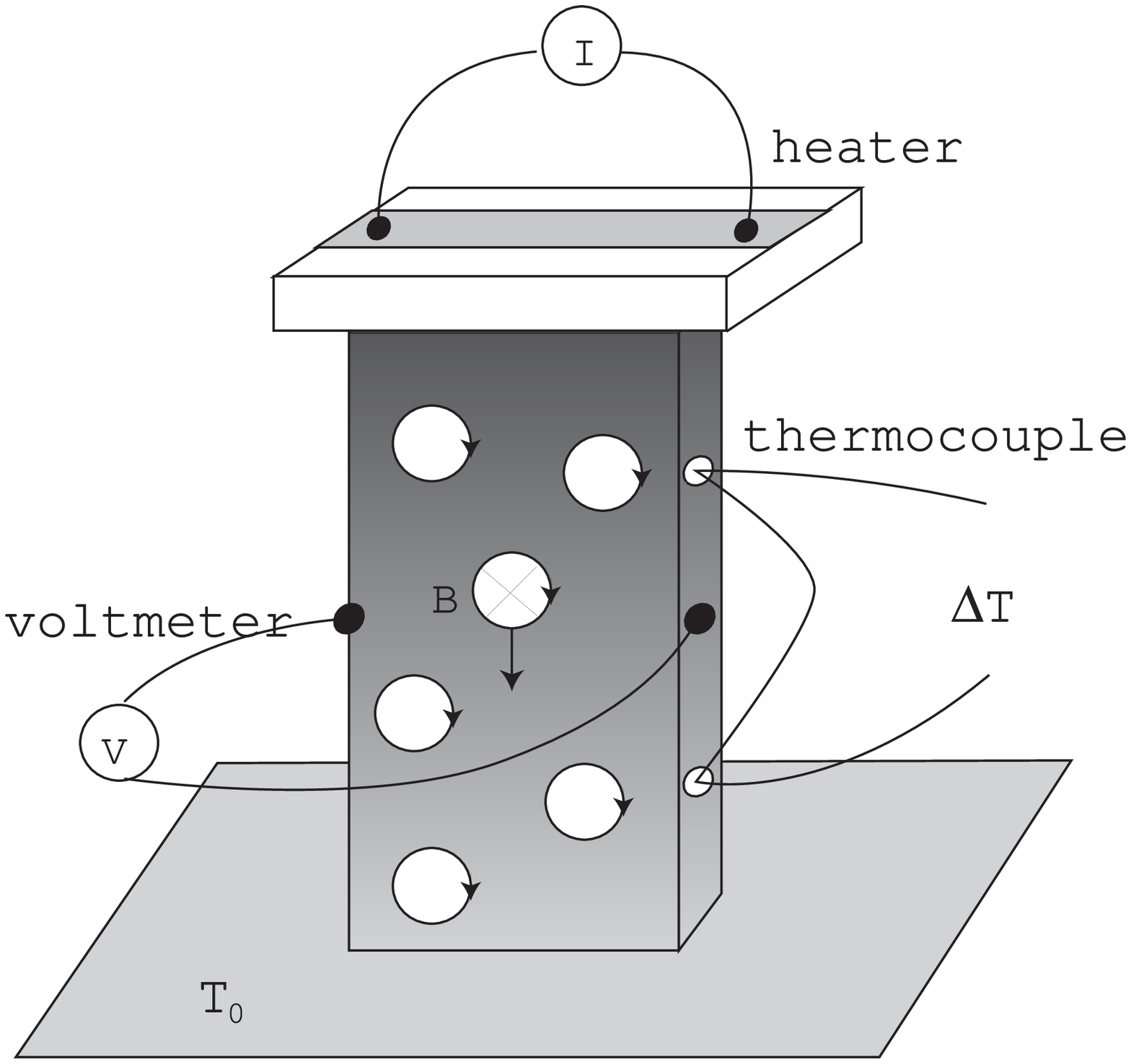}
\caption {\label{setup} Crystal mounting geometry in the Nernst experiment.
}
\efig

In theoretical treatments, the vortex Nernst and Ettingshausen effects are conveniently 
handled together as the current
response of the system to $\bf E$ and $-\nabla T$~\cite{Caroli,Maki68,Hu1,Iddo}.  
Imagine that we apply to a vortex system mutually orthogonal $E$-field and temperature 
gradient in the plane normal to $\bf H$, and observe the response of the charge and 
heat current densities, ${\bf J}$ and ${\bf J}^h$, respectively.  As mentioned, the 
peculiar symmetry of vortex currents dictates that ${\bf J}$ is $\perp$ 
to $-\nabla T$ but $||\,{\bf E}$, whereas ${\bf J}^h$ is $\perp$ to ${\bf E}$ but $||\,(-\nabla T)$.
(These are the \emph{large} current responses; the flux-flow Hall effect and 
thermopower related to vortex diffusion produce much weaker currents 
which we neglect, along with qp contributions).

In linear response, ${\bf J}$ 
and ${\bf J}^h$ are given by~\cite{Caroli,Hu1}
\begin{eqnarray}
J_y &=& \sigma E + \alpha_{yx}^{s} (-\partial T)\label{J1}\\
J_x^h &=& \tilde{\alpha}_{xy}^{s} E + \kappa(-\partial T)\label{J2},
\end{eqnarray}
with $\bf E||\bf\hat{y}$, $-\nabla T||\bf\hat{x}$.  Here $\sigma = 1/\rho$ and $\kappa$ is 
the thermal conductivity.  
The 2 off-diagonal Peltier terms are related by $T\alpha_{xy}^{s} = \tilde{\alpha}_{xy}^{s}$ 
by Onsager reciprocity.  Setting $J$ to zero, we find that 
\be
e_N = \rho \alpha_{xy}^{s}
\label{ealpha}
\ee
which is to be compared with Eq. \ref{sf}.  Similarly with $J^h$ = 0, the Ettingshausen coefficient
is ${\cal Q}_E = \tilde{\alpha}_{xy}^s/H\sigma\kappa$.

In Hartree-Fock approximation, Caroli and Maki~\cite{Caroli,Maki68} found that 
$\alpha_{xy}^{s}$ is proportional to the magnetization $M$ for $H$ 
close to $H_{c2}$ and $T\rightarrow T_c$.  Their coefficient of proportionality 
had an error which was later corrected~\cite{Hu1,Iddo}.  Here we express this relationship as
\be
\alpha_{xy}^{s} = \tilde{\beta} M \quad\quad (H\rightarrow H_{c2}^{-})
\label{beta}
\ee
with the parameter $\tilde{\beta}$ to be determined by experiment ($\tilde{\beta}$ has dimensions
of $1/T$).

The charge carriers also produce a Nernst signal which we refer to as 
the quasiparticle (qp) contribution.  The current density produced by an applied gradient
is given by the Peltier tensor $\alpha_{ij}^{n}$, viz. $J_i=\alpha_{ij}^{n}(-\partial_j T)$.
In terms of $\alpha_{ij}^{n}$ and the Hall resistivity
$\rho_{xy}^{n}$, the qp Nernst signal is~\cite{WangPRB}
\be
e_N^{n} = \rho^{n}\alpha^{n}_{xy} - \rho_{xy}^{n}\alpha^{n},
\label{eNn}
\ee
where $\rho^{n}$ is the qp resistivity and $\alpha^{n} \equiv\alpha_{xx}^{n}$ is related to 
the thermopower by $S = \rho^{n}\alpha^{n}$.

The observed Nernst signal is the sum of the vortex and qp terms, viz.
\be
e_N = e_N^{s} + e_N^{n},
\label{eNs}
\ee
with the caveat that in Eq. \ref{eNn} for $e_N^{n}$, the total (observed)
$\rho$, $\alpha$ and $\rho_{xy}$ are used instead of the strictly qp terms.

In the hole-doped cuprates, the qp term is very small for $T<T_c$.  For the
purpose of determining the onset temperature $T_{onset}$ of $e_N^{s}$, however,
the qp term has to be carefully resolved.  This involves measuring the
thermopower $S = \rho\alpha$, Hall angle $\tan\theta = \rho_{xy}/\rho$ and
resistivity $\rho$ in addition to $e_N$.  As this procedure has been described
in detail in Ref. \cite{WangPRB}, we will not repeat it here.  In what follows,
we will not usually distinguish between $e_N$ and $e_N^{s}$, except when
discussing NCCO in Sec. \ref{ncco}.

Figure~\ref{setup} shows the setup in our Nernst experiment.  The 
samples used are high-quality cuprate single crystals of typical size of $1.2 \times 0.8 \times 0.05$ mm$^3$. 
One end of the crystal is glued with silver epoxy onto a sapphire substrate, which is heat-sunk to a 
copper cold finger.  A thin-film 1-k$\Omega$ heater, silver-epoxied to the top edge of the crystal, 
generates the heat current flowing in the $ab$ plane of the crystal.  The temperature 
difference $\Delta T$ (0.3-0.5 K) is measured by a pair of fine gauge Chromel-Alumel thermocouple. 
A pair of ohmic contacts are prepared on the edge of the sample by annealing different kinds 
of conductive materials.

After the bath temperature is stabilized (to within $\pm$ 10 mK), the gradient
is turned on.  The Nernst voltage is pre-amplified and measured by 
a nanovoltmeter as the magnetic field is slowly ramped up.  The 
uncertainty in $V$ is $\pm$5 nV.  To remove stray longitudinal signals due to 
mis-alignment of the contacts, the magnetic field is swept in both directions.  Only 
the field-asymmetric part of the raw data is taken as the Nernst signal. 

As expressed in Eq. \ref{eNs}, the Nernst signal is comprised of a large component $e_N^{s}$
associated with the pair condensate and a qp component $e_N^{n}$ from carriers.
To date, the existence of $e_N^{s}$ in the Nernst region above $T_c$ has been confirmed in 
Bi 2212, Bi 2223, LSCO and YBCO, and $\rm Bi_2Sr_{2-y}La_yCuO_6$ (Bi 2201).  
The interesting case of electron-doped
$\rm Nd_{2-x}Ce_xCuO_4$ (NCCO) is deferred to Sec. \ref{ncco}.


\section{Vortices and phase-disordering transition at $T_{c}$}\label{vortex}
In this section, we sketch the phase-disordering scenario associated with the
appearance of thermally created vortices, which has heavily informed the analyses of 
our Nernst-effect experiments.  In the superconducting state, the pair condensate 
described by the macroscopic wave function $\hat{\Psi} = |\Psi|\mathrm{e}^{i\theta}$ 
exhibits long-range phase coherence.  The phase spontaneously selects a particular value 
which is rigidly maintained throughout the volume~\cite{Anderson58,Anderson66}.  
The energy cost of local variations in $\theta({\bf r})$ is given by
$H_{\theta} = \frac12 \int d^2r K_s(\nabla\theta)^2$, where the phase stiffness 
$K_s = \hbar^2 n_s/4m^*$ arises from the kinetic energy of the superfluid electrons, with
$n_s$ the 2D density and $m^*$ the effective mass~\cite{Halperin}.

In the Kosterlitz-Thouless problem~\cite{KT,Villain,Beasley,Doniach,Halperin} --
the prototypical example of the phase-disordering scenario -- vortex-antivortex unbinding at 
the KT transition temperature $T_{KT}$ destroys long-range phase coherence and superfluidity, 
even though the pair amplitude $|\Psi|$
remains finite.  Random $2\pi$ jumps in $\theta({\bf r})$ caused by (anti) vortex motion 
drive the thermally averaged order parameter to zero, viz.
\be
\langle\hat{\Psi}({\bf r})\rangle = |\Psi|\langle\mathrm{e}^{i\theta({\bf r})}\rangle = 0.
\label{gapzero}
\ee
Generally, the phase-disordering transition 
$T_{\theta}$ is proportional to the superfluid stiffness, viz.~\cite{Halperin,Emery,PALee}
\be
k_BT_{\theta} = {\cal A}K_s.
\label{TKs}
\ee
where ${\cal A} = \pi/2$ in the KT problem, but can vary from
0.9 for the XY model to 1.5 in the limit of large vortex core energy.

From $\mu$SR experiments, Uemura and collaborators~\cite{Uemura} found that, in 
UD cuprates, $T_c$ follows a universal, linear dependence on $n_s/m^*$.  Although
originally discussed in terms of boson condensation at $T_c$, the Uemura plot -- if reinterpreted 
as confirming Eq. \ref{TKs} -- provided initial evidence for the phase-disordering scenario.  Very early, 
Baskaran \etal~\cite{Baskaran} noted that proximity 
to the Mott insulator implies that $T_c$ in UD cuprates must be controlled by loss of phase coherence.  
The first detailed examination of this issue was provided by Emery and 
Kivelson~\cite{Emery} who found that the ratio $K_s/k_BT_c$ falls in the range 1-2 
in most hole-doped cuprates (compared with $10^3$-$10^5$ in low-$T_c$ superconductors).  
This implies that phase fluctuations are of crucial importance in determining $T_c$.  
Corson \etal~\cite{Corson} measured the complex conductivity $\hat{\sigma}(\omega)$ 
in two thin-film samples of Bi 2212 with $T_c$ = 74 K and 33 K at THz frequencies and 
found that the kinetic inductance persists to $\sim$25 K above $T_c$ in both samples.

Fluctuations in $\theta$ are of two types: analytical spin-wave fluctuations $\Delta\theta_a$ and singular 
vortex-induced fluctuations $\Delta\theta_v$~\cite{Villain}.  The singular 
fluctuations $\Delta\theta_v$ are of specific interest here.  

As $T$ rises bove $T_{KT}$, the density of spontaneous vortices (anti-vortices) $n_{+}$ ($n_{-}$)
increases exponentially but the net vorticity $n_{+}-n_{-}$ stays at zero if ${\bf H} = 0$.  The
applied ${\bf H}$ increases $n_{+}$ (say) to produce a net induction field $B = (n_{+}-n_{-})\phi_0$~\cite{Doniach}.
A detailed calculation of the KT magnetization $M = B\mu_0^{-1} - H$ over a broad interval of $T$ 
was recently reported~\cite{Oganesyan}.  Above $T_{KT}$, both $M$ and the vorticity remain observable
despite the vanishing of $\langle\hat{\Psi}({\bf r})\rangle$.

\bfig
\incl[width=7cm]{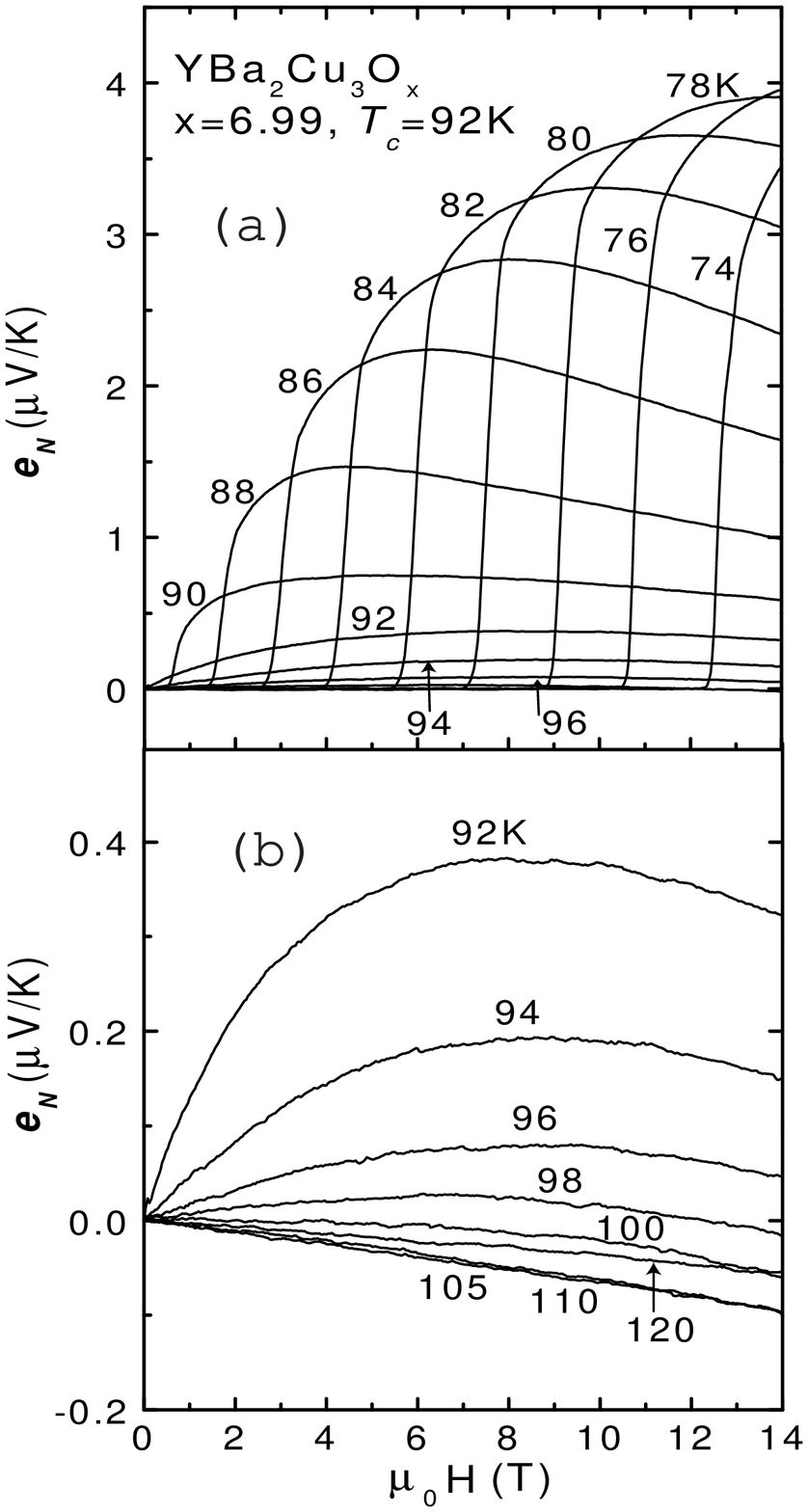}
\caption{\label{NH-Y92K} Curves of the Nernst signal $e_N$ vs. $H$ 
measured in slightly OV YBCO ($x$ = 6.99, $T_c$ = 92 K) at temperatures 
below $T_c$ (Panel a) and above $T_c$ (Panel b).  Below $T_c$, $e_N$ rises
nearly vertically at the melting field $H_m$.  Above $T_c$ (b), the negative
$H$-linear contribution of the qp term $e_N^{n}$ becomes quite apparent.}  
\efig

In experimental studies of the KT transition in layered ferromagnets (e.g. $\rm K_2CuF_4$), the KT 
transition is pre-empted at $T_c > T_{KT}$ by a 3D Curie transition because 
of a weak interlayer exchange $J_{\perp}$ ($J_{\perp}\ll J$, the intralayer exchange)~\cite{Hirakawa}.  
Nevertheless, over a broad interval above $T_C$, KT physics prevails.  Analogously, in bulk cuprates, 
we expect the weak interlayer coupling to induce a 3D transition that pre-empts the KT transition.  
However, the KT description of vortex proliferation is valid over a broad interval 
above $T_c$ (the Nernst region).  

Close to $T_c$, the 3DXY model is the appropriate description for bulk crystals.  
Extensive numerical simulations by Nguyen and Sudb\o~\cite{Sudbo}
of the 3DXY model with moderate anisotropy $\alpha_J = J_{\perp}/J\sim \frac13$
clearly show that the helicity modulus $\Upsilon = (\hbar/m)^2\rho_s$ (
where $\rho_s$ is the superfluid density) 
is destroyed at $T_c$ by the spontaneous appearance of vortex loops.  
Even with $\alpha_J\sim \frac13$, the simulations show a $\sim$10 K interval 
above $T_c$ where vorticity exists.  These results were tentatively compared with YBCO, but
simulations in the limit $1/\alpha\gg 1$ are desirable.  

In the Nernst experiment, the flow of vortices and anti-vortices down 
the gradient generates signals of opposite signs.  Hence the observed $e_N^{s}$ picks up
the difference in population $n_{+}-n_{-}$, i.e. the vorticity.  

\bfig
\incl[width=7cm]{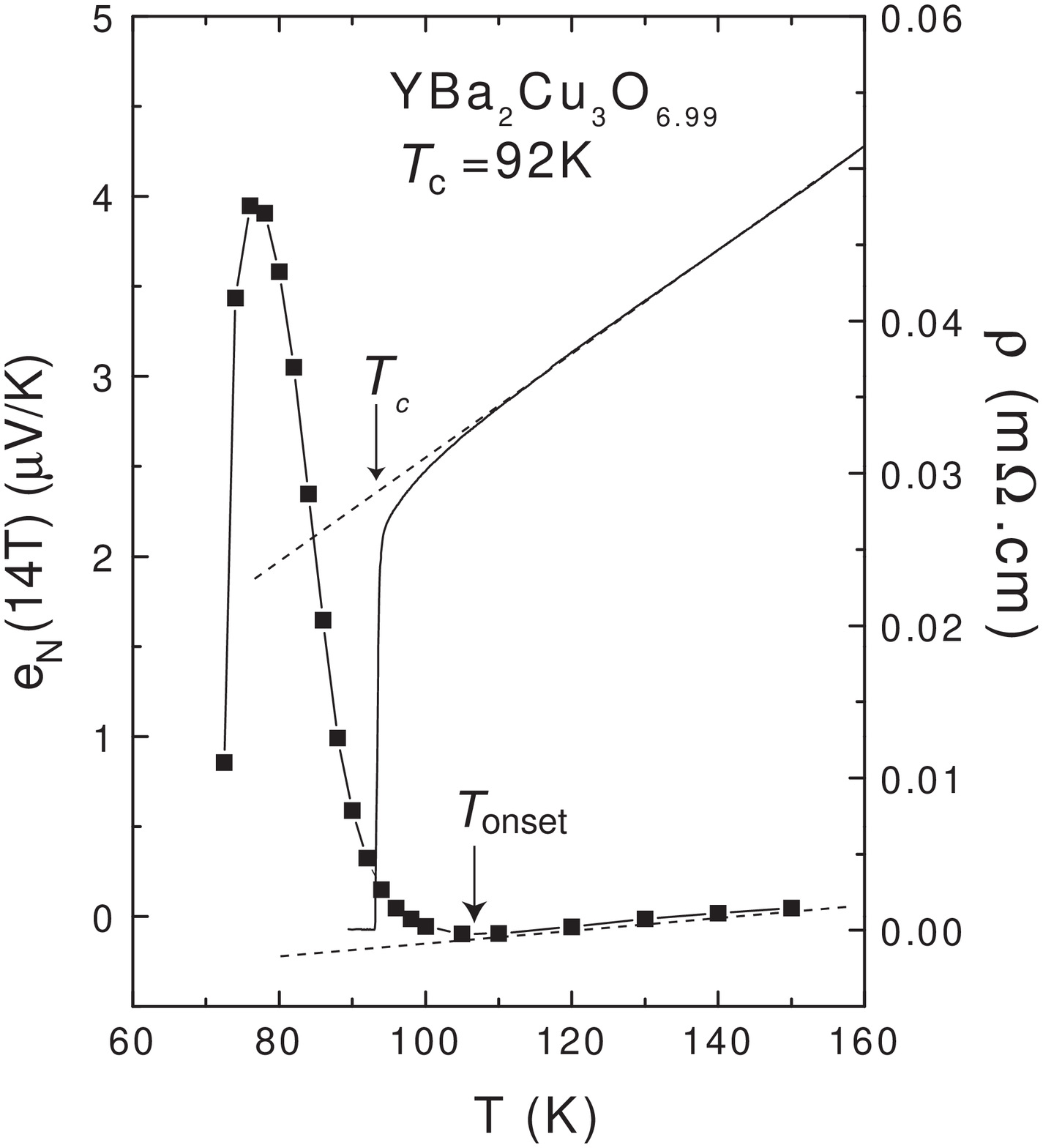}
\caption{\label{NT-Y92K} Temperature dependence of $e_N$ measured at 14 T
compared with the resistivity $\rho$ measured on OV YBCO ($x$ = 6.99, $T_c$ = 92 K).
The onset temperature $T_{onset}$ and $T_c$ are indicated by arrows.  The dashed line indicates
the negative qp contribution.}  
\efig

\bfig
\incl[width=6cm]{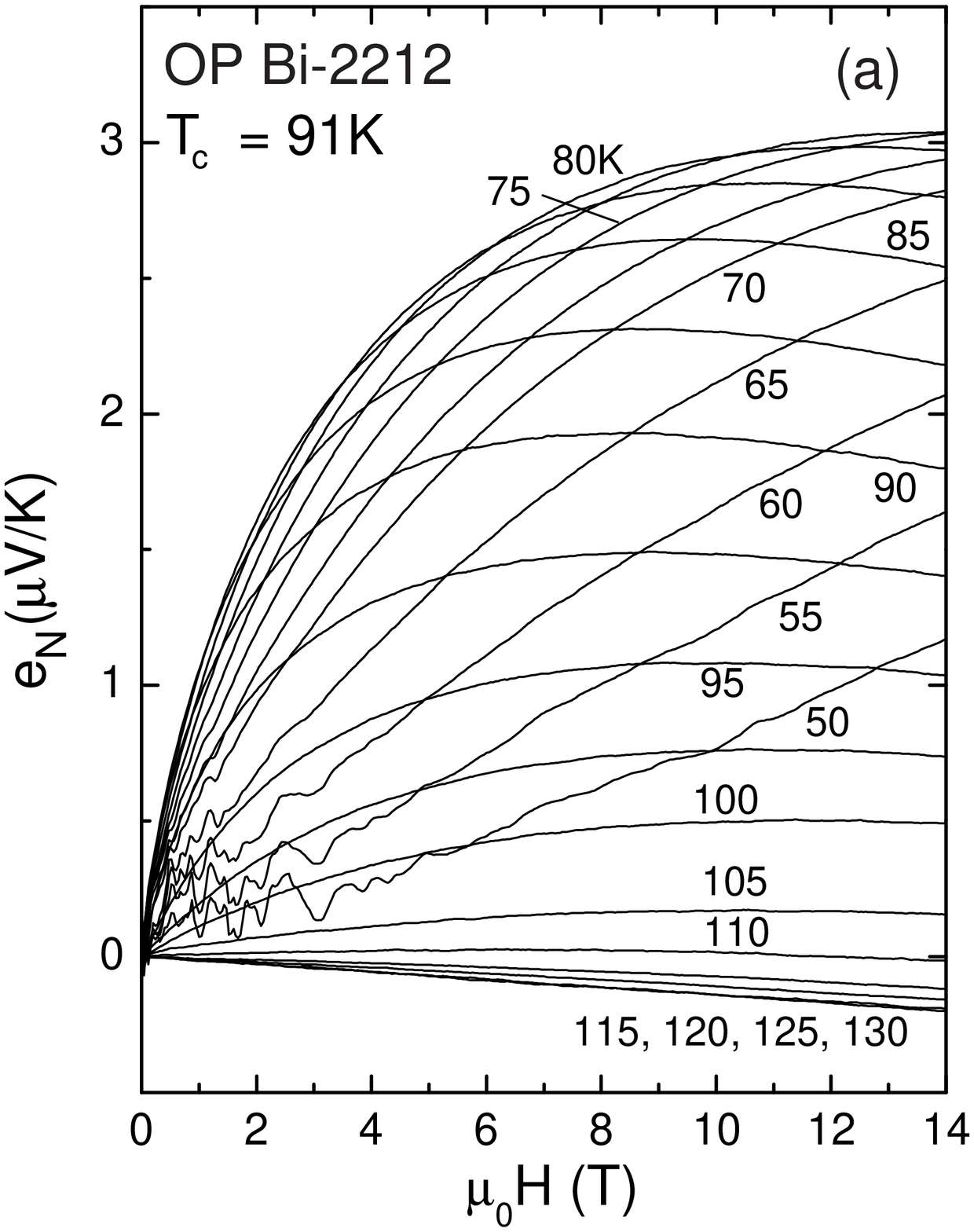}
\incl[width=6cm]{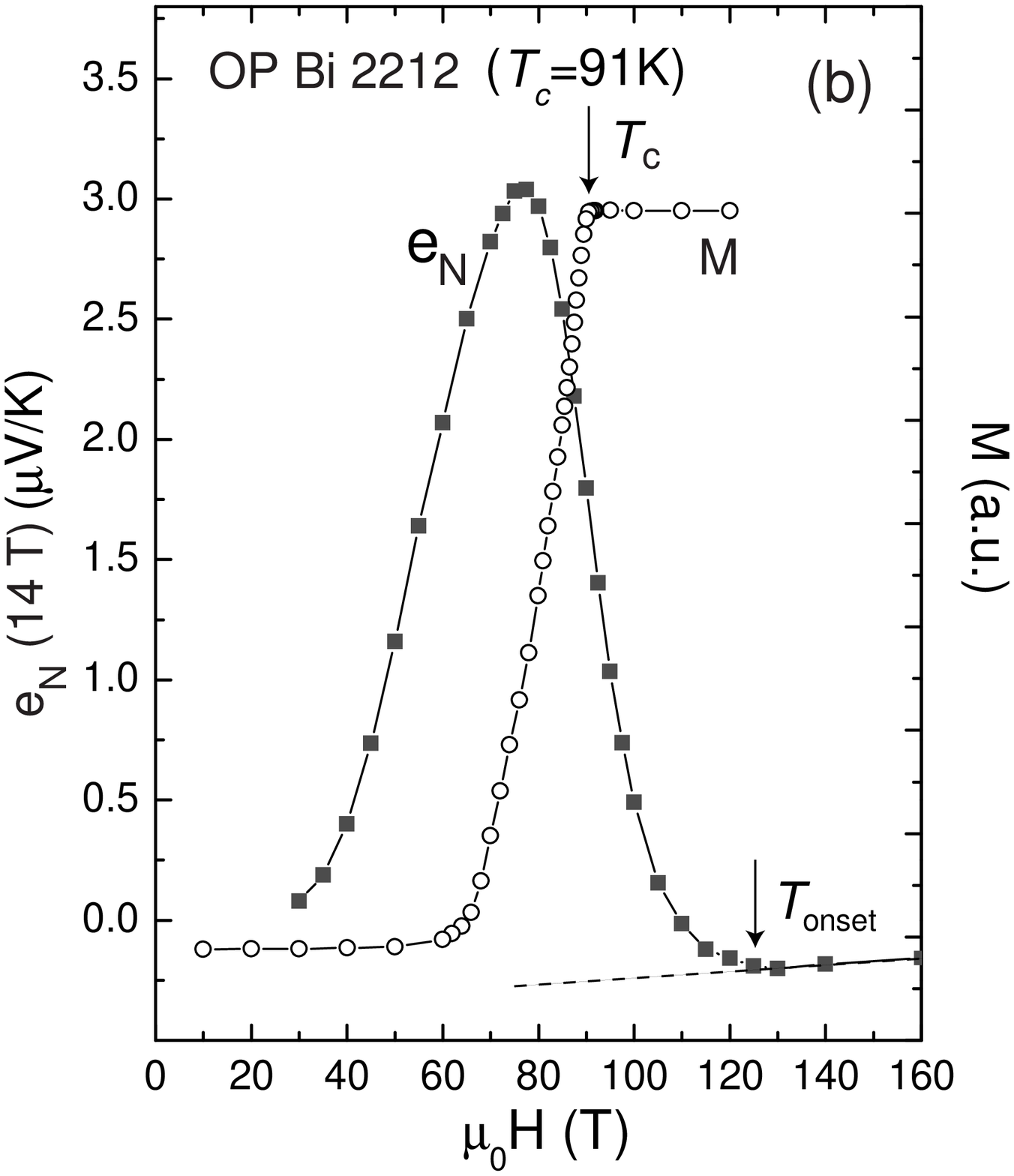}
\caption{\label{Opt-B91K} (a) The Nernst signal $e_N$ vs. $H$ in OP Bi 2212 ($T_c$ = 91 K) 
at temperatures 50-130 K.  The oscillations in $e_N$ in weak $H$ are reproducible, and
may be caused by plastic flow of the vortices. (b) The $T$ dependence of $e_N$ 
measured at 14 T (solid squares) and the Meissner curve (magnetization $M$ 
measured at $H$ = 10 Oe) in OP Bi 2212. The dashed line indicates the
estimated negative qp contribution.}  
\efig

In the BCS scenario, fluctuations of the order parameter $\hat{\Psi}$ are (predominantly) 
fluctuations of the amplitude $|\Psi|$ (from zero).  The Gaussian approximation,
in which only terms in $|\Psi(\bf x)|^2$ are retained in the 
action $S$ of the partition function ${\cal Z}$,
provides a good description of fluctuation diamagnetism measured in low-$T_c$ 
superconductors~\cite{Gollub}.  However, if singular phase fluctuations $\Delta\theta_v$
are dominant in the destruction of superfluidity (as the evidence shows is 
appropriate in the Nernst region), the valid description is inherently non-Gaussian.


\section{Optimally-doped cuprates}\label{optimal}

In the cuprates, early Nernst~\cite{Ri,Hagen,Clayhold} and 
Ettingshausen~\cite{Batlogg} experiments were restricted to OP, 
bilayer cuprates.  The results were largely confined to
the vortex-liquid state below $T_{c}$, and analyzed to extract 
vortex parameters such as $s_{\phi}$ (Eq. \ref{sf}).  However, the existence of unusually
large ``fluctuation signals" extending 10-20 K above $T_{c}$ was noted 
in OP Bi 2212 and YBCO~\cite{Batlogg,Ri}.

We start the description of our data from the OV/OP end of the doping window. 
Figure~\ref{NH-Y92K}a displays the traces of $e_N$ versus $H$ at fixed $T$, in a crystal 
of $\rm YBa_2Cu_3O_y$ (YBCO), with $y$ = 6.99 and $T_c$ = 92.0 K (this sample is slightly OV).  
As the melting field $H_m$ is exceeded, the abrupt motion of a large number 
of vortices leads to a nearly vertical rise in $e_N$ (Panel a).  The signal reaches a broad maximum 
and then decreases slowly.  The envelope of all these curves represents the maximum value that $e_N$
attains in the temperature interval shown.

As $T$ rises above $T_c$, the maximum values of $e_N(T)$ decrease markedly and the 
profiles become broader (Panel b).  However, an abrupt transition is not observed in the Nernst signal.
Instead, it retains its nonlinearity up to $\sim$105 K.  This is analogous to the 
Ettingshausen fluctuation signal reported in OP YBCO~\cite{Batlogg}.  Above 110 K, 
the curve of $e_N$ is linear in $H$ with a slope that changes mildly with $T$,
which we identify with the qp contribution $e_N^{n}$.

To show the fluctuation regime more clearly, we plot the $T$ dependence of $e_N$ measured
at 14 T (Fig.~\ref{NT-Y92K}) together with its zero-field resistivity $\rho$.  Clearly, $e_N$ deviates from 
the qp background at $\sim$107 K, or 15 K above $T_{c}$ = 92 K.  
Similar measurements on OP bilayer $\rm Bi_2Sr_2CaCu_2O_{8+\delta}$ (Bi 2212, $T_{c}$ = 91 K)
are shown in the Fig.~\ref{Opt-B91K}a.  Owing to its extreme anisotropy, 
the vortex-solid in Bi 2212 has a very small shear modulus $c_{66}$.  
The melting field $H_m$ remains small even $T\ll T_{c}$ ($\sim$50 K).  We also
remark that, near $T_{c}$, $e_N$ displays a non-analytic $H$ dependence in weak $H$.  
Above $T_{c}$, $e_N$ rapidly becomes much smaller in amplitude.  Figure~\ref{Opt-B91K}b 
displays the $T$ dependence of $e_N$ measured at 14 T together with the Meissner 
signal measured at $H$ = 10 Oe in a SQUID magnetometer.  The onset 
temperature $T_{onset}$ of the vortex-Nernst signal is 
$\sim$125 K, or 30 K above $T_{c}$.  The broader Nernst region in Bi 2212 
(compared to OP YBCO) reflects its extreme anisotropy.

\bfig
\incl[width=6cm]{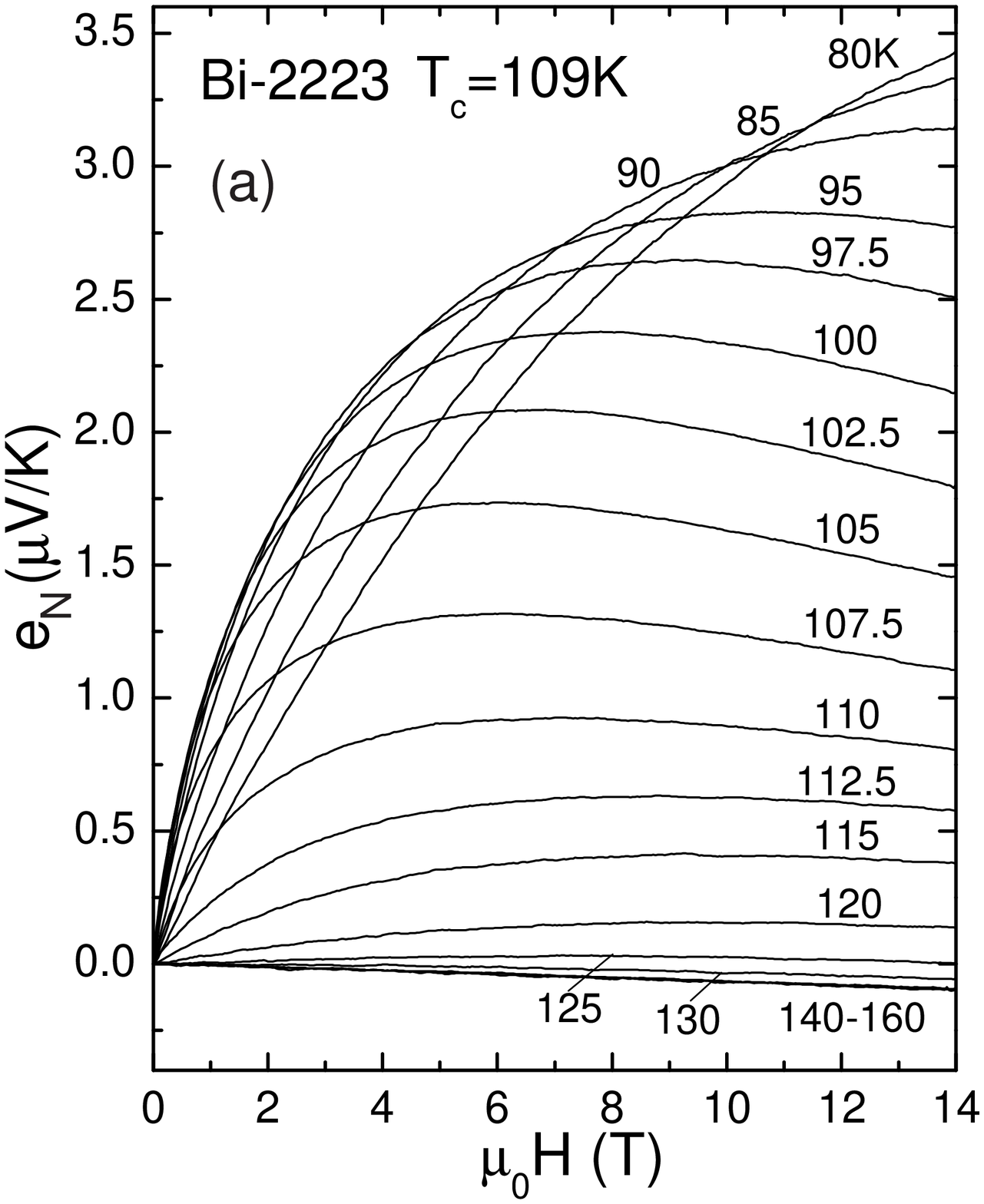}
\incl[width=6cm]{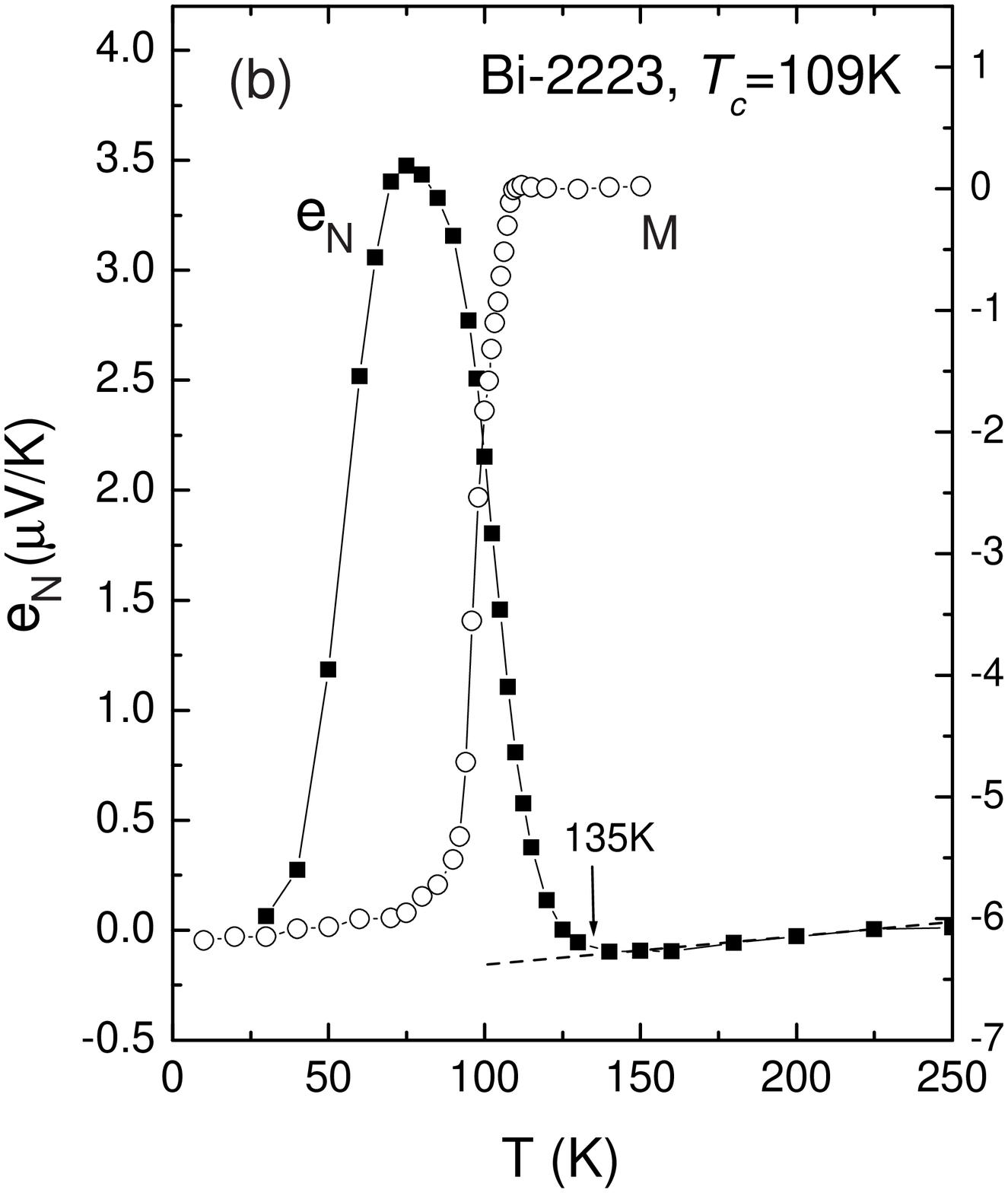}
\caption{\label{B2223} (a) Curves of $e_N$ vs. $H$ in OP Bi 2223 ($T_c$ = 
109 K) measured at selected $T$. (b)  The $T$ dependence of $e_N$ at 14 T compared with the
Meisnner curve ($M$ measured at 10 Oe) in OP BI 2223.  Dashed line is the negative qp contribution.}  
\efig

The trilayer cuprate $\rm Bi_2Sr_2Ca_2Cu_3O_{10+\delta}$ (Bi-2223) also shows similar 
extension of the Nernst signal above its $T_c$ = 109 K (Fig.~\ref{B2223}a).  
The overall behavior of $e_N$ vs. $H$ is strikingly similar to that of the bilayer system (Fig.~\ref{Opt-B91K}a). 
The plots in Fig.~\ref{B2223}b show that the Nernst onset temperature is around 
135 K, $\sim$ 25 K above the $T_{c}$. 

The early results on OP cuprates in the ``fluctuation" regime~\cite{Ri,Batlogg} were deemed 
compatible with the prevailing expectation that, although flucutations 
are strongly enhanced in cuprates~\cite{Huse}, the data appeared to be adequately
described by conventional Gaussian fluctuation theory~\cite{Welp,UllahDorsey}.

\bfig
\incl[width=6cm]{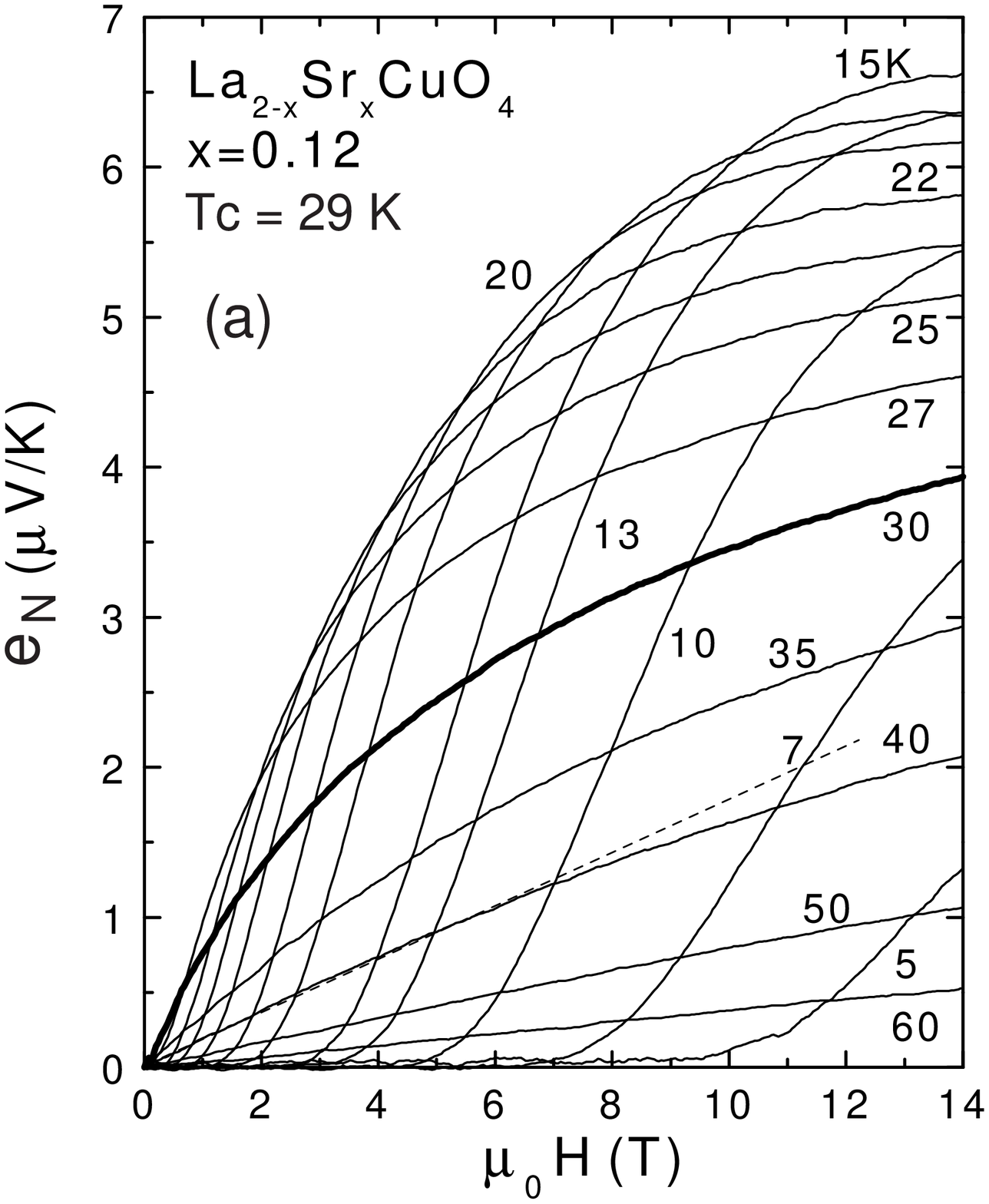}
\incl[width=6cm]{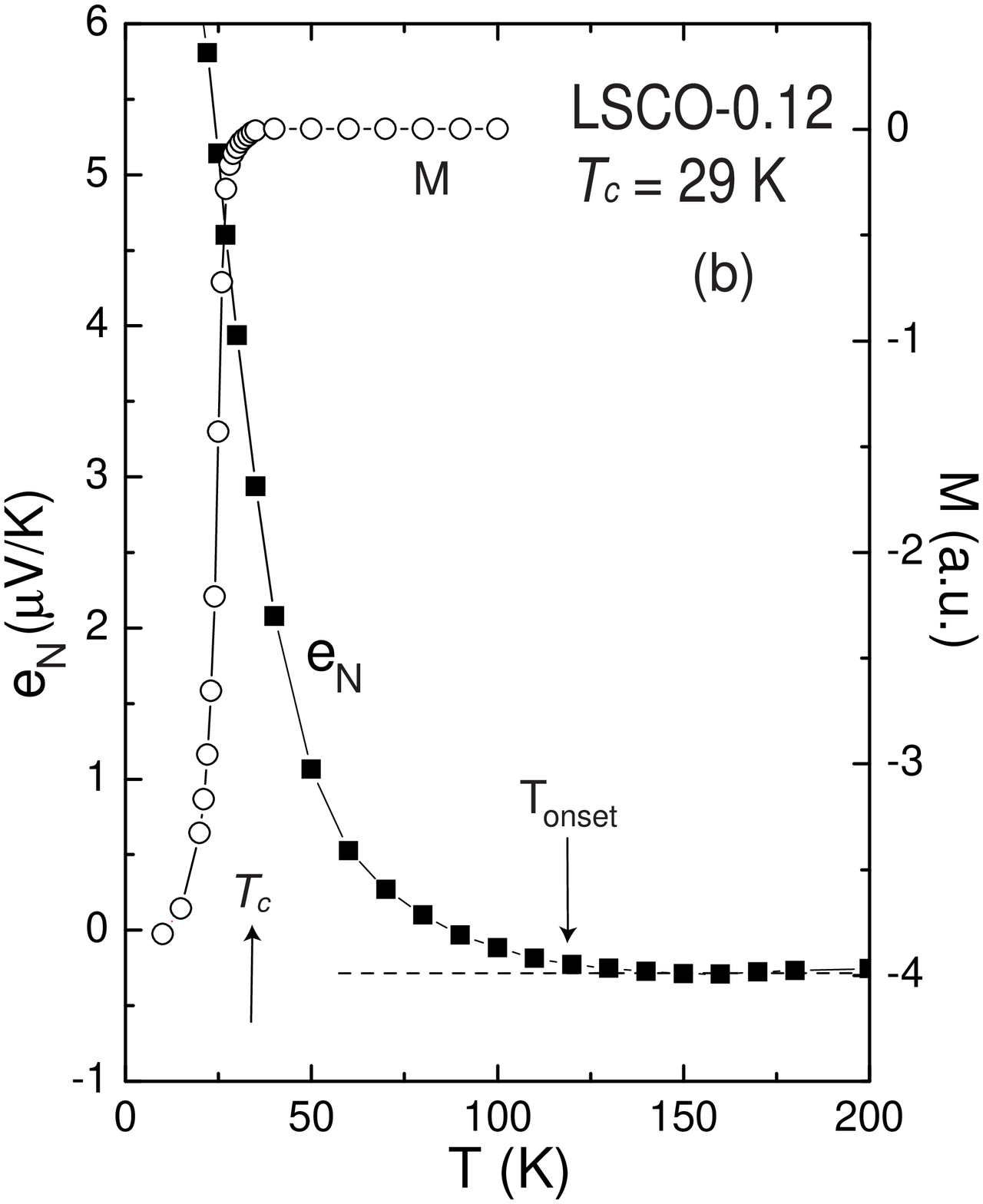}
\caption{\label{LSCO12}  (a) Curves of $e_N$ vs. $H$ in UD LSCO ($x$ = 0.12, $T_{c}$ = 29 K). 
The bold curve is nominally at $T_{c}$.  The dashed line at 40 K gives the initial slope used to find $\nu$.
(b) The $T$ dependence of $e_N$ measured at 14 T in UD LSCO (solid squares) compared with 
the Meissner curve ($M$ measured at 10 Oe, open circles). Dashed line is the negative qp contribution.
}
\efig


\section{Underdoped cuprates}\label{under}
The Nernst signal above $T_c$, already considerable in OP cuprates, becomes even larger
in the UD regime.  We first discuss $\rm La_{2-x}Sr_xCuO_4$ (LSCO)~\cite{Xu}. 
Figure~\ref{LSCO12}a displays the Nernst traces in an UD crystal 
with $x$ = 0.12 and $T_{c}$ = 28.9 K.  The results seem quite similar to that in optimally-doped YBCO 
(Fig.~\ref{NH-Y92K}).  Below $T_c$, the curves, which display the characteristic profile of a `tilted hill' 
associated with vortex motion, are all enclosed within a smooth envelope curve.
On closer examination, however, the data reveal an important difference.
In OP YBCO, the maximum value of the $e_N$--$H$ curve taken at $T_{c}$ = 92 K is $\sim$ 
0.38 $\mu$V/K, less than 10\% the maximum value attained by the envelope 
curve ($\geq$ 4 $\mu$V/K) below $T_c$.  Above $T_{c}$, $e_N$ rapidly falls 
to a negligible fraction of 4 $\mu$V/K.  By contrast, UD LSCO shows a different pattern of 
behavior.  The bold curve in Fig.~\ref{LSCO12}a is taken at $T$ = 30 K, slightly 
higher than $T_{c}$.  Its value at 14 T, $\sim$ 4 $\mu$V/K, is more than 50 \% 
of the maximum of the envelope, and still increasing with field.  Even at 
50 K, more than 20 K above $T_{c}$, the signal is a sizable fraction of the maximum 
envelope value.  Pronounced non-linearity in field is apparent in these curves.  The Nernst 
signal decays quite slowly with temperature, becoming indistinguishable 
from the qp Nernst signal only above 100 K.

Figure~\ref{LSCO12}b displays the $T$ dependence of $e_N$ taken at 14 tesla on underdoped 
LSCO, $x$ = 0.12, together with the magnetization curve measured at 10 Oe. 
The anomalous Nernst signal starts to deviate from the small qp background 
at $T_{onset} \sim$ 120 K.  The $T$ dependence of $e_N$ measured at 14 T shows 
a long ``fluctuation'' tail that extends to $T_{onset}$.

\bfig
\incl[width=6cm]{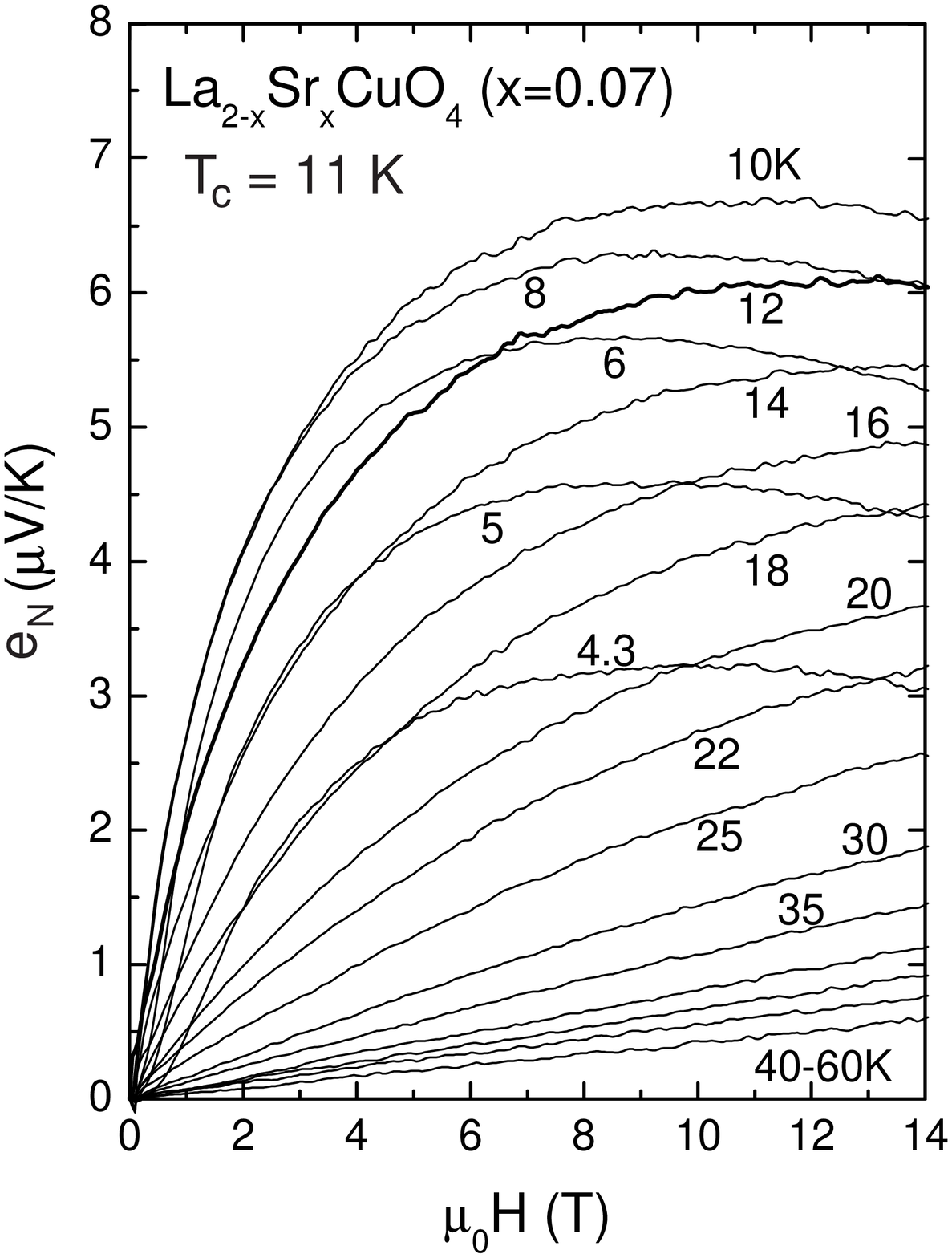}
\caption{\label{NH-L07}  Curves of $e_N$ vs. $H$ in heavily UD LSCO ($x$ = 0.07, $T_{c}$ = 11 K). 
The bold curve is taken at 12 K, 1 K above $T_{c}$.}
\efig

Moving to LSCO with smaller $x$ (0.07, $T_c$ = 11 K), we find that these anomalous features become
enhanced (Fig.~\ref{NH-L07}).  The curve at $T=$ 12 K -- 1 K above $T_c$ -- displays a Nernst 
signal that is similar in overall magnitude to any of the curves below $T_c$.  Indeed, curves taken at 20 K are comparable 
in magnitude with many of those below $T_c$.  Similar results have been obtained by Capan \etal~\cite{Capan}.
Hence from the perspective of the Nernst effect, the 
boundary between the superconducting state and the normal state in UD cuprates is truly blurred. 
This poses a serious challenge to the conventional notion of `fluctuations'.

\bfig
\incl[width=6cm]{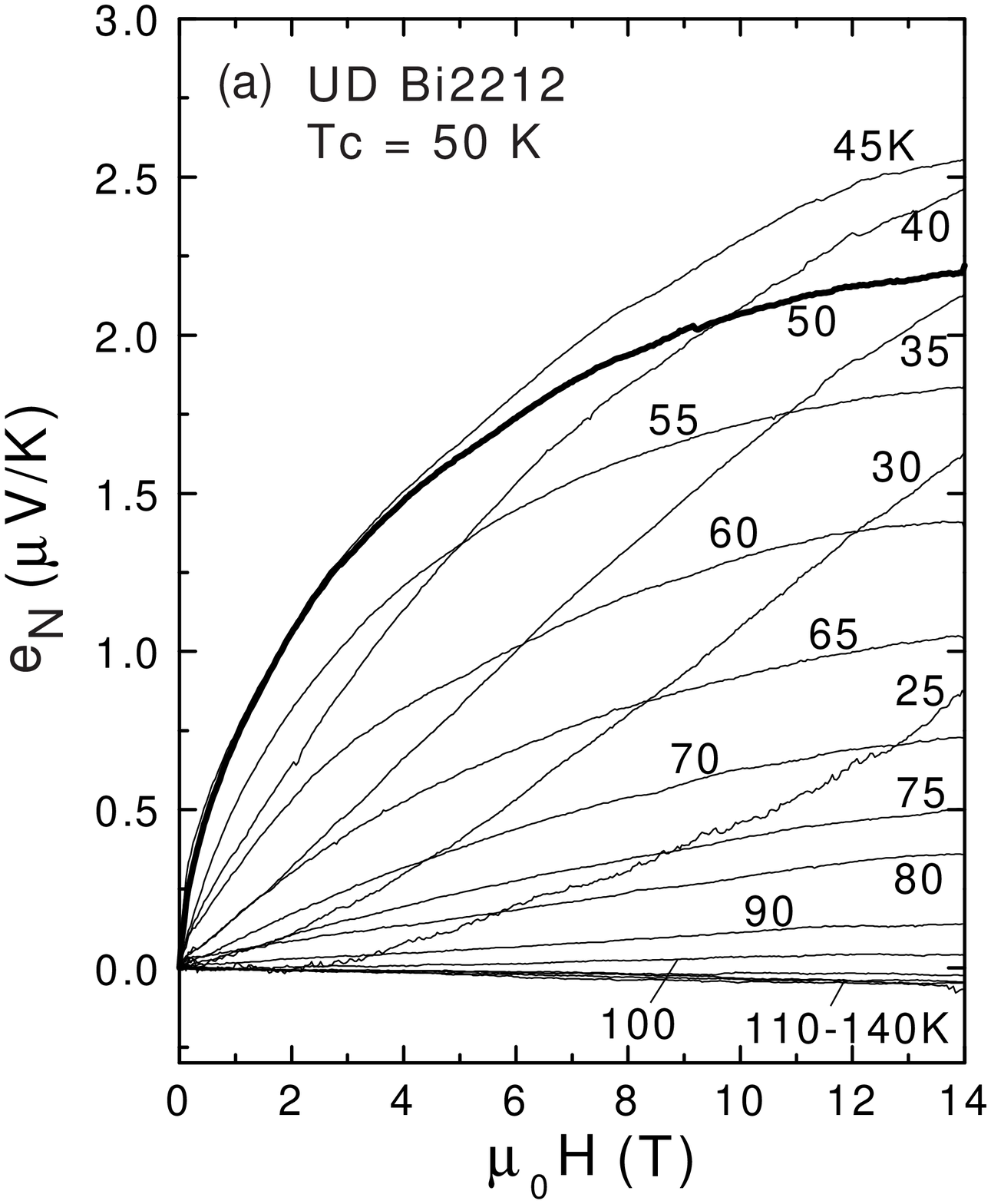}
\incl[width=6cm]{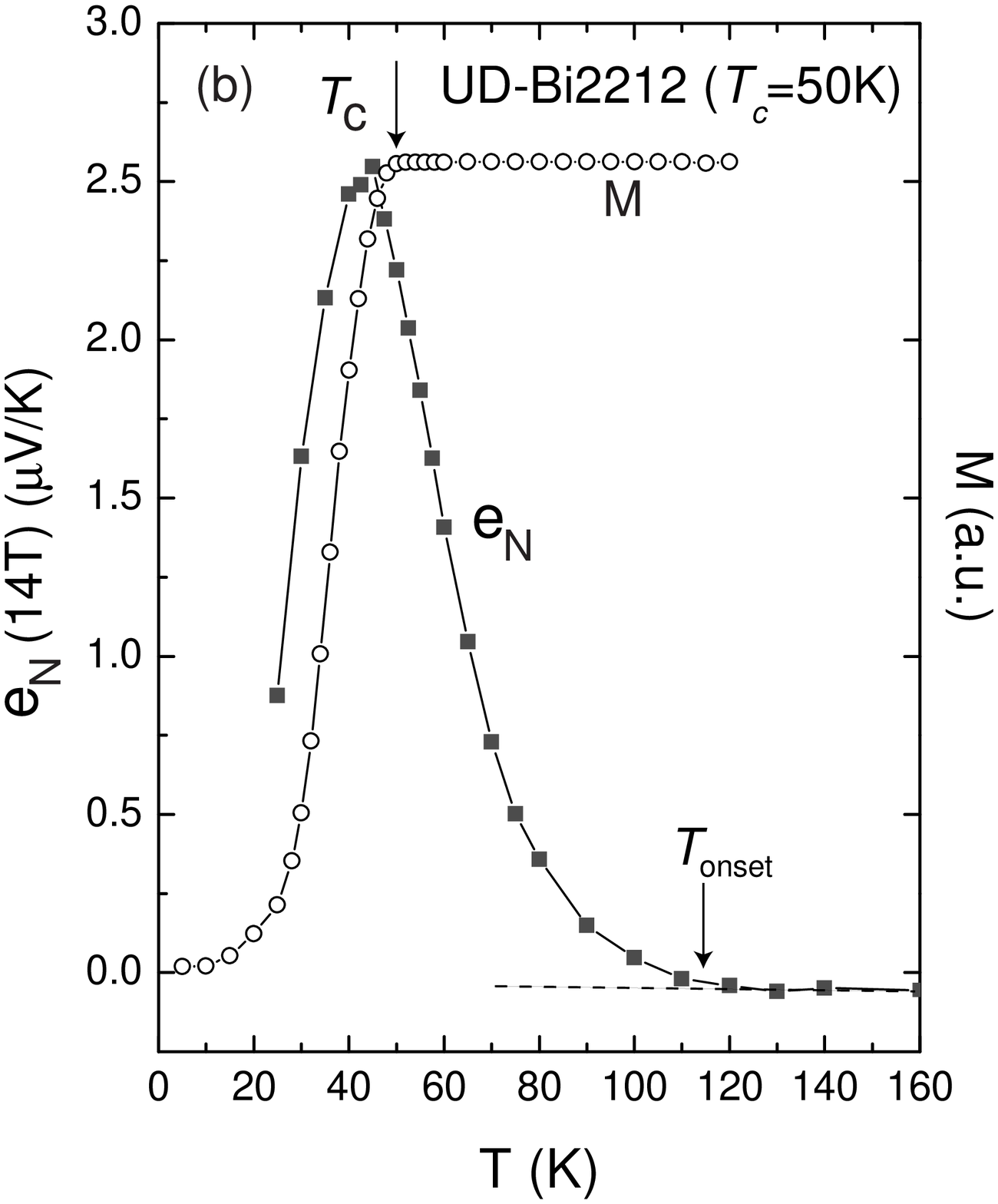}


\caption{\label{B50K} (a) Curves of $e_N$ vs. $H$ in UD Bi 2212 from 25 to 140 K.  The bold curve is
taken at $T_c$ = 50 K.  The curves remain strongly nonlinear up to 70 K.
 (b) Comparison of the $T$ dependence of $e_N$ measured at 14 T and the Meissner
 curve ($M$ at 10 Oe) in UD Bi 2212.  The vortex signal deviates from the qp background
 at $T_{onset}\sim$ 118 K.}  
\efig

We next turn to the underdoped Bi 2212, which has been intensively studied 
by ARPES (angle-resolved photoemission spectroscopy)~\cite{Ding,Shen} and 
STM (scanning tunneling microscopy)~\cite{Seamus,Yazdani} because the crystals 
can be cleaved cleanly.  Nernst results on this system are particularly valuable.  
In Fig.~\ref{B50K}, we show Nernst results on a very underdoped Bi 2212 crystal 
with $T_{c}$ = 50 K and hole density $x \sim$ 0.09.  In close similarity
with very underdoped LSCO ($x$ = 0.07), the curve measured at $T_{c}$ (bold line) has a peak value 
close to the maximum of the envelope (Fig.~\ref{B50K}a).  Traces taken at $T$ higher 
than $T_c$ remain very large in magnitude and possess the strong curvature characteristic 
of the vortex-state.  An important feature of Bi 2212, both UD and OP, is the very 
small magnitude of the qp Nernst signal (the ``background'').  This allows the onset 
temperature $T_{onset}$ to be determined unambiguously.  The $T$ dependence 
of $e_N$ in this UD sample indicates that $T_{onset}$ = 115 K, or about 
65 K above $T_c$ (Fig.~\ref{B50K} b).


\section{The vortex-Nernst profile and continuity across $T_{c}$}\label{profile}
As described in Secs. \ref{optimal} and \ref{under}, the vortex-Nernst signal versus $H$
displays a characteristic peaked form which we call a ``tilted-hill'' profile.  
This is quite apparent in electron-doped NCCO 
where the depairing field (10 T) is readily attained (Sec. \ref{ncco})~\cite{WangSci}. 
In hole-doped cuprates, however, $H_{c2}$ is very large.  The maximum field employed (14 T) in 
earlier experiments was barely enough to reach the peak of the profile.  More recent measurements 
to fields of 45 T now provide a more complete view of the hill profile in Bi 2201 and LSCO.  
The similarity with the profile in NCCO is striking.  

As discussed, $e_N$ rises steeply when $H$ exceeds $H_m$.  
The vanishing of the shear modulus $c_{66}$ in the vortex solid 
allows the vortices to move down the gradient $-\nabla T$, to generate the Nernst 
signal as discussed in Sec. \ref{nernst}.  The steep rise
in $e_N$ above $H_m$ is primarily driven by the increase in
vortex velocity ${\bf v}$, but also reflects the increase in the vortex density $n_v$.  
However, with increasing $H$, the magnitude of the 
magnetization $M$ decreases monotonically as $M\sim-\log H$ over 
a broad interval of field above the lower critical field $H_{c1}$.  
In high fields, $|M|$ decreases to zero at the upper critical field $H_{c2}$,
as $|M|\sim (H_{c2}-H)$ (see Sec. \ref{hc2}).
From Eq. \ref{beta}, $\alpha_{xy}^{s}$ should scale like $M$ in high fields~\cite{Caroli}, so
that it should also approach zero as $\sim (H_{c2}-H)$.  Combining the weak field
and high field trends, we may understand the tilted-hill profile.  Just above $H_m$,
the steep increase of $\bf v$ in the liquid state leads to a rapid increase in $e_N$ until 
$\alpha_{xy}^{s}$ encounters the ceiling set by a decreasing $M$.  In high fields,
the $H$ dependence of $e_N = \rho\alpha_{xy}^{s}\sim M(H)$ 
follows that of $M(H)$ since $\rho$ becomes nearly $H$-independent once
$H$ exceeds $\sim$ 2 $H_m$ (in LSCO and YBCO; in Bi 2201 and Bi 2212
the saturation in $\rho$ is much more gradual).

A similar hill profile is observed in the Ettingshausen signal in
the superconductor PbIn (Fig. \ref{PbIn})~\cite{Vidal}.  The signal rises steeply to a maximum
when the vortex lattice is depinned.  As $H$ approaches $H_{c2}$, the signal decreases
to zero $\sim(H_{c2}-H)$.  The small ``tail'' above $H_{c2}$ is due to amplitude fluctuations. 
As discussed in Sec. \ref{nernst}, the Ettingshausen signal $\sim\tilde{\alpha}_{xy}^{s}$
has the same $H$ dependence as $\alpha_{xy}^{s}$.

\bfig
\incl[width=4cm]{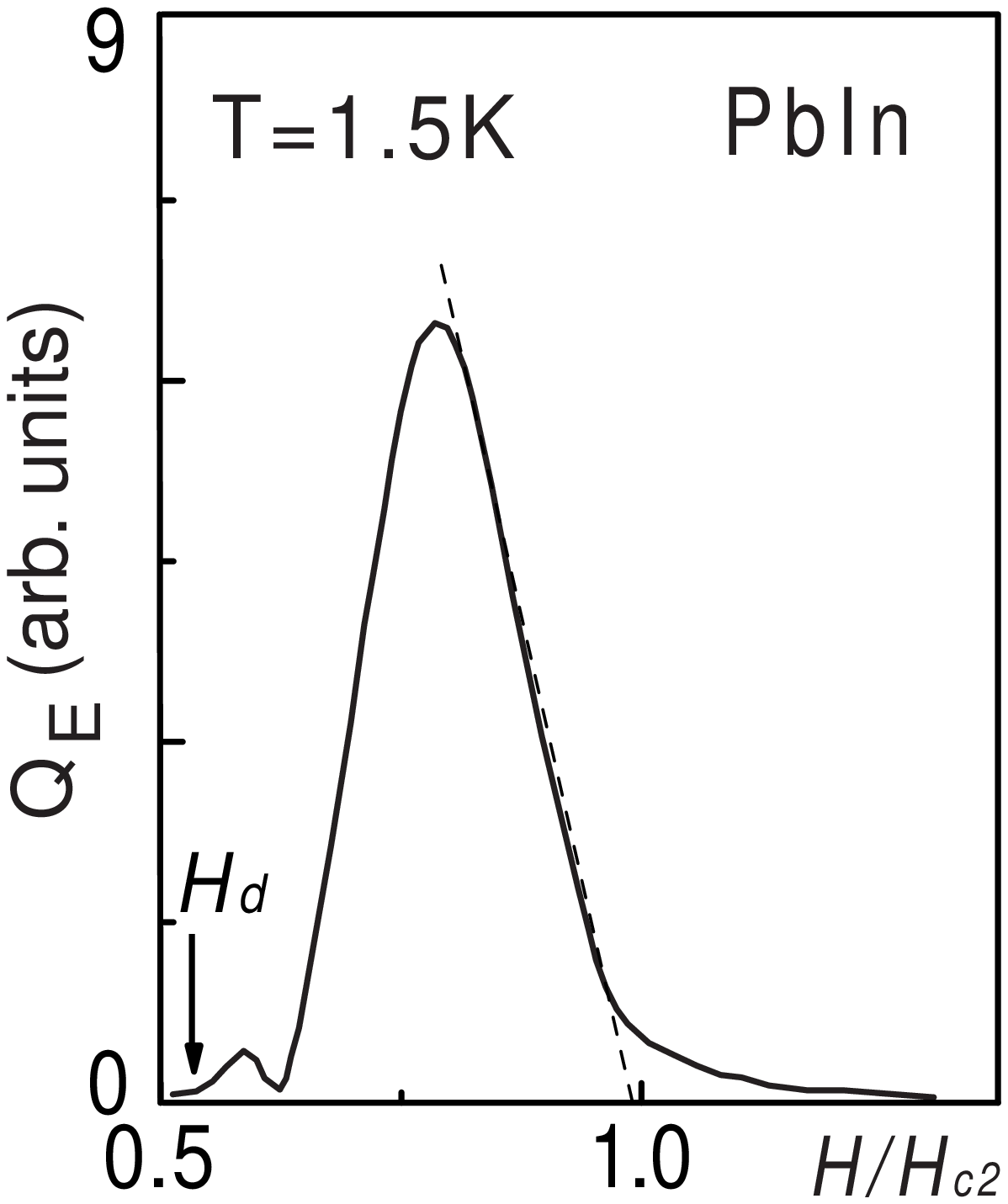}


\caption{\label{PbIn} The Ettingshausen signal ${\cal Q}_E$ vs. $H$ in the 
type II superconductor PbIn (adapted from~\cite{Vidal}).  The $H$ dependence 
of ${\cal Q}_E$ has the characteristic peak profile of the vortex response.  
The dashed line locates $H_{c2}$ determined by $\rho$ vs. $H$.  
Above $H_{c2}$, the signal displays a weak fluctuation tail.}  
\efig

\bfig
\incl[width=6cm]{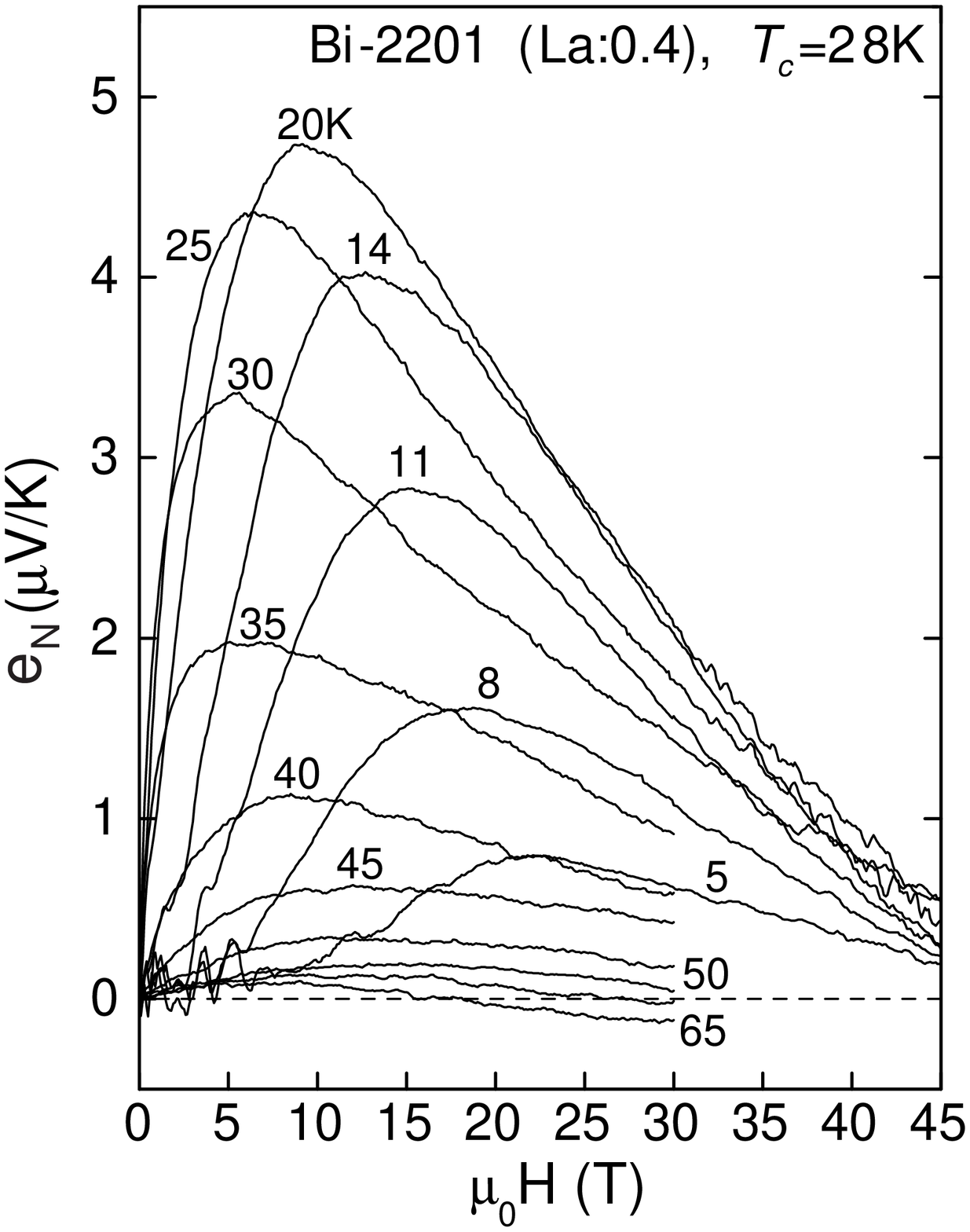}
\caption{\label{NH-Bi04} Curves of $e_N$ vs. $H$ in OP $\rm Bi_2Sr_{2-y}La_yCuO_6$ 
(with La content $y_{La}$ = 0.4, $T_c$ = 28 K) measured to intense fields (45 T below 35 K;
32 T above 35 K).  The depairing field $H_{c2}$ is determined by extrapolation of $e_N$  to zero.  
Above 60 K, the weak negative qp contribution becomes apparent.}  
\efig

By contrast, the qp signal $e_N^{n}$ is nominally linear in $H$ with a small $H^3$ correction
observable only in high fields ($>$20 T) at low $T$ ($<$10 K).  The short qp mean-free-path $\ell$
in hole-doped cuprates ($\ell\sim$ 80 \AA~at 80 K) precludes any possibility of observing 
a hill-type profile in $e_N^{n}$ even in fields 20-40 T.

\bfig
\incl[width=6cm]{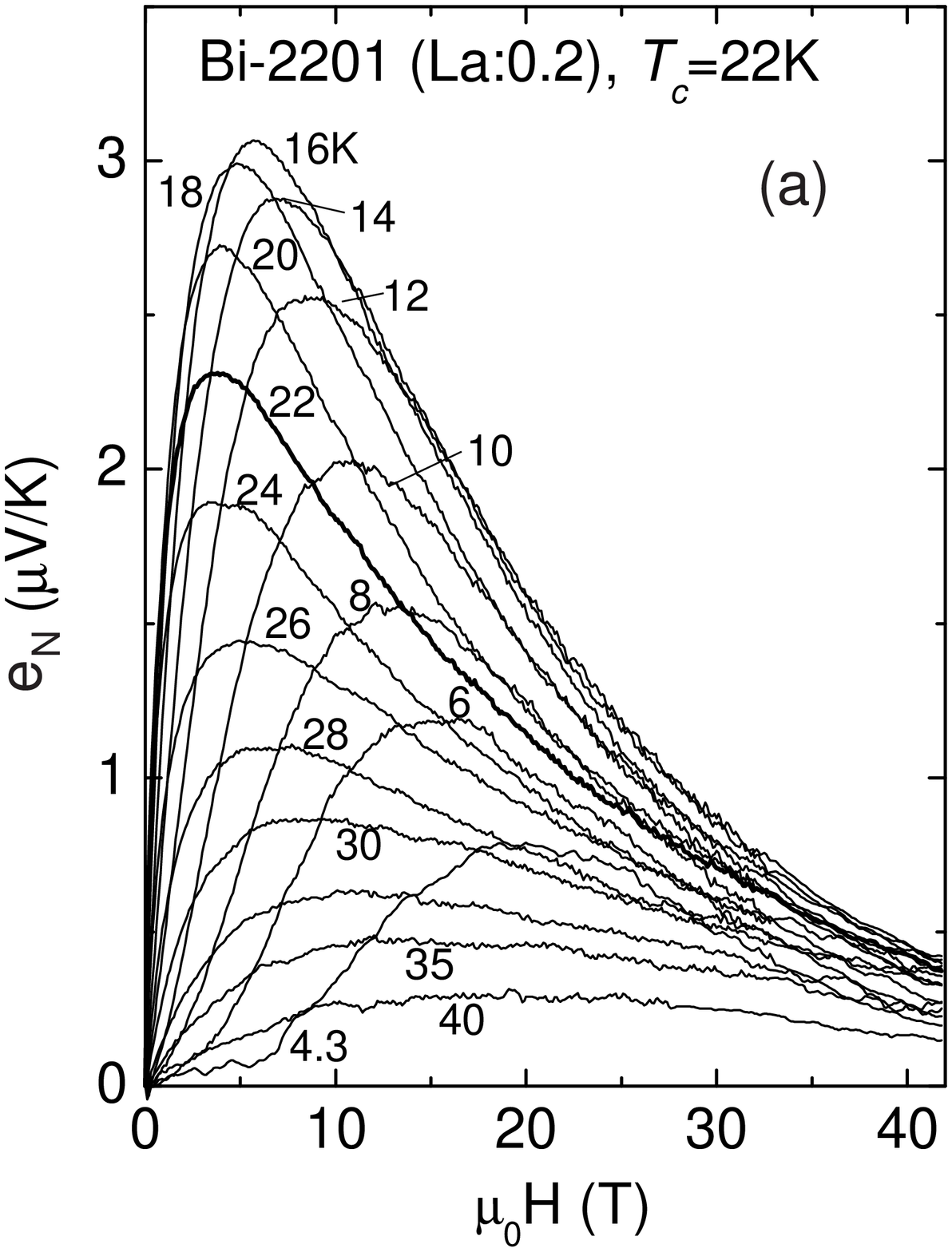}
\incl[width=6cm]{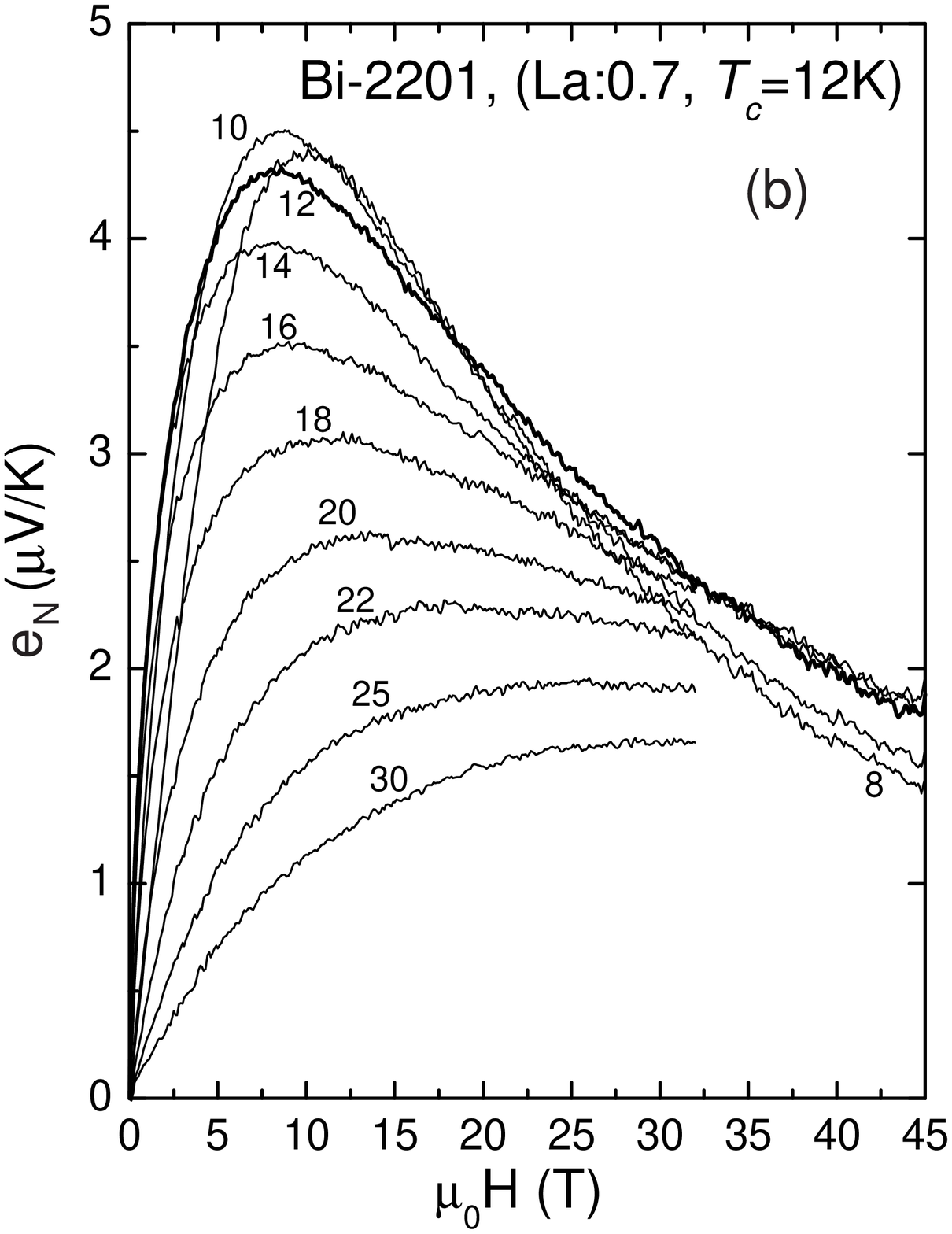}
\caption{\label{NH-Bi02} (a) Curves of $e_N$ vs. $H$ in 
OV $\rm Bi_2Sr_{2-y}La_yCuO_6$ ($y_{La}$ = 0.2, $T_c$ = 22 K) measured to 
intense fields from 4.3 to 40 K.  At all $T$, the curves appear to extrapolate 
to zero at the same nominal field scale (42-45 T).
(b) Curves of $e_N$ vs. $H$ in 
UD $\rm Bi_2Sr_{2-y}La_yCuO_6$ ($y_{La}$ = 0.7, $T_c$ = 12 K) measured to
45 (32) T for temperatures below (above) 12 K.  
}  
\efig

Figure~\ref{NH-Bi04} shows Nernst curves measured in OP Bi 2201 in $H$ up to 
45 T.  The curves taken below $T_c$ (28 K) all display the tilted-hill profile.
As $H$ is increased above $H_m$, $e_N$
rises steeply to a prominent maximum and then falls more slowly in high fields with a
slope that is only weakly $H$-dependent.  Significantly, when we exceed $T_c$,
the curves retain the same hill profile (see curves at 30--45 K). 
In fact, the curves up to 65 K show the same nominal profile except that
the maximum is quite broad.  However, above $\sim$50 K, the negative qp term $e_N^{n}$ 
grows in significance and pulls $e_N$ towards negative values at large $H$.

Similar results are also seen at other dopings.  In Fig.~\ref{NH-Bi02}a we show Nernst 
results on OV Bi 2201 ($y_{La}$ =  0.2) with $T_c$ = 22 K.  Even at $T$ = 40 K or $\sim 2T_c$, 
the curve retains the characteristic hill profile of the vortex signal.  In underdoped 
Bi 2201 ($y_{La}$ = 0.7) shown in Fig.~\ref{NH-Bi02}b, the effect is even more dramatic.  
The curves of $e_N$ measured above $T_c$ (= 12 K)
continue to show a vortex profile up to our highest temperature 30 K.
The extension of the tilted-hill profile to $T$ high above $T_c$ implies that 
the same mechanism generating $e_N$ below $T_c$ -- vortex flow -- must be operating
above $T_c$.

\bfig
\incl[width=8cm]{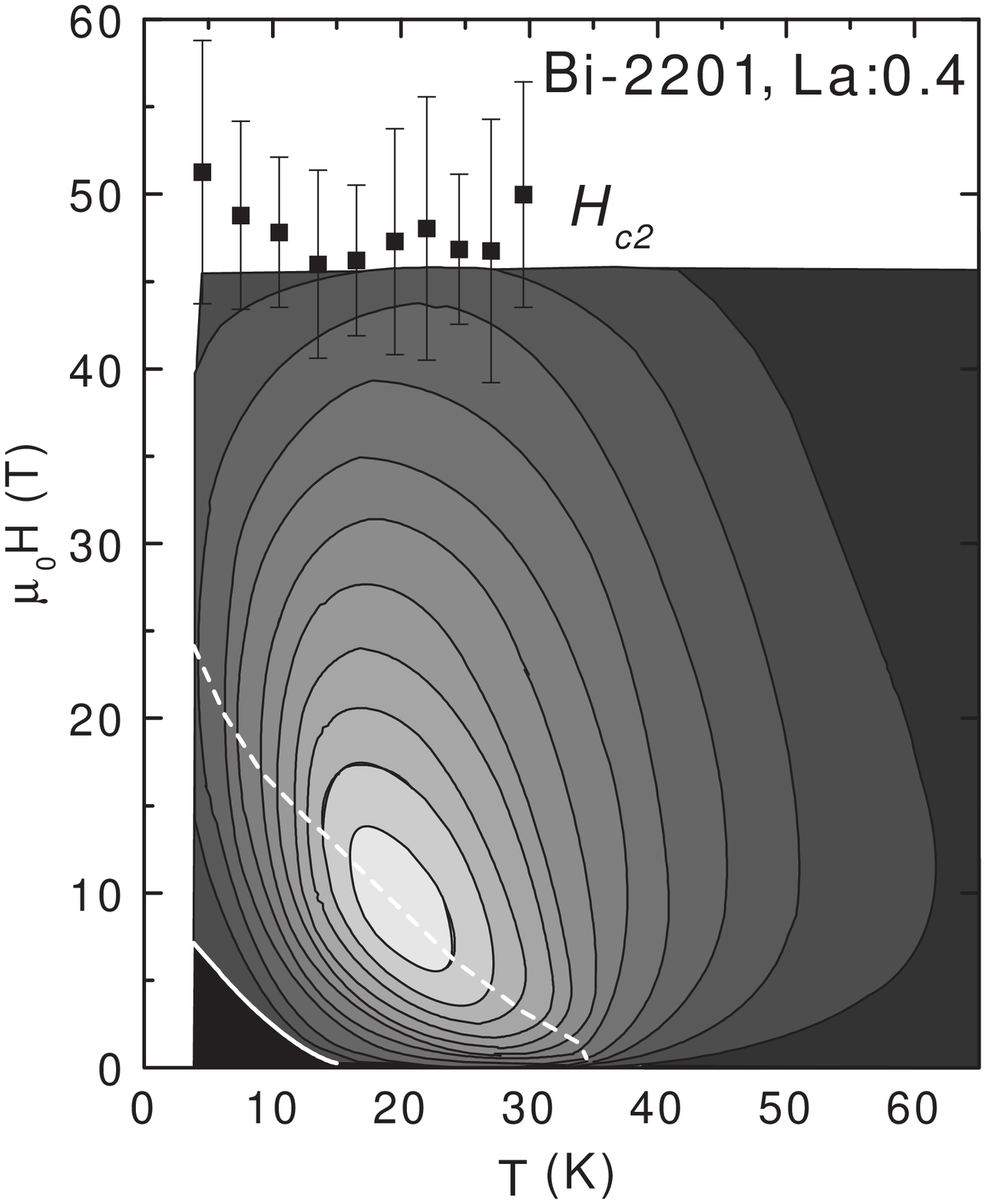}
\caption{\label{contourBi} Contour plot of $e_N(T,H)$ in OP Bi 2201 ($y_{La}$ = 0.4).
The value of $e_N$ is highest in the light-gray region
and zero in black regions.  The white curve terminating at 15 K is $H_m(T)$.  The dashed
curve is the ridge-line joining points of maxima of $e_N$ vs. $H$.  Solid squares are values of $H_{c2}$
estimated from extrapolation of the curves in Fig. \ref{NH-Bi04}.  The plot emphasizes the smooth
continuity of the vortex signal to temperatures high above $T_c$ (28 K).
}
\efig

A graphic way to represent the continuity between the Nersnt region above $T_c$ 
and the vortex-liquid state below $T_c$ is the contour plot in the $T$-$H$ plane 
~\cite{WangPRL}.  In Fig. \ref{contourBi}, the grey scale
represents regions with successively higher values of $e_N$ in Bi 2201 ($y_{La}$ = 0.4).  
The black area ($e_N$ = 0) is the vortex solid phase below $H_m(T)$.  
If we increase $H$ at fixed $T<T_c$ (= 28 K), $e_N$ climbs rapidly just above $H_m$
and attains a maximum (lightest shade), before dropping gradually towards zero at $H_{c2}$ 
(solid squares are $H_{c2}(T)$ values discussed in the next section).  All the preceding 
figures displaying curves of $e_N$ vs. $H$ are vertical cuts in the $T$-$H$ plane.  
The maxima in the $e_N$-$H$ curves define the ridge-field  $H_{ridge}(T)$ (dashed 
curve in Fig. \ref{contourBi}). As the pair condensate remains very large 
above $H_{ridge}$, it clearly lies far below the depairing field $H_{c2}$.  We 
discuss $H_{ridge}$ in relation to $\rho$ in Sec. \ref{phase}.

The contour plot provides a global view of the tilted-hill profiles shown 
in Fig. \ref{NH-Bi04}.  The strong curvature of the contour lines at high $T$
implies that the hill profile is also observed above $T_c$, as noted above.
From the contour plot, it is clear that the vortex liquid state just above $H_m$ smoothly
extends into the Nernst region above $T_c$.  There is no phase boundary discernible between
the 2 regions; the vortex liquid state below $T_c$ spreads continuously to temperatures above $T_c$.  
Overall, the magnitude of $e_N(T,H)$ changes very smoothly over the whole
$T$-$H$ plane.  The only indication of $T_c$ is the approach of the shallow contour minima 
towards 28 K as $H\rightarrow$ 0.  We refer the reader to Ref. \cite{WangPRL,OngRio,OngAnn}
for contour plots of LSCO and YBCO.  
The contrasting contour plot in NCCO is discussed below (Sec. \ref{ncco}).


\section{The upper critical field}\label{hc2}
In the hole-doped cuprates, the upper critical field (or depairing field) $H_{c2}$ is a
rather poorly established quantity compared with the other parameters of the superconducting state.  
On the one hand, flux-flow resistivity experiments have given a very low estimate 
of $H_{c2}$~\cite{Mackenzie,Alexandrov96} ($\rho$ is discussed at the end of this section).  
On the other, it was widely believed that $H_{c2}$ in the cuprates is an 
inherently \emph{unmeasurable} quantity.

A powerful advantage of the Nernst experiment is that it provides a 
direct determination of $H_{c2}$ that remains sensitive in intense magnetic fields.
In type II superconductors, as $H\rightarrow H_{c2}$ from below, 
the packing of vortex cores steadily reduces the volume fraction
of the condensate in the interstitial ``puddles'' between cores.  The coherence length $\xi$
is related to the upper critical field by $H_{c2} = {\phi_0}/(2\pi\xi^2)$.
For fields just below $H_{c2}$, the supercurrent is~\cite{deGennes} 
\be
{\bf J}_s = -\frac{e\hbar}{m}\nabla\times |\Psi|^2\bf\hat{z}
\label{Js}
\ee
(with $\bf H||\hat{z}$).  The (diamagnetic) circulation of the supercurrent ${\bf J}_s$ 
around each of the interstitial condensate puddles generates a magnetization $\bf M$ 
that is greatly reduced from its value near $H_{c1}$ and given by Abrikosov's expression 
\be
{\bf M} = -\frac{(H_{c2}-H){\bf\hat{z}}}{\beta_A(2\kappa^2-1)},
\label{M}
\ee
with $\beta_A\sim 1$ and $\kappa = \lambda/\xi$, where $\lambda$ is the penetration length.  
Traditionally, the curve of $M$ vs. $H$ has provided the most reliable method for finding $H_{c2}$.

However, in the cuprates, where $\kappa\sim 100$ 
and $H_{c2}\sim$ 50-150 T, resolving the greatly suppressed $M$ in high fields 
has been a formidable challenge.  Recently, though,
rapid progress is being made using high-field torque magnetometry (some of the new
$M$ vs. $H$ results are discussed in Sec. \ref{magnetization}.)

The Nernst experiment provides an alternative way to measure $H_{c2}$.  As stressed in Sec. \ref{profile}, 
the curve of $e_N$ vs. $H$ has a characteristic peaked profile.  On the high-field side, 
$e_N$ is driven inexorably to zero in proportion to the magnetization (Eq. \ref{M}).
This is clearly seen in the Ettingshausen curve in the low-$T_c$ superconductor PbIn 
discussed in Sec. \ref{nernst} (Fig.~\ref{PbIn}).  
The signal peaks near 0.8 $H_{c2}$ and then decreases to zero as $(H_{c2}-H)$
(ignoring the high-field tail caused by amplitude fluctutations).
The high-field end-point of $\alpha_{xy}^{s}$ (or $\tilde{\alpha}_{xy}^{s}$)
may be used to locate $H_{c2}$.

In most cuprates, the values of $H_{c2}$ exceed the 45-Tesla maximum available in current $dc$ magnets.
In analogy with the low-$T_c$ case, we assume that linear extrapolation of the high-field Nernst signal
to zero gives a reliable determination of the scale of $H_{c2}$.  Adopting this assumption, 
we have broadly applied high-field Nernst experiments to estimate $H_{c2}$ in several cuprate families.

\bfig
\incl[width=6cm]{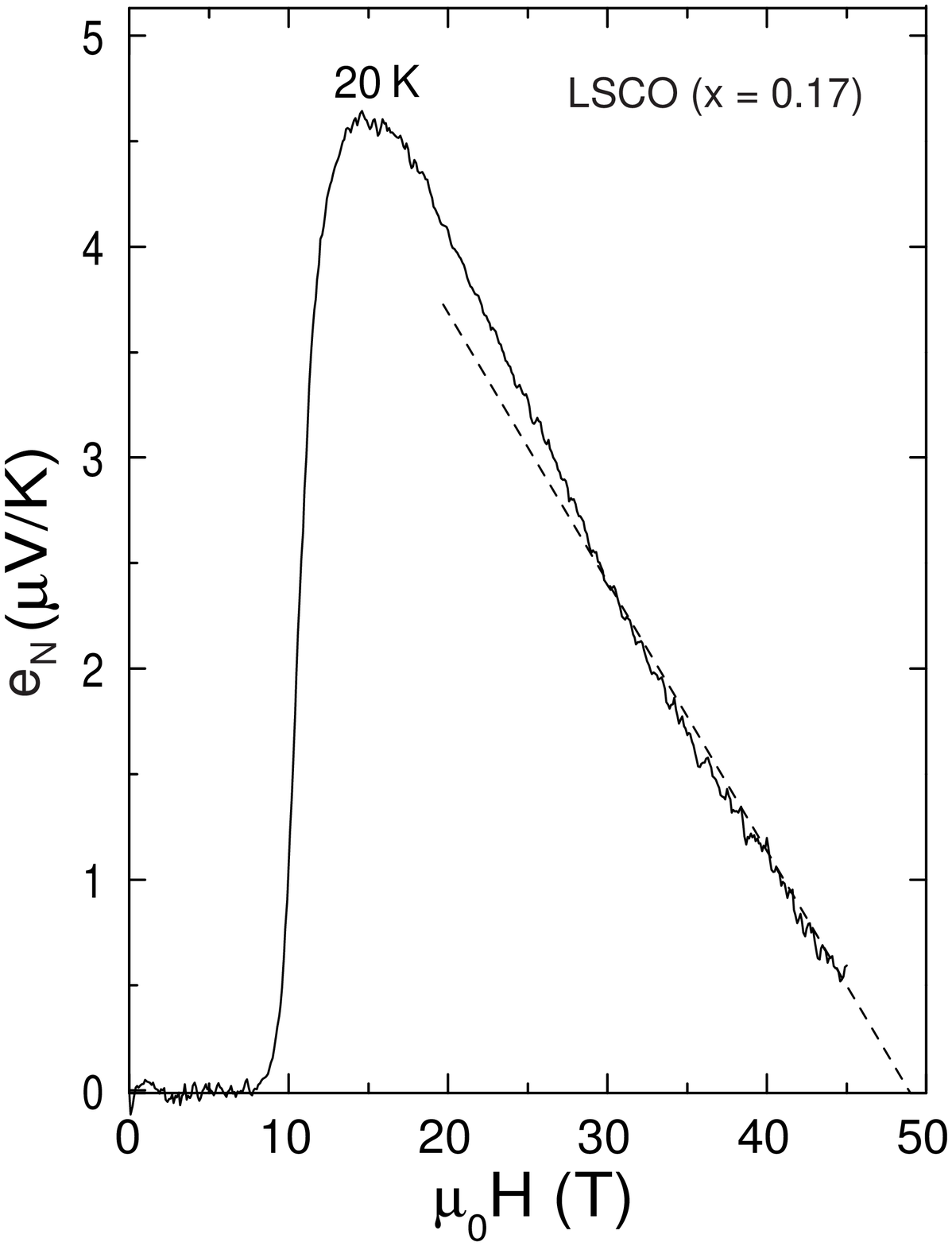}
\caption{\label{L17-20K}  The curve of $e_N$ vs. $H$ measured at $T$ = 20 K in 
OP LSCO ($x$ = 0.17, $T_c$ = 36 K).  The value of $H_{c2}\sim$ 50 T is estimated by 
the dashed-line extrapolation of the high-field data to zero.
}
\efig

\bfig
\incl[width=6cm]{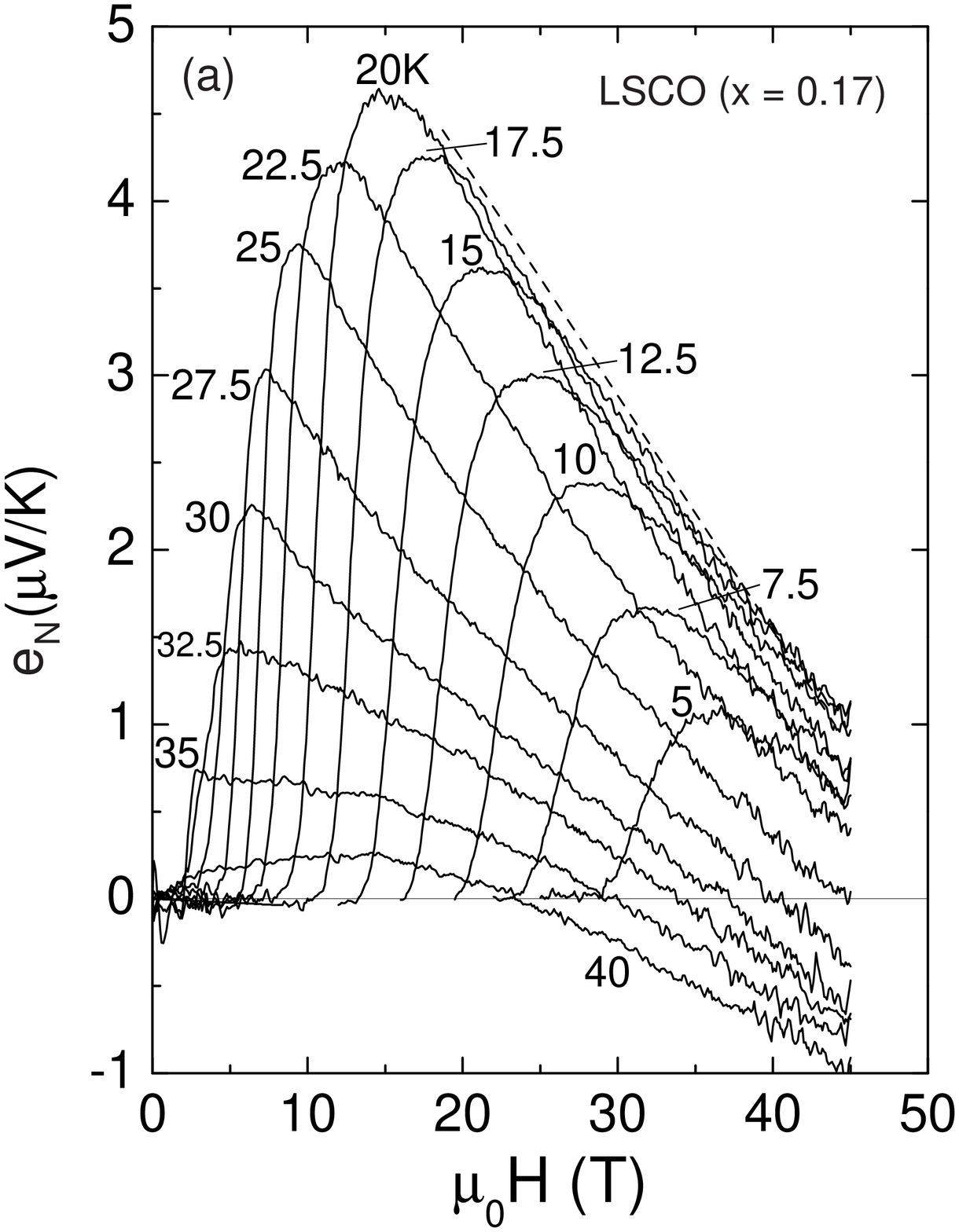}
\caption{\label{L17-Hc2}  The high-field Nernst curves in optimally-doped LSCO ($x$ = 0.17) 
from 5 to 40 K.  Below 20 K, all curves merge to the dashed line at high fields.  As $T$ rises above 20 K,
the qp contribution increasingly ``pulls'' the curves of $e_N$ to negative values in high fields.  
This effect is much less pronounced at lower $T$.
}  
\efig

First, we consider optimally-doped LSCO ($x$= 0.17, $T_{c}$
= 36 K).  In Fig. ~\ref{L17-20K}, the extrapolation of $e_N$ at 20 K gives $H_{c2}\sim$ 50 T.
Figure~\ref{L17-Hc2} displays $e_N$ vs. $H$ in this sample at selected $T$.  An
interesting trend that is immediately apparent is that, in high fields, all the curves below 20 K 
merge to a common line (dashed line).  With $H_{c2}$ determined by linear extrapolation, 
we obtain the conclusion that $H_{c2}$ is almost $T$-independent 
from 5 K to 20 K.  
Unfortunately, in OP and OV LSCO, the qp negative background is 
moderately large above 20 K.
As shown in Fig. \ref{L17-Hc2}, the $H$-linear qp term added to the diminishing 
vortex term ``pulls'' the observed $e_N$ to negative values.  This prevents 
$H_{c2}$ from being readily estimated above 20 K in OP and OV LSCO.  
However, because $\alpha_{xy^n}$ is an entropy current, the qp term must 
decrease to zero as $T\rightarrow$ 0 (this is shown 
for UD LSCO in Ref. \cite{WangPRB}).  Hence it doesn't affect our estimate of
$H_{c2}$ at low $T$.

\bfig
\incl[width=6cm]{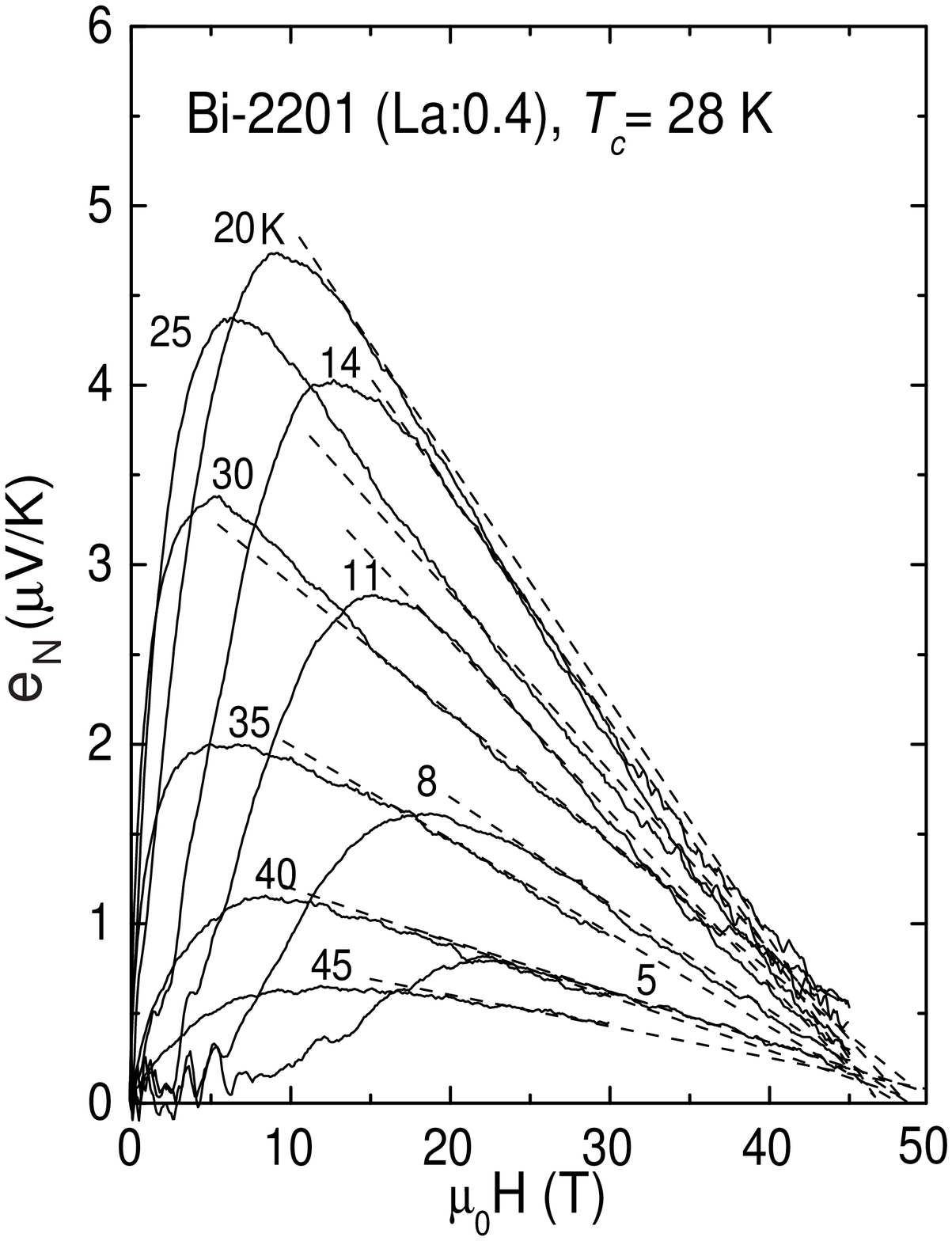}
\caption{\label{Bi04-Hc2}  Extrapolation of the curves of $e_N$ (dashed lines) to determine
$H_{c2}$ values in OP Bi 2201 ($y_{La}$ = 0.4).  The estimated values of $H_{c2}$ are 
virtually $T$ independent.}  
\efig

Determination of $H_{c2}$ is most reliably carried out in single-layer Bi 2201 
where the qp contribution is very small and $H_{c2}$ values slightly more accessible.  
Figure~\ref{Bi04-Hc2} displays curves of $e_N$ in optimally-doped Bi 2201 ($y_{La}$ = 0.4).  
We show the linear extrapolations as dashed lines.  Again, we see 
that the curves all extrapolate to zero at nearly the same $H$.  This implies 
a $T$-independent $H_{c2}$ value of 48 $\pm$ 4 T for temperatures from 5 to 45 K.  
We remark that, quite independent of the extrapolations, 
the convergent behavior of the measured curves already reveals this surprising result.  
This trend is also seen at other dopings.  The Nernst traces in overdoped Bi 2201 ($y_{La}$ = 0.2) 
also exhibit this convergence (Fig.~\ref{NH-Bi02}).

Curves of $H_{c2}$ vs. $T$ in optimally-doped and overdoped Bi 2201 are displayed in 
Fig.~\ref{Hc2-T}.  Within the uncertainty of the data, the $H_{c2}$ values in both samples are $T$ 
independent from 4 K to well above $T_c$. This behavior is 
in sharp contrast with low-$T_c$ type-II superconductors where $H_{c2}$ decreases
linearly as $H_{c2}\sim(T_c-T)$ near $T_c$.  The data show that the $H_{c2}$ values 
continue nearly unchanged for a significant interval above $T_c$.

\bfig
\incl[width=5cm]{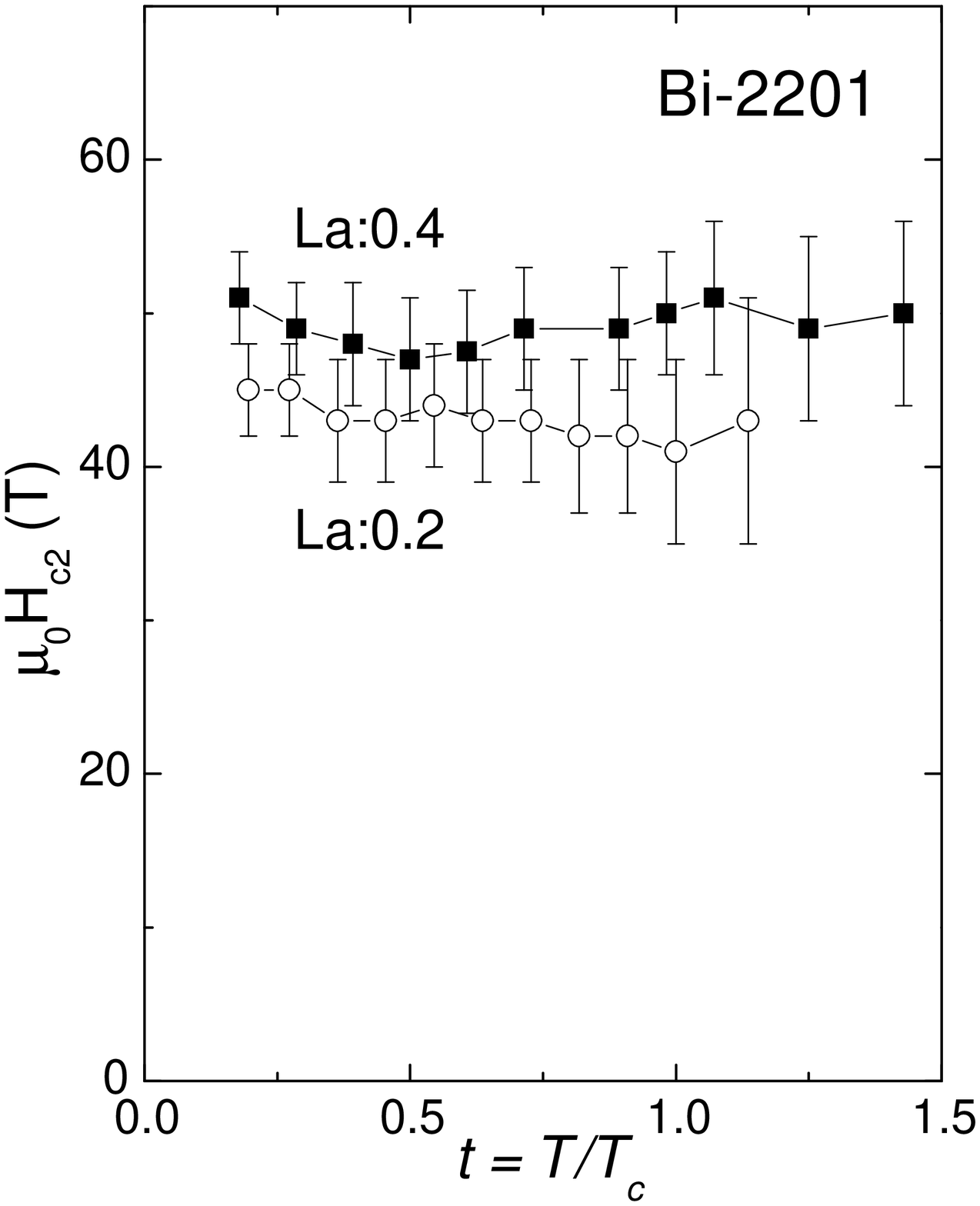}
\caption{\label{Hc2-T} Plot of $H_{c2}$ versus the reduced 
temperature $t = T/T_c$ in OP and OV Bi 2201 (with $y_{La}$ = 0.4 and 0.2, respectively).
At each $T$, $H_{c2}$ is estimated as in Fig. \ref{Bi04-Hc2}.
}  
\efig

This anomalous result is closely related to findings from ARPES~\cite{Ding,Harris} 
that the gap amplitude $\Delta_0$ in Bi 2212 is nearly $T$ independent below $T_c$, and varies
only weakly above $T_c$.  Tunneling experiments on Bi 2201 have also shown that the gap does 
not close at $T_c$ but remains finite above
$T_c$~\cite{Renner}.  We may relate $\Delta_0$ to the Pippard length $\xi _P$ by 
\be
\xi _P = \frac{\hbar v_F}{a\Delta_0},
\ee
where $a\sim$1.5 for a $d$-wave gap (compared with $\pi$ for $s$-wave) 
and $v _F$ the Fermi velocity.  Assuming that $\xi\sim\xi_P$ in our samples, we have
$H_{c2}\sim \Delta_0^2$.  Hence the constancy of the gap amplitude across $T_c$ implies
our finding that $H_{c2}$ is constant across $T_c$.  In Ref. \cite{WangSci}
the doping dependence of $\xi_P$ infered from ARPES is shown to be quantitatively similar to 
that of $\xi$ obtained from the Nernst experiment.  Finally, recent measurements of $M$ vs. $H$ to 33 T
have confirmed this anomalous constancy of $H_{c2}$ across $T_c$ in UD and OP Bi 2212 (Sec. \ref{magnetization}).

As reported in Ref. \cite{Ding-PRL}, the gap amplitudes
$\Delta_0$ in a slightly UD and an OV Bi 2212 are nominally $T$-independent from low 
$T$ to well above $T_c$.  It is instructive to compare these curves with our inferred $H_{c2}$ vs. $T$ 
plot in Fig.~\ref{Hc2-T}.   Both experiments imply that, in the cuprates, $H_{c2}$
is nearly unchanged from low $T$ to above $T_{c}$. 

As mentioned, the constancy of $H_{c2}$ across $T_c$ is strikingly inconsistent with the mean-field 
BCS scenario in which $|\Psi|^2$ vanishes at $T_c$.  By contrast, it supports strongly the scenario that 
the collapse of the Meissner state at $T_c$ is caused by the loss of long-range phase coherence,
with $|\Psi|$ remaining finite above $T_c$.  The loss of phase coherence arises from the spontaneous
generation of vortices and the resultant rapid phase-slippage caused by their motion.
The constancy of $H_{c2}$ up to $T_c$ implies that it 
actually goes to zero only at a much higher temperature (the mean-field transition $T^{MF}_c\gg T_c$).  
In the 2D KT transition, Doniach and Huberman~\cite{Doniach} have noted that $H_{c2}$ (or the depairing field)
must remain at a high value across the KT transition temperature $T_{KT}$.
In summary, the anomalous behavior of $H_{c2}$ in the hole-doped cuprates strongly supports our vortex
interpretation of $e_N$.  

\bfig
\incl[width=6cm]{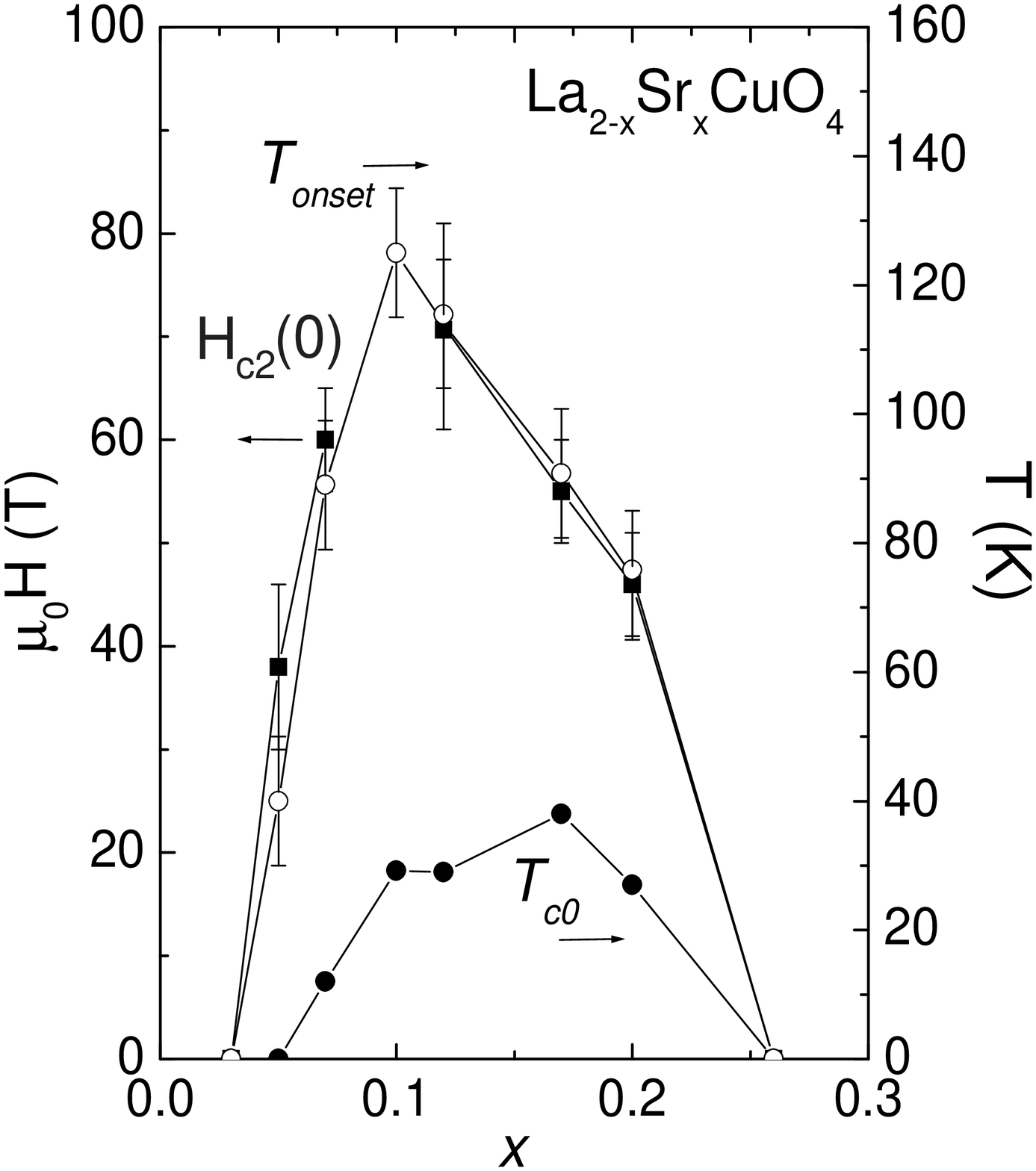}
\caption{\label{Hc2x}
Variation of the low-$T$ upper-critical field $H_{c2}(0)$ estimated at 4.2 K versus $x$ in LSCO (solid squares).
The values are estimated by extrapolation of $e_N^{s}$ to zero from measurements
in $H$ to 45 T.  For comparison, we also plot $T_{onset}$ (open circles) and $T_c$ (solid circles).
Lines are guides to the eye.}
\efig

The doping dependence of $H_{c2}$ was initially investigated from the slightly UD 
regime to the OV regime.  In Ref.~\cite{WangSci}, it was found that $H_{c2}$ decreases
systematically from slightly UD to OP to OV samples in Bi 2212, Bi 2201 and LSCO.
The trend agrees with that of $\Delta_0$ measured by ARPES in Bi 2212 by Harris \etal~\cite{Harris}
and Ding \etal~\cite{Ding-PRL}.  Further, the values of $\xi$ inferred from $H_{c2}$ 
vs. $x$~\cite{WangSci} are consistent with $\xi_P$ obtained from $\Delta_0$ and 
with the vortex core size observed by STM~\cite{Pan-STM}. 

We have now extended these $H_{c2}$ estimates over the whole doping range
in LSCO, in 5 crystals with $x$ = 0.05, 0.07, 0.12, 0.17 and 0.20. 
The values of $H_{c2}$ determined at our lowest $T$
(4.2 K) -- which we identify as $H_{c2}(0)$ -- are plotted as 
solid squares in Fig. \ref{Hc2x}.  As shown, the $x$ dependence of 
$H_{c2}(0)$ is nominally similar to that of $T_{onset}$ (open circles).  As $x$
increases from 0.03, $H_{c2}(0)$ rises very steeply to peak near 0.10, and
then decreases more gradually towards 0 as $x\rightarrow$ 0.26.  (The values 
for $x>$ 0.10 are in agreement with the estimates reported in Ref.
\cite{WangSci}, but the steep fall on the low-$x$ side was not investigated in that study.)

\bfig
\incl[width=6cm]{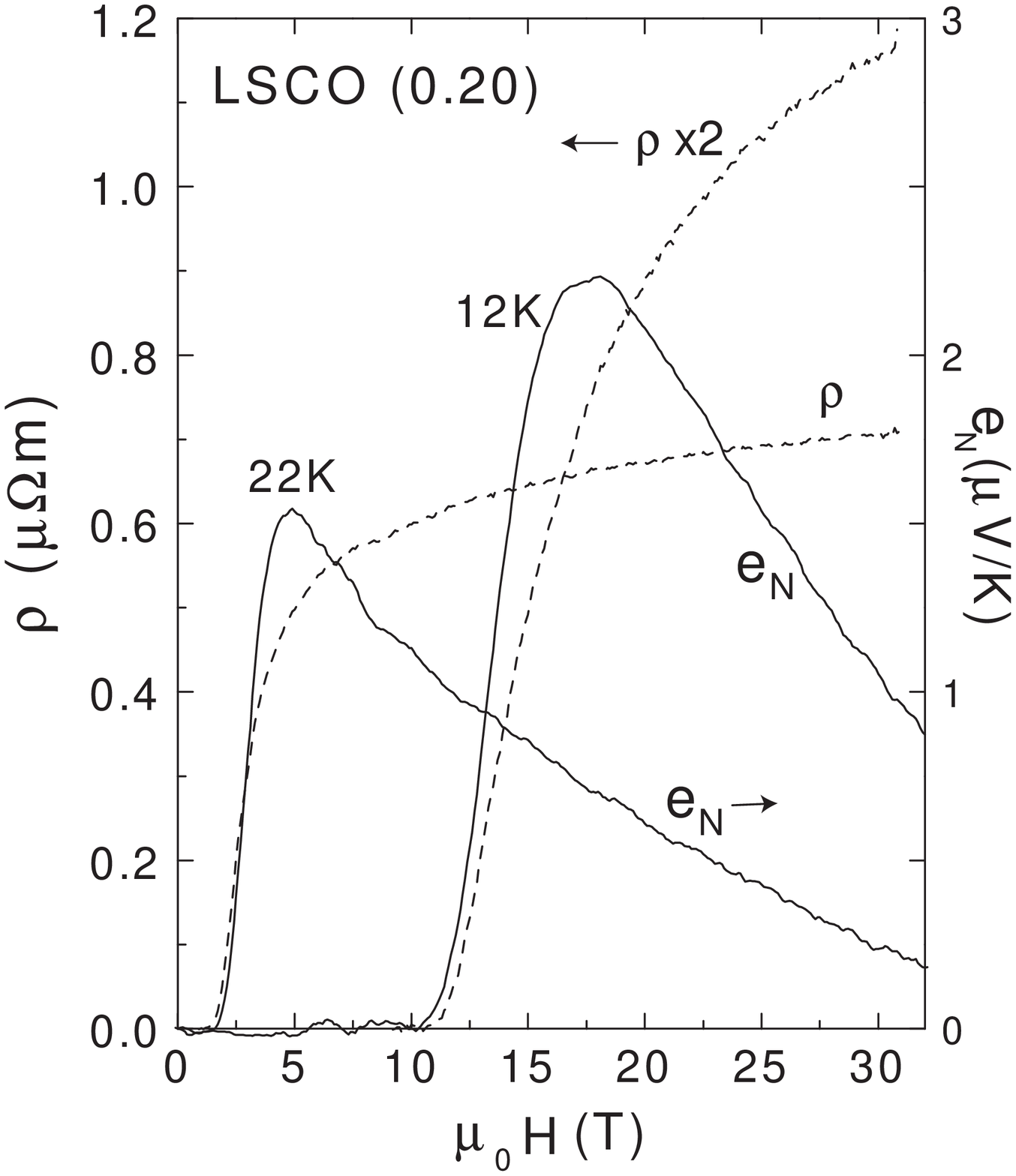}
\caption{\label{rho}  Comparison of the field profiles of the flux-flow resistivity
$\rho$ and the Nernst signal $e_N$ measured on the same sample, an overdoped 
crystal of LSCO ($x$ = 0.20) at $T$ = 22 and 12 K.  
Above $H_m$, $\rho$ quickly approaches saturation to the resistivity value extrapolated 
from above $T_c$ (this occurs near $H_{ridge}$ defined by the peak in $e_N$).  
However, $e_N$ decreases to zero at the depairing field $H_{c2}$ which lies much higher
($H_{c2}\sim$ 45 T).
}
\efig

\emph{Resistivity}
In low-$T_c$ type-II superconductors, the flux-flow resistivity $\rho$ provided a 
convenient means to determine $H_{c2}$.  If the applied current $I$ is high enough
to depin the vortex lattice, $\rho$ increases linearly as $B\phi_0/\eta = (H/H_{c2})\rho_N$ 
to reach the normal-state value $\rho_N$ at $H_{c2}$ (the Bardeen-Stephen Law~\cite{Kim}).
Conveniently, $H_{c2}$ is often flagged by a sharp notch minimum in $\rho$ (the peak effect).  The Bardeen
Stephen law is rarely -- if ever -- observed in cuprates.
Just above the melting field $H_m(T)$, $\rho$ rises very rapidly to saturate 
at a value close to that of the pre-transition $\rho$ suitably extrapolated below $T_c$.
Two examples of $\rho$-$H$ profiles are shown in Fig. \ref{rho} in OV LSCO ($x$ = 0.20).
Previously, attempts were made to identify the ``knee'' feature corresponding
to this saturation with ``$H_{c2}$''~\cite{Mackenzie,Alexandrov96}.  
The inferred curve of $H_{c2}$ vs. $T$ invariably displayed the wrong curvature, 
together with anomalously low depairing fields (0.01-0.1 T) near $T_c$.  They are strikingly incompatible with
the $H_{c2}$ values obtained from the Nernst results.

The field profiles of $\rho$ and $e_N$ at 22 K are compared in Fig. \ref{rho}.
As mentioned, the knee feature in $\rho$ occurs near $H_{ridge}\sim$ 5 T.
However, the vortex signal remains quite large above the knee, eventually decreasing
to zero at the much higher $H\sim$48 T.  At 12 K, the knee feature in $\rho$ is much broader, but it
occurs at $\sim$20 T, still considerably below 48 T.  The comparison emphasizes the
fallacy of identifying the saturation of $\rho$ with a depairing field scale.  The condensate 
amplitude remains robust up to considerably higher fields.  We argue that the
knee feature instead reflects the shrinking with increasing field of the length scale over which phase \emph{stiffness} 
holds.  This loss occurs in the field interval between $H_m$ and the $H_{ridge}$ curve (dashed line
in Fig. \ref{contourBi}).  In Bi-based cuprates this loss is quite gradual, whereas in 
OP/OV YBCO and LSCO it is abrupt (Figs. \ref{NH-Y92K} and \ref{L17-20K}, respectively).
Further, above $H_m$, the dissipation climbs much more rapidly than prescribed by the Bardeen-Stephen law.  
This rapid increase implies a very weak damping viscosity $\eta$ and is 
known as the fast-vortex problem (Sec. \ref{discussion}).


\section{Phase diagram, onset temperature and magnitude}\label{phase}
In the phase diagram of the cuprates, superconductivity occupies a 
dome-shaped region defined by the curve of $T_c$ vs. $x$.  The pseudogap 
temperature $T^*$ decreases monotonically from the scale 300-350 K to 
terminate at the end-point $x_p$ 
(the Nernst experiments along with many experiments indicate that $x_p\sim$ 0.26, 
but other groups~\cite{Tallon} favor $x_p$ = 0.19).  
As reported previously~\cite{WangPRB,OngAnn}, 
in the phase diagram of LSCO, the onset temperature of the Nernst signal $T_{onset}$ 
falls between $T^*$ and $T_c$.  As $x$ increases from 0.03, $T_{onset}$ rises 
steeply to a maximum value 
of 130 K at 0.10 and then falls more
gradually to a value near zero at $\sim$0.27 (Fig. \ref{LSCO-phase}).

\bfig
\incl[width=6cm]{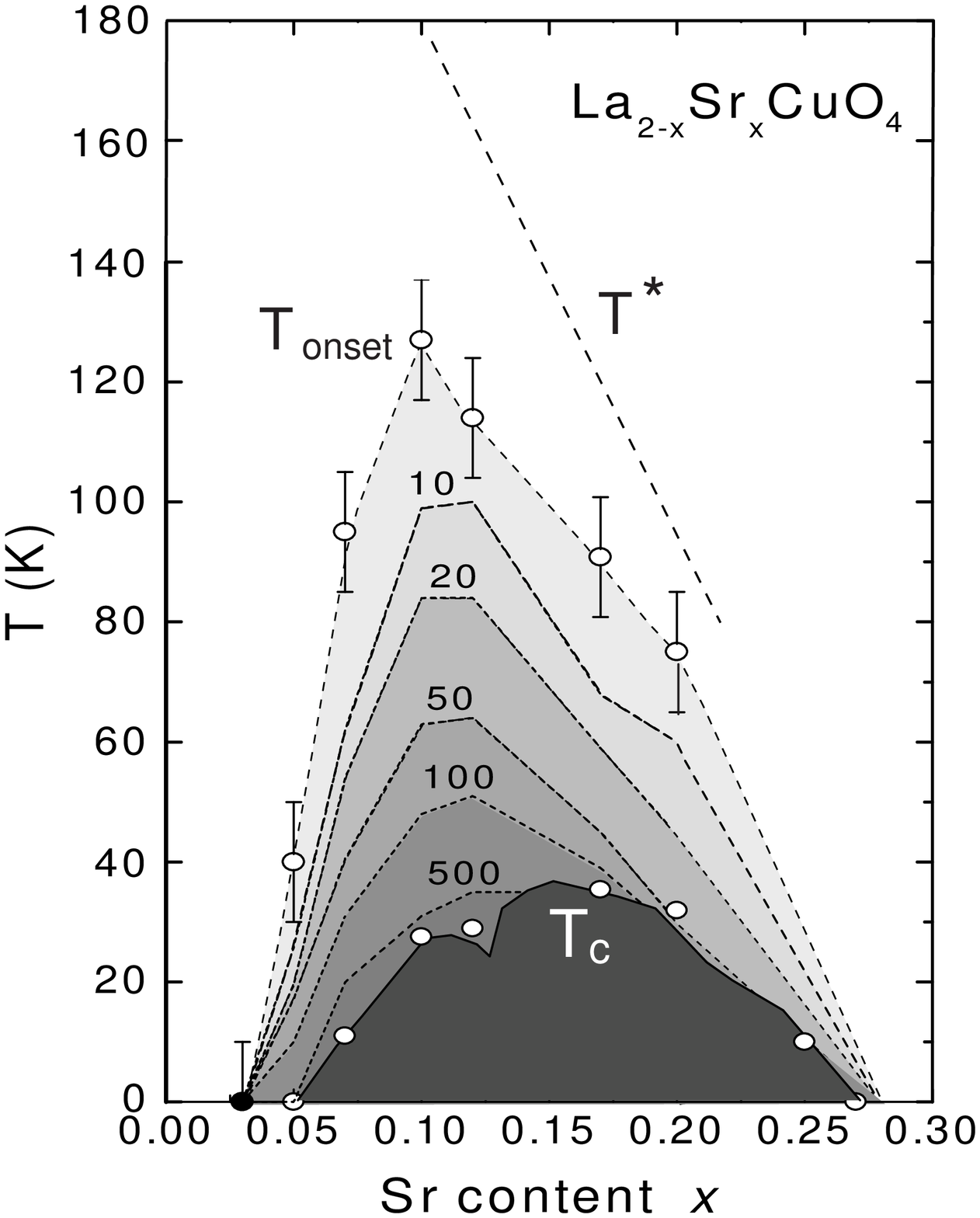}
\caption{\label{LSCO-phase} The phase diagram of LSCO showing the Nernst region
between $T_c$ and $T_{onset}$ (numbers on the contour curves indicate the value of 
the Nernst coefficient $\nu$ in $\mu$V/KT).  The curve of $T_{onset}$ vs. $x$ has end-points
at $x=0.03$ and $x=0.26$, and peaks conspicuously near 0.10.  The dashed line is $T^*$
estimated from heat capacity measurements.
}  
\efig

\bfig
\incl[width=6cm]{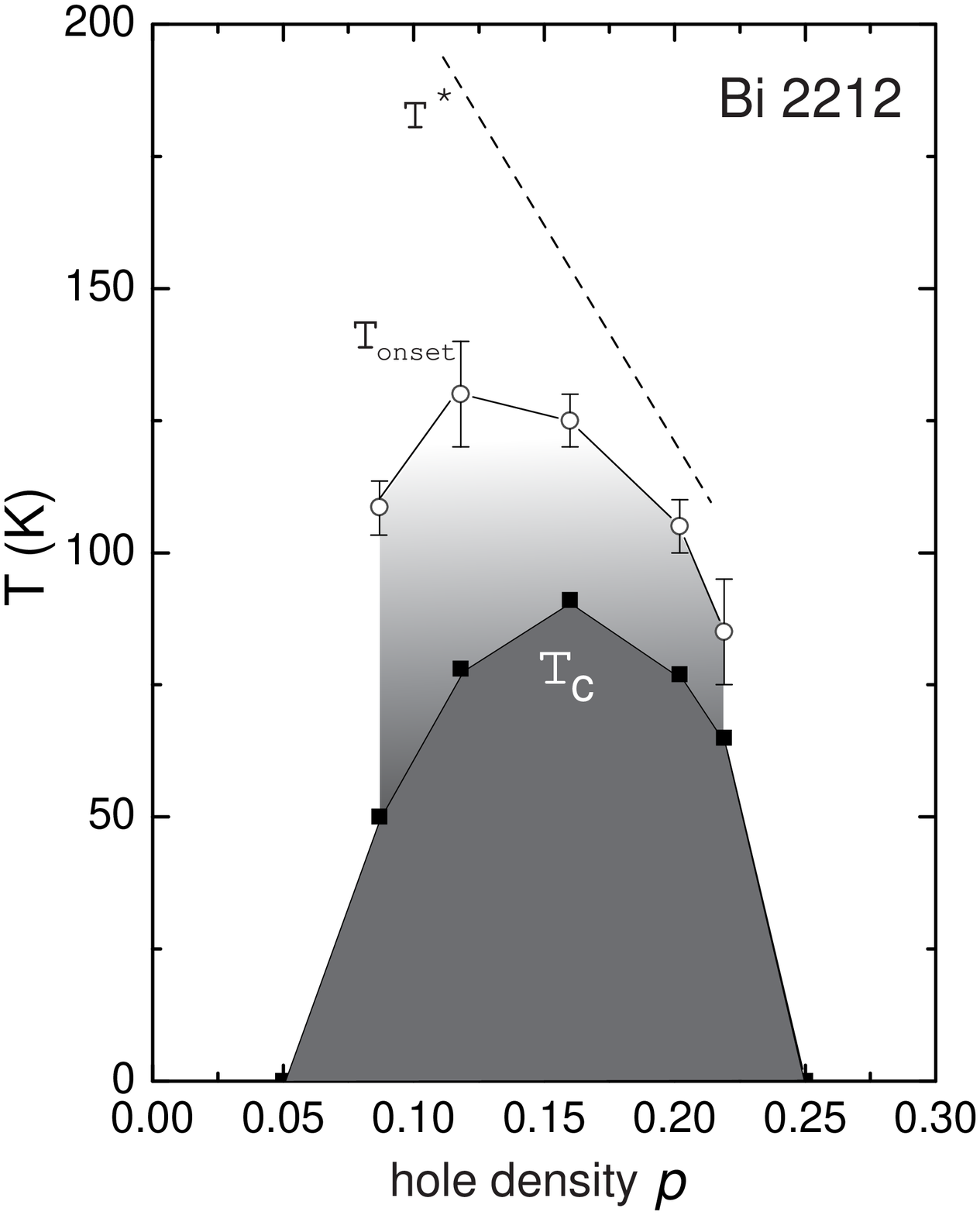}
\caption{\label{Bi-phase} The phase diagram of Bi 2212 showing the Nernst region
between $T_{onset}$ and $T_c$ (based on Nernst measurements on 5 crystals).  As in
LSCO (Fig. \ref{LSCO-phase}), the Nernst region does not extend to the pseudogap
temperature $T^*$ on the OP and OV side.  In the UD regime, $T_{onset}$ shows a
decreasing trend as $x$ decreases below 0.15.
}  
\efig

We turn next to $T_{onset}$ in bilayer Bi 2212.  
In Fig.~\ref{Bi-phase}, we display the variation of $T_{onset}$ in the 5 crystals investigated to date.
The hole density $x$ is estimated from the empirical formula $T_c(x)=T_{c,max}[1-82.6(x-0.16)^2]$, 
with $T_{c,max}$ = 91 K~\cite{oxygen}.  The curve of $T_{onset}$ shares key features with that found in LSCO.
As in LSCO, the superconducting dome in Bi 2212 is 
nested inside the curve of $T_{onset}$ vs. $x$ which lies under the curve of $T^*$.  Whereas $T^*$
appears to continue to increase as $x$ falls below 0.10, $T_{onset}$ deviates downwards in
qualitative similarity with LSCO.  The interval between $T_{onset}$ and $T_c$ becomes
systematically narrower towards the OV side, but it remains quite broad on the UD side.
Interestingly, the maximum value of $T_{onset}$ ($\sim$130 K) is close to the maximum in LSCO,
despite the large difference in maximum $T_c$ in the 2 families.  The maximum value in YBCO is
$\sim$ 130 K as well.  However, in the Hg-based cuprates, evidence from torque magnetometry
suggests that $T_{onset}$ lies higher~\cite{Naughton}.  

In the phase diagrams in Figs. \ref{LSCO-phase} and \ref{Bi-phase}, 
the nesting of the $T_c$ dome within the curve of $T_{onset}$ underscores once more the continuity of 
the region in which the vortex-Nernst signal is observed with the region under the
superconducting dome.  The high-temperature $e_N$ associated with vortices is observed only
inside the superconducting dome.  Once we move outside (either on the UD or OV side), $e_N$
becomes very small.  In LSCO with $x$ = 0.03 and 0.26, the tilted hill profile characteristic of vortex flow
is completely absent.  Instead, the observed $e_N$ is small and $H$-linear 
to fields as high as 33 T, which is characteristic of the qp current.

\bfig
\incl[width=6cm]{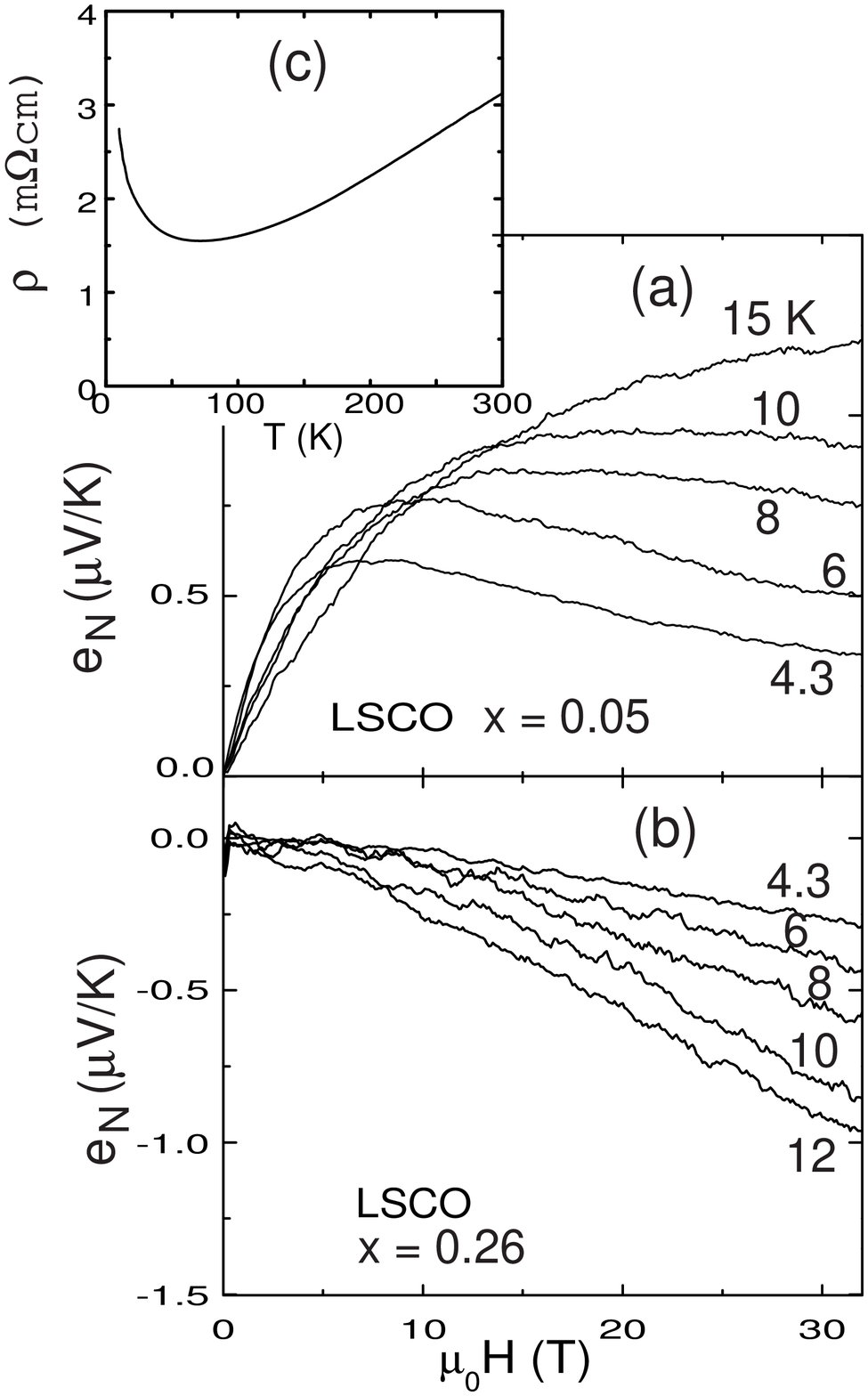}

\caption{\label{NH-05-26} Comparison of the Nernst curves in 
severely UD LSCO ($x$ = 0.05, Panel a) and heavily OV LSCO 
($x$ = 0.26, Panel b).  In both samples, $T_c <$ 2 K and the 
observed Nernst signal is weak. 
However, in the UD sample, $e_N$ retains the ``tilted-hill'' profile 
characteristic of vortices, whereas, $e_N$ in the OV sample shows only 
the negative, $H$-linear qp contribution.  The resistivity profile $\rho$ vs. $T$ 
of the UD sample is shown in Panel c.}  
\efig

On the UD side, the rapid vanishing of the vortex-Nernst signal for samples 
with $x$ = 0.03, 0.05 and 0.07 was already analyzed in detail in Ref.~\cite{WangPRB}.  
Because the vortex signal is rapidly decreasing relative to the qp signal, 
it is necessary to measure the Hall angle and thermopower to separate out the 2 contributions
to the off-diagonal Peltier term $\alpha_{xy} = \alpha^{s}_{xy}+
\alpha^{n}_{xy}$~\cite{WangPRB}.  

\begin{figure*}
\incl[width=17cm]{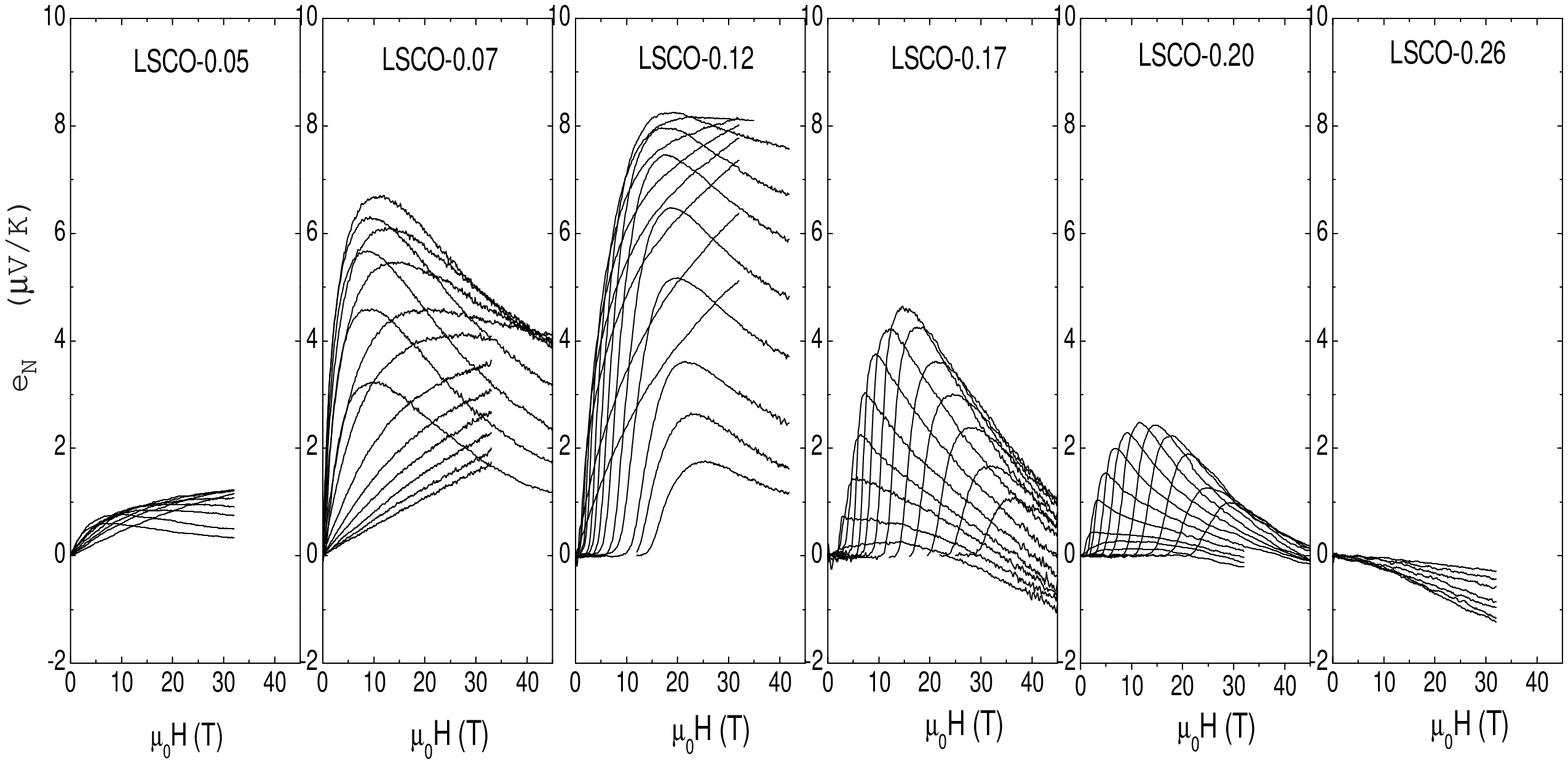}
\caption{\label{LSCO-all} Comparison of curves of $e_N$ vs. $H$ in LSCO 
samples with (from left panel to right) $x$ = 0.05, 0.07, 
0.12, 0.17, 0.20 and 0.26. The same $x$ and $y$ scales are used in all panels. 
In each panel, the envelope curve provides a measure
of the overall magnitude of the vortex signal.
}  
\end{figure*}

It is interesting to compare the Nernst signals at the two extremes of 
the $T_c$ dome.  Figure~\ref{NH-05-26} shows $e_N$ measured in UD LSCO
($x$ = 0.05, Panel a) and in OV LSCO ($x$ = 0.26, Panel b).  In both samples, $T_c<$ 2 K. 
As shown in Fig.~\ref{LSCO-all}, the Nernst signals in these 
two samples are about 10 times smaller than the largest signals observed 
in superconducting LSCO.  
However, these two samples exhibit strongly contrasting Nernst behaviors. 
In the UD sample, $e_N$ is strongly nonlinear in $H$, displaying the ``tilted-hill'' 
profile characteristic of the vortex signal, whereas $e_N$ in the OV sample
shows only the negative, $H$-linear contribution from quasiparticles.  
In the UD sample, phase disordering caused by vortex motion destroys
superconductivity (in $H$ = 0).  However, the pair condensate is robust to intense fields
(the Nernst curves show that $H_{c2}$ is larger than 40 T).  
By contrast, in the OV sample, superconductivity above 2 K is absent 
because the pair condensate is absent altogether.  Moreover, the resistivity 
profile in the UD sample shows an insulating trend below $\sim$80 K (Panel c), 
whereas the OV sample remains metallic down to 2 K.  These differences
reflect the presence of the pseudogap on the UD extreme of the $T_c$ dome,
and its absence on the OV extreme.

The doping dependence of the magnitude of the Nernst signal also reveals an interesting pattern that
complements the previous point (that the vortex $e_N$ is confined to within the dome).  
Figure~\ref{LSCO-all} displays the high-field Nernst results of 6 LSCO samples at various $T$.
The hole density $x$ and $T_c$ values of these samples are
0.05 (0 K), 0.07 (11 K), 0.12 (29 K), 0.17 (36 K), 0.20 (28 K) and 0.26 (0 K) 
respectively (the $x$ and $y$ scales are the same in all panels).

In each sample, the Nernst curves are nested within an envelope which has a peak value. 
In the doping range $0.12 \leq x \leq$ 0.20, the envelope peaks at $\sim \frac12 T_c$, 
whearas in very UD samples ($x$ = 0.05 and 0.07), it peaks above $T_c$.  
As $x$ increases, the peak value rises to the value $e_N ^{max} \sim 8.3\;\mu$V/K near 0.12, 
and falls rapidly as $x$ reaches 0.26.  The variation of the peak 
value $e^{max} _N$ with $x$ is summarized in Fig.~\ref{LSCO-x}, together with the curves of 
$T_{c}$ and $T_{onset}$ in the same samples.

\bfig
\incl[width=6cm]{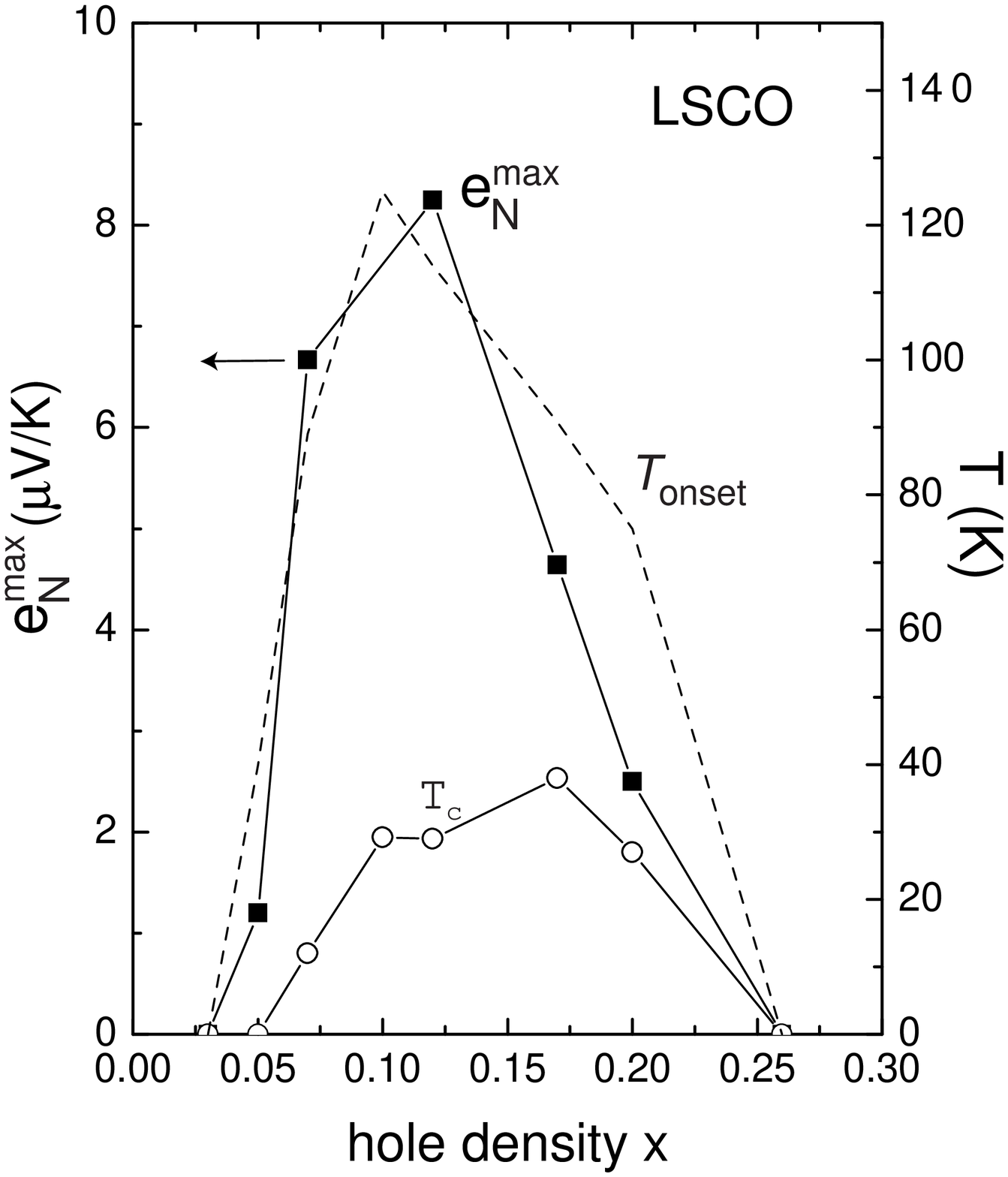}
\caption{\label{LSCO-x} The variation with $x$ of the maximum vortex-Nernst signal $e_N^{max}$ 
(solid symbols), $T_c$ (open circles) and $T_{onset}$ (dashed line) in LSCO.  Solid lines are
guides to the eye.
}  
\efig

The plots in Figs. \ref{LSCO-all} and \ref{LSCO-x} show that the large Nernst signal observed in
LSCO crystals are intimately related to the $T_c$ dome.  When we go beyond the dome
on either side, the peak value $e_N^{max}$ falls rapidly towards zero.  Examination of the panels in
Fig. \ref{LSCO-all} shows that, in each sample, $e_N^{max}$ also dictates the overall scale of 
the Nernst signal both above and below $T_c$.  Hence we deduce that large Nernst signal derives
from the superconducting pairs.  In the UD limit where a large pair density cannot be sustained, 
or in the OV limit when pairing is absent, the Nernst signal vanishes apart from the usual qp term.
The tight correlation between the overall amplitude of the signal and the $T_c$ dome plays an important
role in refuting theories that interpret the large Nernst signal as caused by quasiparticles in some exotic 
state that abuts the superconducting dome (we discuss this in Sec. \ref{discussion}).


\section{Enhanced diamagnetism above $T_c$}\label{magnetization}
The evidence for vortices above $T_c$ described in the preceding sections would
seem to be sufficiently compelling.  However, for reasons already listed (Sec. \ref{vortex}),
it was desirable to seek evidence from non-transport experiments.
In searching for other probes of phase fluctuations, we reasoned that, even if long-range 
phase coherence is destroyed by vortex motion, the large supercurrent
${\bf J}_s$ circulating around the condensate puddles (see Eq. \ref{Js}) should persist on 
length-scales slightly larger than the average vortex spacing $a_B\sim (\phi_0/B)^{\frac12}$.  
Hence, above $T_c$, the magnetization must retain a weak diamagnetic term analogous to Eq. \ref{M}.  
This diamagnetism should be non-Gaussian
and \emph{survive} to 33 T and beyond, if it is to be related at all to $e_N$.  However, in 
prior reports of ``fluctuation diamagnetism" in cuprates~\cite{Bergemann,Lascialfari}
found features that were anomalous and difficult to understand, but one report 
claimed~\cite{Vidal2000} a good fit to conventional Gaussian theory.  
Naughton then drew our attention to his report~\cite{Naughton} of an unexplained
diamagnetic signal in Hg 2212 that persists to 33 T and 200 K.  

Because the diamagnetic component of $\bf M$ is strictly $||\,\bf\hat{c}$ in the phase-fluctuating regime above $T_c$ 
(${\bf J}_s$ is confined within the $\rm CuO_2$ layers), torque magnetometry 
with $\bf H$ tilted slightly away from $\bf\hat{c}$ is ideally suited for resolving a very small $M$ from the
background torque signal (the background arises from the anisotropic spin 
susceptibility $\Delta\chi_p$)~\cite{Wang05}.
In collaboration with Naughton, we have performed extensive measurements of
$M$ above $T_c$ in several crystals of LSCO and Bi 2212 using a 
sensitive Si cantilever magnetometer~\cite{Wang05,Lu05}.

\bfig
\incl[width=6cm]{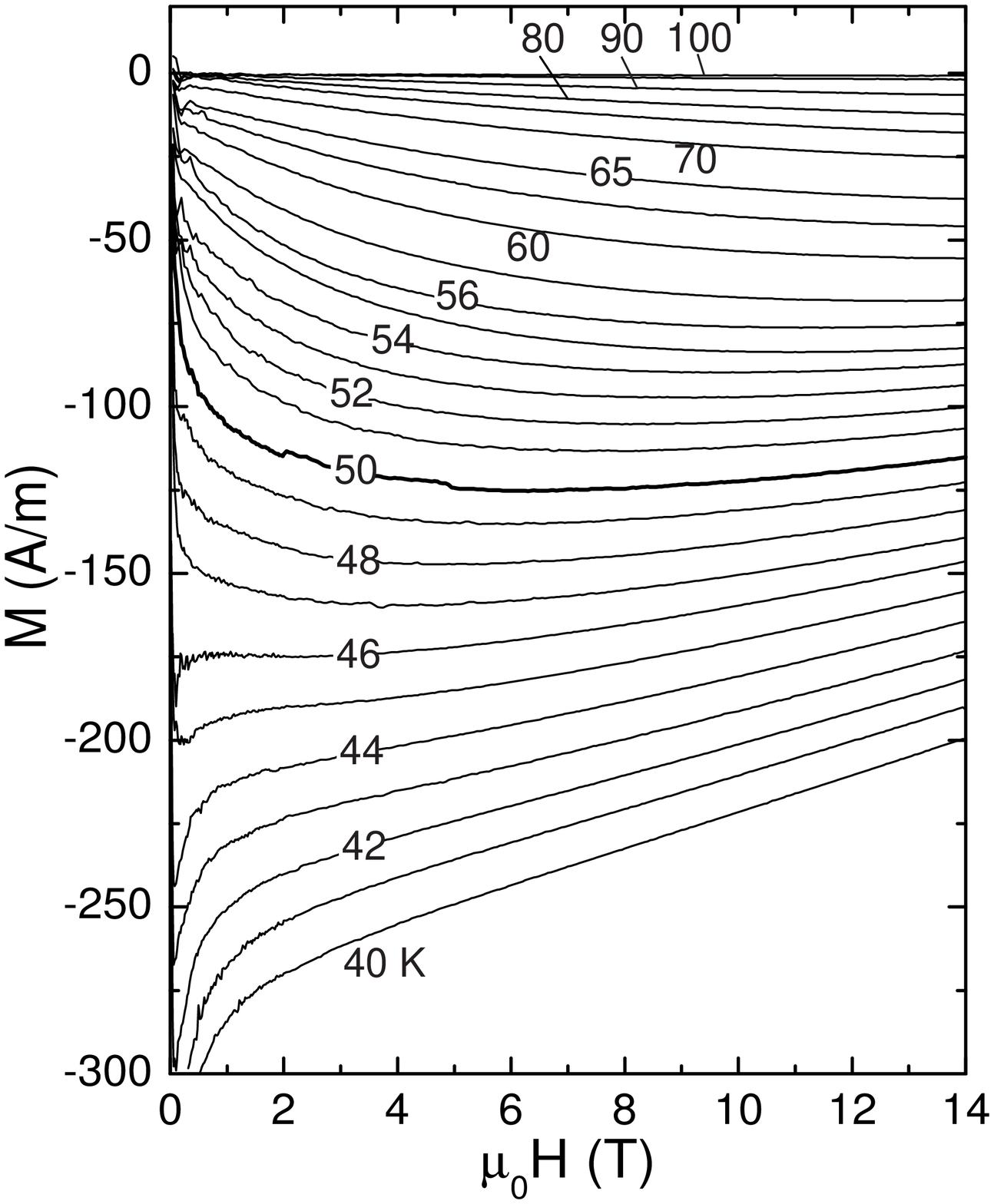}
\caption{\label{MH-B50K} Magnetization curves $M$ vs. $H$ 
in UD Bi 2212 ($T_{c}$ = 50 K) at temperatures from 40 to 200 K.  The
magnetization is extracted as a strongly $T$-dependent diamagnetic
contribution to the total torque signal in a Si cantilever~\cite{Wang05}.
The bold curve is taken at $T_c$ = 50 K.  Even at 70 K, $M(H)$ 
displays non-linearity in $H$.
}  
\efig

Here we briefly report results that are closely relevant to this paper 
(see Ref. \cite{Wang05} for details).  Figure~\ref{MH-B50K} displays $M$ vs. $H$ 
curves of an underdoped Bi 2212 ($T_{c}$ = 50 K) obtained from torque magnetometry.  
The background term $\Delta\chi_p$ which is weakly $T$-dependent has been subtracted.  
The diamagnetic signal starts to appear near 120 K and increases in magnitude over 
a broad 70-K interval above $T_{c}$.   At $T_{c} = 50$ K (bold curve), the diamagnetic signal 
attains the value of -120 A/m.  At even lower $T$, the rapid growth of the 
Meissner effect becomes apparent at low field, and the $M$-$H$ curves resemble those 
in low-$T_c$ type II superconductors.  The Nernst curves measured on this 
crystal was shown in Fig.~\ref{B50K}.

\bfig
\incl[width=6cm]{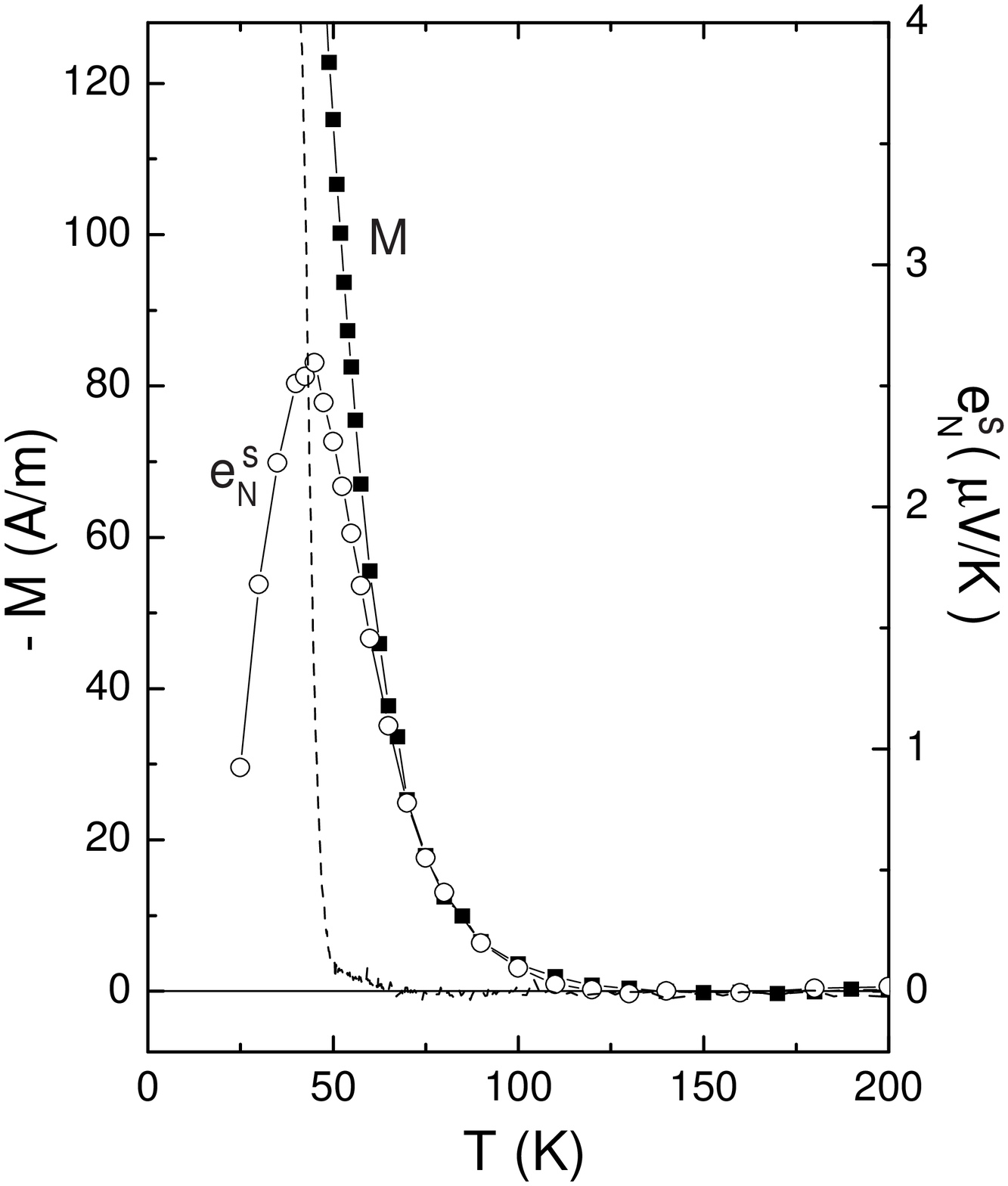}


\caption{\label{M-N-B50K} The $T$ dependence of $e_N^{s}$ (open circles) and the 
diamagnetic signal $-M$ (solid squares) in UD Bi 2212.  Both 
quantities were measured at 14 T in the same crystal ($e_N^{s}$ is the vortex-Nernst signal 
obtained after subtracting the small negative qp contribution $e_N^{n}$ from the total Nernst signal).
The dashed line is the weak-field Meissner effect of this sample measured at $H$ = 10 Oe measured by SQUID.}  
\efig

The magnetization measurements on underdoped Bi 2212 are remarkably consistent with 
the Nernst results. In Fig.~\ref{M-N-B50K}, the temperature dependence of the 
vortex-Nernst signal $e_N^{s}$ and the diamagnetic signal measured at $H$ = 14 T are plotted 
together.  Both deviate from the flat normal background at $T \sim 120$ K, indicating 
the onset of strong fluctuations well above the $T_{c}$ defined by the sharp 
Meissner transition (dashed line).  The 2 signals track each other closely 
over a broad interval of $T$ from 120 K down to $\sim$65 K below which
$e_N^{s}$ attains a peak before sharply falling to zero (as the vortex-solid phase is 
approached).  Scaling between $e_N^{s}$ and $M$ is also observed in OP and 
OV Bi 2212~\cite{Wang05}.  

The scaling observed confirms that the vortex-Nernst effect is accompanied by a weak diamagnetic
response within the CuO$_2$ layers, which we identify with local supercurrents within the interstitial puddles
between mobile vortices (Sec. \ref{hc2}).  The scaling observed above $T_c$ suggests that the linear 
relation between $\alpha_{xy}^{s}$ and $M$ extends beyond the restricted regime of Eq. \ref{beta} 
found by Caroli and Maki, and may have rather broad generality.
We also remark that even at such high $T$, $M$ is robust 
in field, surviving to above 33 T~\cite{Wang05} (the curves in Fig. \ref{MH-B50K}
are displayed to 14 T).  As noted in Ref. \cite{Wang05}, the robustness $M$ here distinguishes it
from ``fluctuation'' diamagnetism arising from amplitude fluctuation familiar in low-$T_c$ 
superconductors~\cite{Gollub}.  We emphasize the importance of the magnetization results
in providing thermodynamic evidence that confirms the transport Nernst results, and refer the 
reader to Ref.~\cite{Wang05}.


\section{Electron-doped cuprate}\label{ncco}
The electron-doped cuprate $\rm Nd_{2-x}Ce_xCuO_{4-y}$ (NCCO) provides
an interesting counter example to the hole-doped cuprates.
Although NCCO shares the layered structure comprised of CuO$_2$ planes, its phase diagram 
differs from that of the hole-doped cuprates.  The 3D antiferromagnetic (AF) state extends up to $x\sim$ 0.15
and the superconducting region is confined to a narrow doping range (0.15--0.17) abutting the AF state.  
A pseudogap phase has not been detected above $T_c$ (below $T_c$ a residual gap is detected if
superconductivity is completely suppressed by a field; whether this is simply the AF gap is
still an open question).

Figure~\ref{NH-NCCO} shows curves of $e_N$ versus $H$ in 
optimally-doped NCCO ($x = 0.15$ and $T_{c}$ = 24.5 K) between 
5 and 30 K~\cite{WangSci,Greene2}.  In this cuprate, the qp contribution to the Nernst signal is large.  
From early Hall-effect experiments~\cite{ZZWang,Greene}
as well as ARPES, it is known that both electron-like and hole-like
bands contribute to the qp current.  The presence of both bands leads to a change in sign
of the normal-state Hall coefficient $R_H$ in OP crystals~\cite{ZZWang,Greene}.
As given in Eq. \ref{eNn}, the qp Nernst signal is the difference of two terms $\rho^{n}\alpha_{xy}^{n}$
and $\rho_{xy}^{n}\alpha^{n}$~\cite{WangPRB}.  For a 1-band system, cancellation between the 2 terms
(dubbed~\cite{WangPRB} Sondheimer cancellation) greatly reduces $e_N^{n}$.  
However, if both holes and electrons are present, this cancellation is
suppressed and $e_N^{n}$ becomes enhanced.  A clear example was recently demonstrated
by Behnia's group in NbSe$_2$~\cite{NbSe2}.

A similar suppression of the cancellation exists in NCCO~\cite{WangSci,Greene2}. 
At 30 K, the qp Nernst coefficient $\nu^{n}\sim$ 0.26 $\mu$V/KT, which is $\sim$10 times larger 
than in that in OP and OV LSCO (and $\sim$50 times larger than in Bi 2201 and Bi 2212).
The observed Nernst signal is the sum of the vortex and qp terms (Eq. \ref{eNs}).
In sharp contrast with results in hole-doped cuprates, the qp term actually 
dominates the Nernst signal at 6 K, far below $T_c$.  Nevertheless, as evident 
in (Fig.~\ref{NH-NCCO}a), the vortex term retains its characteristic tilted-hill profile 
which is easily distinguished from the monotonic qp term.  By fitting the latter to the form 
$e_N^{n} = c_1H + c_3H^3$, with $c_i$ as fit parameters at each $T$ (dashed curves), 
we have extracted $e_N^{s}(T,H)$, which are displayed in Fig. \ref{NH-NCCO}b.

\bfig
\incl[width=6cm]{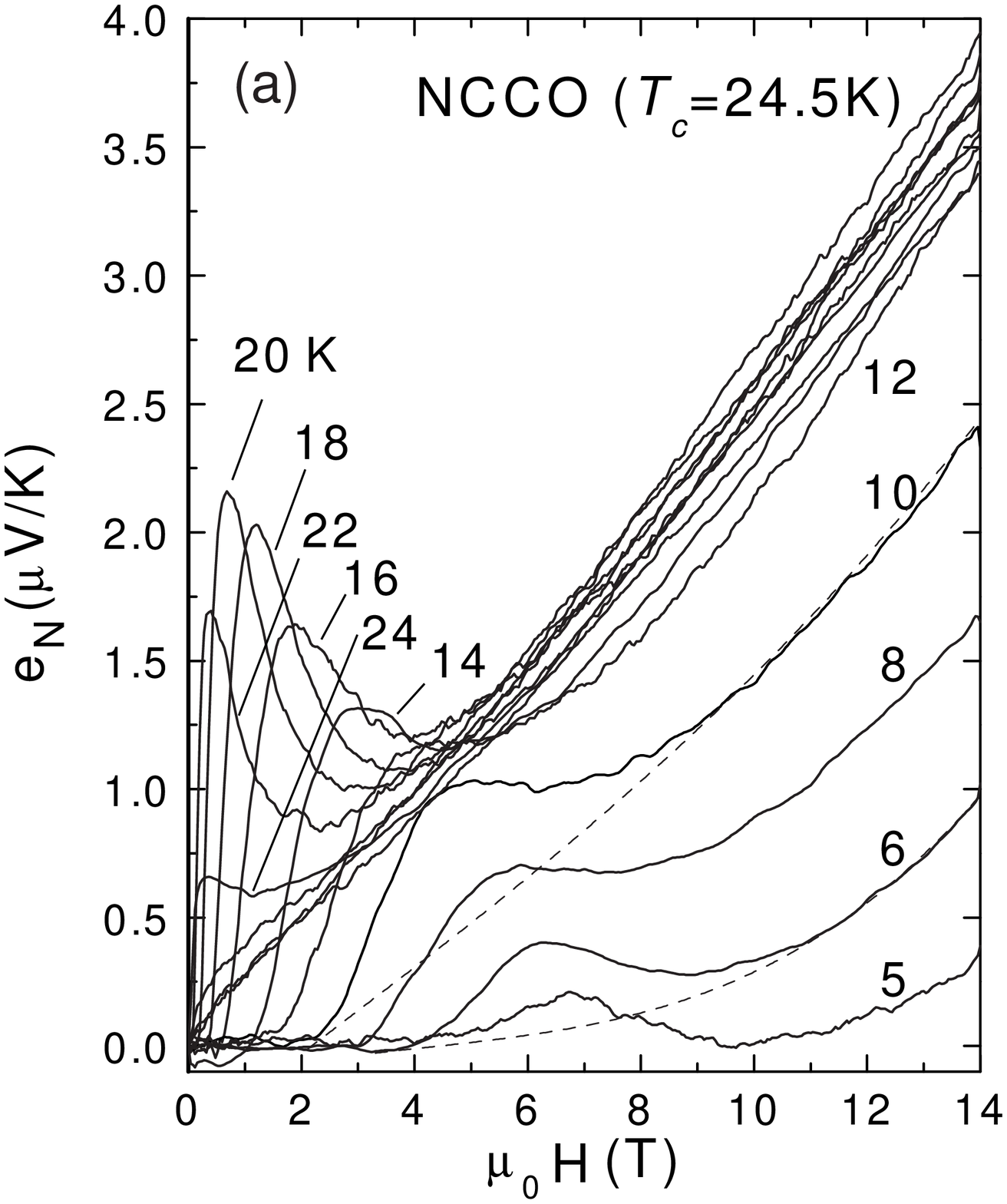}
\incl[width=6cm]{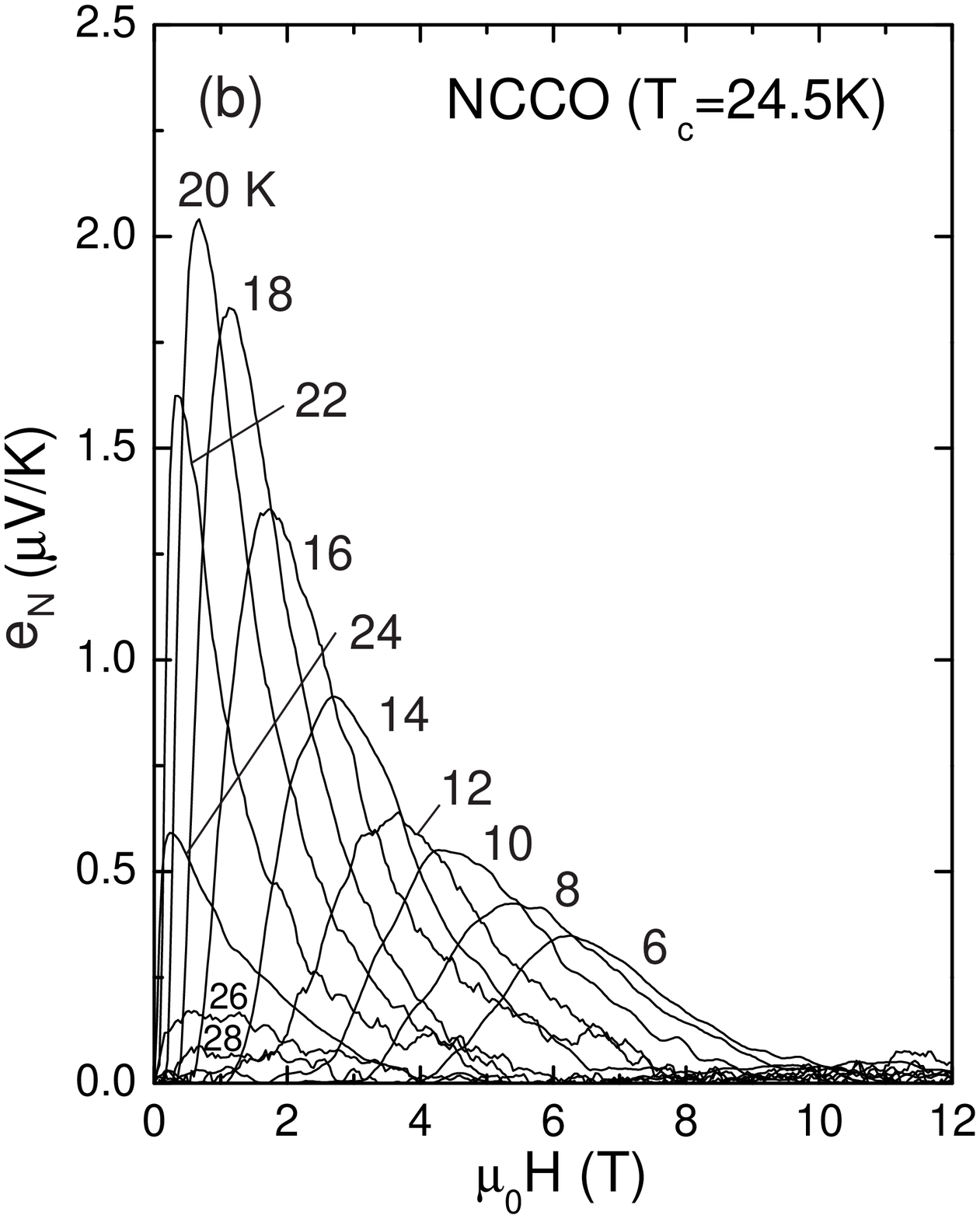}
\caption{\label{NH-NCCO} (a) Curves of the observed Nernst 
signal $e_N$ vs. $H$ in OP NCCO ($x$ = 0.15 and $T_c$ = 24.5 K) at temperatures
5 K to 30 K. The dashed lines are fits of the high-field segments to a qp 
term of the form $e_N^{n}(T,H) = c_1H + c_3H^3$.  
(b) The vortex-Nernst signal $e_N^{s}$ extracted from Panel a
by subtracting $e_N^{n}(T,H)$ from the observed signal.  The contour plot of
this vortex signal is displayed in Fig. \ref{contourNCCO}.
}  
\efig

\bfig
\incl[width=6cm]{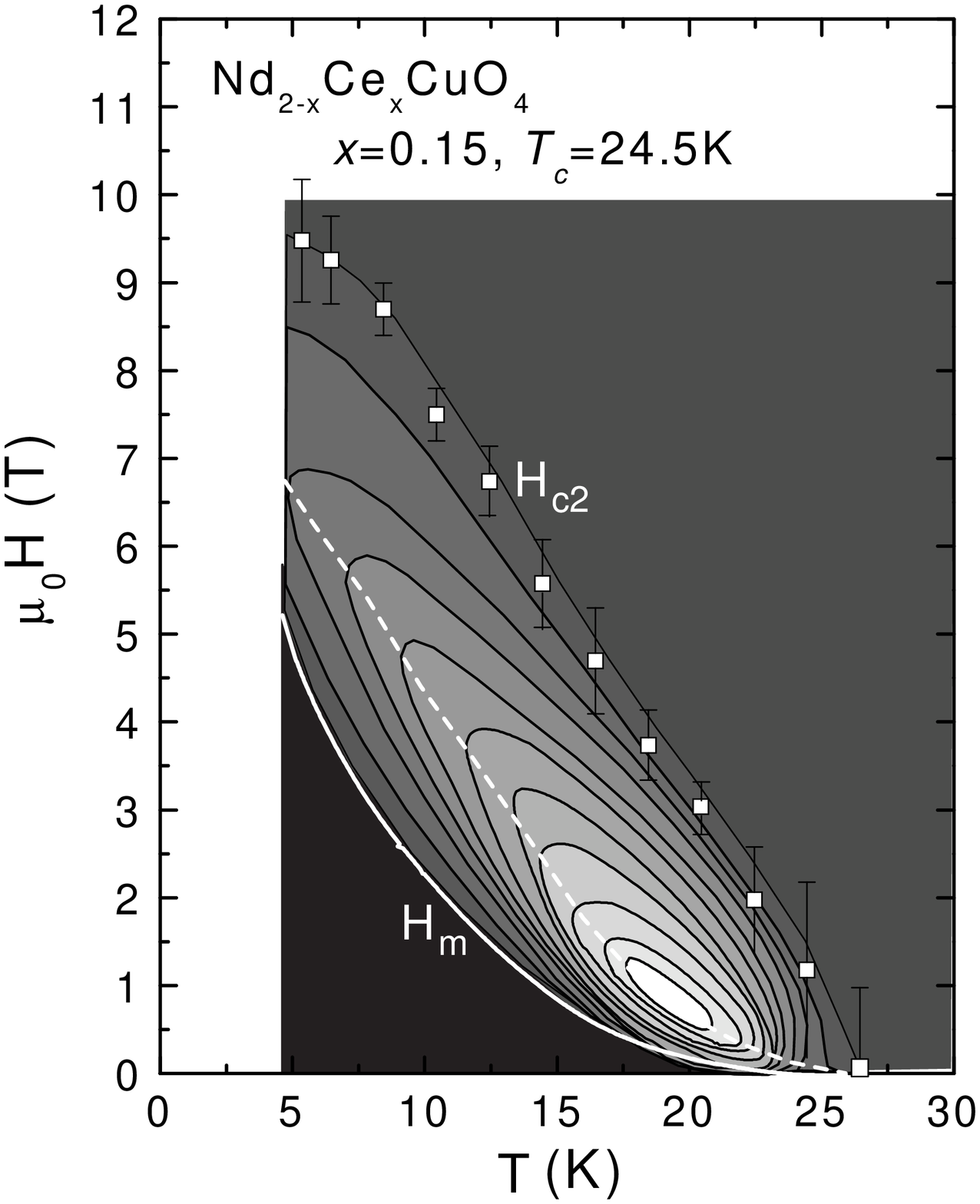}
\caption{\label{contourNCCO}  The contour plot of the vortex-Nernst signal $e_N^{s}(T,H)$ 
in NCCO ($x$ = 0.15, $T_c$ = 24.5 K).  The magnitude of $e_N^s$ is highest in the light-grey
region and zero in the black region below the melting-field curve $H_m(T)$ (white curve).  The 
dashed curve is the ridge joining the maxima of the curves of $e_N^s$ vs. $H$.  The upper critical field values $H_{c2}$
estimated from where $e_N^s\rightarrow 0$ are shown as white symbols.  The absence of
a pseudogap correlates with the vanishing of the vortex-Nernst signal at $T_c$, and the termination
of the $H_{c2}$ curve at $T_c$ (contrast with Fig. \ref{contourBi} for Bi 2201).
}  
\efig

The field profiles of the vortex signal are remarkably similar to those in OP Bi 2201 except that the
depairing field scale is much lower here ($H_{c2}\sim$ 10 T compared with $\sim$50 T).  The
Nernst curves in Fig. \ref{NH-NCCO}b are also nested within an envelope curve that peaks sharply
at 20 K.  

However, there is an interesting difference above $T_c$.  As $T\rightarrow T_c$ = 24.5 K 
from below, the individual hill profiles become narrower in field while the peak value 
decreases.  The decrease in the peak value
of $e_N^{s}$ is so rapid that at 28 K, the vortex contribution falls below our resolution.  
The convergence to zero at $T_c$ of the amplitude is similar to the behavior of $e_N$ observed 
in low-$T_c$ superconductors~\cite{Huebener}, but contrasts sharply with 
that in the UD hole-doped cuprates.
  
The narrowing of the hill profiles also implies that $H_{c2}(T)$ decreases rapidly with 
increasing $T$.  In Fig. \ref{contourNCCO} we plot as open squares $H_{c2}$ 
determined as the field at which $e_N^{s}\rightarrow 0$ at each $T$.  
We find that it goes to zero linearly as
$H_{c2}\sim (1-t)$ with the reduced temperature $t = T/T_c$.  Again, this is 
similar to $H_{c2}(T)$ in BCS theory.  As discussed at length in Sec. \ref{hc2}, 
$H_{c2}$ and $e_N^{s}$ remain at very large values in all hole-doped cuprates as we
cross $T_c$.

Figure \ref{contourNCCO} also shows the contours of the \emph{vortex} signal $e_N^{s}$ in NCCO.
The contrast with the contour plot of Bi 2201 (Fig. \ref{contourBi}) is instructive.  
Instead of spreading outwards to temperatures high above $T_c$, the regions of 
finite $e_N^{s}$ here are confined to the vortex-liquid state between $H_{c2}(T)$ and $H_m(T)$, 
with the long axes of the contour ellipses roughly parallel to $H_{c2}$.  Above $H_{c2}(T)$, $e_N^{s}$ cannot be resolved.  
We determined the melting field $H_m(T)$ (solid white curve) as the line at which $e_N^{s}$ first becomes 
detectable.  The vortex-liquid state occupies a large fraction of the phase below 
the $H_{c2}$ curve.  Aside from this unusual feature, the phase diagram of NCCO is similar to that in a
conventional low-$T_c$ superconductor.

The Nernst results in NCCO are valuable in two aspects.  The modest depairing scale
($H_{c2}\le$10 T) allows the full hill profile of $e_N^{s}$ to be easily distinguished 
even though it is riding atop a larger qp term.  The juxtaposition shows unequivocally 
that there exist 2 very different contributions to the total Nernst signal of cuprates, each with
its distinct $H$ profile.  The close similarity of the profile of $e_N^{s}$ to that of
$e_N$ in Bi 2201 provides evidence that our procedure for extracting $H_{c2}$ 
in the latter is sound.  

More significantly, the comparison reinforces the point that the persistence of
$e_N^{s}$ and $H_{c2}$ above $T_c$ in the hole-doped cuprates is closely 
tied to the pseudogap phenomenon.  In NCCO where the pseudogap is absent (above $T_c$),
the vortex Nernst signal is also absent.  Moreover, the curve of $H_{c2}$
terminates at $T_c$.  The comparison shows that the high-temperature Nernst
phase is not generic to any highly anisotropic layered superconductor with a modest
carrier density (the resistivity anisotropy in NCCO is comparable to that in LSCO and UD YBCO).
It is inherently related to the physics of the pseudogap state in hole-doped cuprates.


\section{Discussion}\label{discussion}

\emph{States above and below $T_{onset}$}
In the cuprate phase diagram, the Nernst region represents an extended area in which 
vorticity -- hence charge pairing -- survives above the curve of $T_c$ vs $x$.
In the hole-doped cuprates Bi 2201, Bi 2212, Bi 2223, LSCO and 
YBCO, the Nernst results establish that $T_c$ is primarily dictated by the loss of phase 
coherence due to spontaneous vortex-antivortex unbinding in $H$ = 0.  In the
Nernst region just above the $T_c$ curve, the phase $\theta({\bf r})$ is strongly disordered 
by rapidly diffusing vortices and antivortices, whereas closer to the curve of $T_{onset}$, fluctuations
in the amplitude become equally important.  It is important to note, however,
that, in each cuprate family, the Nernst region does not extend all the way to the pseudogap
temperature $T^*$.  As shown in Figs. \ref{LSCO-phase} and \ref{Bi-phase}, 
$T_{onset}$ lies roughly between $T_c$ and $T^*$ for doping $x>$ 0.10.  In the
very UD regime ($x<$ 0.1), $T_{onset}$ falls steeply as $x\rightarrow 0$, 
whereas $T^*$ seems to continue to increase.
Our data show that the Nernst region is nested within the pseudogap region.  [The data do not 
seem to support the recent proposal~\cite{PALee} that the curves of $T_{onset}$ and $T^*$ actually
cross near 0.19.]  

This implies that, as we cool an UD sample from room temperature, 
the pseudogap state appears first at $T^*$ but is seen only in experiments 
that couple to the spin degrees.  
Cooling below $T_{onset}$ produces the signals $e_N$ and $M$,
which result from the existence of vortices and a diamagnetic response, both distinct
signatures of short-range supercurrents.  It seems that, in order for the high-$T$ pseudogap 
state to coexist with $d$SC ($d$-wave superconductivity) over the broad interval $T_{onset}>T>T_c$, the two
states must be intimately related, and be distinguished by a subtle difference.  The Nernst
region is where the system smoothly evolves or fluctuates between the two states.

Several interesting theories incorporating this subtle change have 
been proposed.  According to Anderson~\cite{Anderson05}, 
the uniform RVB (resonating valence bond) state is stable above $T_{onset}$, 
but the spin triad defined by $\hat{\Delta}$ and $\hat{\zeta}$ (the anomalous and normal
self-energies, respectively) fluctuates strongly relative to the electron charge triad.  At $T_{onset}$, the two
triads lock to produce a vortex-liquid state that is, however, phase disordered (phase coherence 
occurs at $T_c$).  In the SU(2) formulation of RVB~\cite{PALee,LeeWen}, 
the quantization-axis vector ${\bf \hat{I}}(\vartheta,\varphi)$ of the slave-boson spinor 
$\left[\begin{array}{c} b_1 \\ b_2 \end{array}\right]$ 
distinguishes the staggered-flux pseudogap state ($\vartheta \simeq 0,\pi$)
from $d$SC ($\vartheta = \frac{\pi}{2}$).  Above $T_{onset}$, $\bf \hat{I}$ points mostly towards the
poles, but in the Nernst region, $\bf \hat{I}$ fluctuates away from the poles, eventually 
coming to lie in the equatorial plane below $T_c$. 
In the striped model, the competing state involves quasi-1D dynamic stripes~\cite{Kivelson}.  
Vortex excitations are also fundamental to other recent theories of the pseudogap/charge-ordered 
states above the $T_c$ dome~\cite{Tesanovic,Weng,Zhang,Sachdev,Balents}.
We anticipate that detailed experiments on $e_N$ and $M$ in intense fields, in combination with
STM experiments above $T_c$, should allow these theories to be tested.

\emph{Cheap and fast vortices} 
An important issue raised by these results is the energy cost of
creating the vortices.  In the limit $\kappa\gg 1$ in BCS theory, the energy of a vortex 
line of length $d$ arises chiefly from the superfluid kinetic energy and is given by~\cite{deGennes}
$E_V = \phi_0^2d/(4\pi\mu_0\lambda^2) \ln{\kappa} = \pi K_s\ln{\kappa}$.
Here, it is important to add to this the core energy $E_c$. 
The total energy $E_{pc}$ of a vortex pancake is then~\cite{PALee,Honerkamp}
\be
E_{pc} = E_c + 2\pi K_s \ln{L/\xi}.
\label{core}
\ee
As $T\rightarrow T_c$ from below, the superfluid term in $K_s$ vanishes.  However, $E_c$
does not in the phase-disordering scenario.
In BCS theory, $E_c$ is the loss of condensation energy inside an 
area $\xi^2$, viz. $E_c\sim \Delta_0^2\xi^2/(\epsilon_Fa^2)\sim\epsilon_F$, 
in the clean limit~\cite{PALee} (with $\epsilon_F$ the Fermi energy).  Hence, $E_c$ 
is at a very high energy scale relative to $T_c$.  Because the vortex 
unbinding temperature depends primarily on the stiffness term in $K_s$ in Eq. \ref{core}
and is insensitive to $E_c$, this observation does not affect $T_c$.  
However, a large ratio $E_c/k_BT_c$ implies that the 
spontaneous vortex density should remain very small over a broad interval above $T_c$
which is inconsistent with magnetization and transport experiments.  
The inconsistency has been used~\cite{PALee,Lee1,Honerkamp} to argue 
that the state stable inside the core is actually much closer in energy to $d$SC than the 
true normal state (this is known as the cheap-vortex problem).  
Lee and coworkers propose that this is the sF state~\cite{PALee,Honerkamp}.  

As discussed in Sec. \ref{phase}, $\rho$ rises very steeply above $H_m$ to saturate near $H_{ridge}$
long before $H_{c2}$ is reached.  Ioffe and Millis~\cite{Ioffe} have investigated how proximity to the 
Mott insulator influences the coupling between quasiparticles and the supercurrent and dissipation
inside the vortex core.   They propose that a small damping $\eta$ results from the small number of states in the cores.  
The weak damping leads to a high velocity of the vortices transverse to $\bf I$, and a 
large flux-flow resistivity (or small vortex conductivity $\sigma^s$).  
Additivity of the vortex and qp charge currents~\cite{HarrisOng} implies that, eventually, the observed
conductivity $\sigma^s+\sigma^{n}$ is dominated by the qp term $\sigma^{n}$.  This seems to account
for the steep rise of $\rho$ followed by rapid saturation.

\emph{Gaussian limit}
As mentioned in Sec. \ref{vortex}, phase fluctuations are classified as either analytical (spin-wave) 
$\Delta\theta_a$ or singular (vortex) $\Delta\theta_v$.  The Gaussian-fluctuation theory, based on an 
expansion in small $|\Psi|$ of the action $S$, leaves out the essential role of $\Delta\theta_v$ in 
destroying superfluidity.  Ussishkin \etal~\cite{Iddo} have investigated the extent to 
which $e_N^{s}$ measured in LSCO may be described by Gaussian theory applied to a generic 
layered, extreme type-II superconductor.  In the 2D limit, they calculate that, above $T_c$, $\alpha_{xy}^{s}$ 
has the mean-field Aslamazov-Larkin (AL) form familiar from fluctuation diamagnetism, 
viz. $\alpha _{xy}^{s} \sim B (1-t)^{-1}$, which provides a reasonable fit 
to $e_N^{s} = \rho\alpha _{xy}^{s}$ in the OV regime (using the measured 
$\rho(T)$).  However, the Gaussian expression fits poorly in the OP 
and UD regimes even when unrealistically large values are used for the 
in-plane $\xi$.  

Because Gaussian theory does not handle singular phase fluctuations and 
the phase-disordering scenario, it cannot describe the anomalous behavior 
of $H_{c2}$ described above.  The poor fits in the OP and UD regimes are 
perhaps unsurprising.  However, in a restricted range of temperatures just below the curve 
of $T_{onset}$ in Fig. \ref{LSCO-phase} where
amplitude fluctuations must be dominant it serves as a useful quantitative guide~\cite{Iddo2}.
Numerical simulations of the 2D time-dependent Ginzburg Landau (TDGL) equation 
show reasonable fits to the high-field Nernst results in OV LSCO~\cite{Subroto}. 

The separate issue of whether any generic quasi-2D superconductor should display
a large Nernst signal above $T_c$ is interesting. 
The electron-doped NCCO, with an anisotropy and $\rho$ 
comparable to OP LSCO, and $\xi _{ab}$ ($\sim$60 \rm \AA)
2.5 times larger, should display an even larger Nernst signal above $T_c$ (according
to the Gaussian theory).  However, this is not the case (Sec. \ref{ncco}).  The presence
or absence of the pseudogap state is a much more important discriminant in cuprates.

\emph{Quasi-particle models} 
We discuss some of the proposed models in which $e_N$ above $T_c$
is attributed to quasiparticles.  It has been argued~\cite{Kontani} that, if strong 
antiferromagnetic fluctuations exist in a Fermi liquid, vertex corrections cause the qp 
current $\bf J_k$ to deviate from being normal to the Fermi Surface, and a consequent 
enhancement of $\nu$.  Also, an enhanced qp $\nu$ is purportedly obtained in an 
unconventional $d$ density-wave ($d$DW) model~\cite{Dora}, as well as from paired holes 
in ``anti-phase'' domains in an antiferromagnetic state~\cite{Hu2}.

These models introduce a rather exotic qp property or ground state tailored 
to account for the Nernst data in a restricted interval of $T>T_c$, but ignore 
the (known) correlations of the data with other properties over a much larger parameter space.  
For e.g., it is difficult to see how the qp signal can smoothly evolve into the vortex signal 
below $T_c$ (Fig. \ref{contourBi}).  It is equally difficult to imagine how the 
unusual qp states/properties abruptly cease to be effective
once we move out of the $T_c$ dome (Fig. \ref{LSCO-x}).  
The extended high-field results reported here compound these problems.  
The vortex hill-profile which persists above $T_c$ (Fig. \ref{NH-Bi04}),
the scaling of $e_N$ with $M$ (Fig. \ref{M-N-B50K}), the anomalous behavior of $H_{c2}$ 
(Fig. \ref{contourBi}), and the
contrasting case of NCCO (Fig. \ref{contourNCCO})
all present serious challenges for the qp models.
(Further, a recent calculation~\cite{Vadim2} has shown that the qp Nernst signal
in the $d$DW state is actually too small to account for the observed $e_N$.)

Finally, in a proposed ``bipolaron'' model, even the Nernst signal observed \emph{below}
$T_c$ has been identified as coming from (``localized'') quasiparticles~\cite{Alexandrov04}.  
The extreme view is proposed that $H_m$ represents the depairing field~\cite{Alexandrov96},
so that the condensate is destroyed as soon as $\rho$ becomes 
non-zero.  As we stressed in discussing $\rho$ in Sec. \ref{hc2}, loss of phase 
stiffness should be carefully distinguished from the destruction of the condensate.
The ubiquitous ``tilted-hill'' profile observed below $T_c$ and the robustness of $M$ 
observed to intense fields~\cite{Wang05,Lu05} provide simple, direct evidence 
refuting the basic assumption in this model.

\section{Summary and conclusions}\label{summary}
In the hole-doped cuprates, we uncover a large region above the ``superconducting dome" 
in which an enhanced Nernst signal exists.  The upper limit of the Nernst region is defined by $T_{onset}$
which lies nominally half way between $T_c$ and the pseudogap scale $T^*$ (Sec. \ref{phase}).  The Nernst signal 
is consistent in sign and magnitude with the phase-slip 
$E$-field caused by a vortex current driven by the applied gradient
(and incompatible with a ferromagnetic origin~\cite{WLee} by orders of magnitude).
At each $T$ within this region, the Nernst signal $e_N$ is manifestly non-linear in $H$ and
closely similar in shape to the tilted-peak profile that characterizes the vortex-Nernst signal observed below $T_c$
(Sec. \ref{profile}).  This profile is strikingly incompatible with a qp origin, given the very short qp $\ell$.
Overall, the enhanced Nernst signal above $T_c$ displays a smooth continuity with the 
vortex-liquid state below $T_c$, which is best seen in the contour plot of $e_N(T,H)$ in the $T$-$H$ plane
(Sec. \ref{profile}).  

An enhanced magnetization signal is observed above $T_c$
that scales accurately with $e_N$ measured in the same crystal (Sec. \ref{magnetization}).  
The magnitude of $M$ is significantly larger than that anticipated from Gaussian fluctuations.  
Moreover, it remains robust to intense fields like $e_N$ even very near $T_c$.  
With the magnetization result, the vortex liquid above $T_c$ has now been detected 
by both transport and thermodynamic experiments.

The direct implication of these results is that the loss of superfluidity 
and the collapse of the Meissner state at $T_c$ occurs because long-range phase 
coherence is destroyed by the thermal generation of vortices and anti-vortices, 
which implies that the pair amplitude $|\hat{\Psi}|$ persists to temperatures 
much higher than $T_c$.  This phase-disordering is the 3D analog of the KT 
transition in 2D systems.  

For this scenario to be self-consistent, the depairing field $H_{c2}$ must remain at 
a large finite value at $T_c$ -- as previously noted for the KT transition~\cite{Doniach} --
instead of decreasing to zero as $(1-T/T_c)$.   
Utilizing the vortex profile for $e_N^{s}$, we have determined that $H_{c2}$
behaves anomalously, remaining large as $T_c$ is crossed, consistent with the vortex
scenario (Sec. \ref{hc2}).  This implies that, in the plane $(T, H)$, the critical point ($T_c$, 0) serves 
as the termination point of the melting-field curve $H_m(T)$, but not of the depairing 
field scale (Fig. \ref{contourBi}).  This contrasts with the phase diagram in NCCO
(Fig. \ref{contourNCCO}), in which ($T_c$, 0) serves as the termination point of both 
$H_m(T)$ and $H_{c2}(T)$.  The latter is the canonical behavior in the BCS gap-closing scenario.  
The extension of the vortex liquid to above $T_c$, together with the anomalous 
behavior of $H_{c2}$ constitute the most striking signatures of the phase-disordering scenario.


\section*{Acknowledgments}
We thank Z. Xu, Yoichi Ando, S. Ono, S. Uchida, Genda Gu, Y. Tokura, Y. Onose, B. Keimer, 
R. X. Liang, D. A. Bonn and W. N. Hardy for helpful collaborations in these experiments.  We are especially indebted
to M. J. Naughton for drawing our attention to Ref. \cite{Naughton}, and for the loan of 
cantilever magnetometers.  We thank P. W. Anderson, J. C. Davis, D. A. Huse, S. A. Kivelson, P. A. Lee, K. Levin, A. J. 
Millis, V. N. Muthukumar, V. Oganesyan, J. Orenstein, S. Sachdev, S. Sondhi, Z. Tesanovic, I. Ussishkin, Z. Y. Weng, A. 
Yazdani, and S. C. Zhang for many helpful discussions. The high-field measurements were performed at the National High Magnetic Field 
Laboratory, Tallahassee, a facility supported by the U.S. National Science Foundation (NSF) and the State of 
Florida. We are grateful to Scott Hannahs for technical assistance.  This research is supported by 
NSF (Grant DMR 0213706) and by the New Energy and Industrial Technol. Development Org. NEDO (Japan).\\

\noindent $^*$\emph{Present address of YYW: Department of Physics, University of California at Berkeley, Berkeley, CA 
94720-7300}.\\ Electronic addresses: yywang@berkeley.edu, luli@princeton.edu, npo@princeton.edu



\end{document}